\documentclass[pdflatex,sn-standardnature]{sn-jnl-arXiv}
\usepackage{amsmath,amssymb,amsfonts,booktabs,multirow,subcaption}

\usepackage{xr-hyper} 
\usepackage[font={scriptsize}]{caption}
\makeatletter

\usepackage{caption,float}
\usepackage{graphicx}
\usepackage[numbers]{natbib}
\usepackage{textcomp, setspace}
\usepackage{xcolor,arydshln}
\usepackage{amsthm,supertabular}

\usepackage{lmodern}

\newcommand{\bt}[1]{\textcolor{black}{#1}}

\DeclareMathOperator{\sign}{sign}
\newcommand{\ModelName}{JPoNG}

\graphicspath{{Nature Energy/}}

\jyear{2023}%

\theoremstyle{thmstyleone}%
\theoremstyle{thmstyletwo}%

\theoremstyle{thmstylethree}%

\begin{document}

\title[{ }]{Cost-effective Planning of Decarbonized Power-Gas Infrastructure to Meet the Challenges of Heating Electrification}


\author[1,2]{\fnm{Rahman} \sur{Khorramfar}}\email{khorram@mit.edu}
\equalcont{Equal contributions with names in alphabetical order}
\author[1]{\fnm{Morgan} \sur{Santoni-Colvin}}\email{msantoni@mit.edu} 
\equalcont{Equal contributions with names in alphabetical order}
\author[2,3]{\fnm{Saurabh} \sur{Amin}}\email{amins@mit.edu}
\author[4]{\fnm{Leslie K.} \sur{Norford}}\email{lnorford@mit.edu}
\author[2]{\fnm{Audun} \sur{Botterud}}\email{audunb@mit.edu}

\author[1]{\fnm{Dharik} \sur{Mallapragada}}

\affil[1]{\orgdiv{MIT Energy Initiative}, \orgname{Massachusetts Institute of Technology}, \orgaddress{\city{Cambridge}, \state{MA, 02139}}}

\affil[2]{\orgdiv{Laboratory for Information \& Decision Systems}, \orgname{Massachusetts Institute of Technology}, \orgaddress{\city{Cambridge}, \state{MA, 02139}}}

\affil[3]{\orgdiv{Civil and Environmental Engineering}, \orgname{Massachusetts Institute of Technology}, \orgaddress{\city{Cambridge}, \state{MA, 02139}}}

\affil[4]{\orgdiv{Department of Architecture}, \orgname{Massachusetts Institute of Technology}, \orgaddress{\city{Cambridge}, \state{MA, 02139}}}


\abstract{Building heat electrification is central to economy-wide decarbonization efforts and directly affects energy infrastructure planning through increasing electricity demand and {reducing the building sector's} use of gas infrastructure that also serves the power sector. Here, we develop a modeling framework to quantify end-use demand for electricity and gas in the buildings sector under various electrification pathways and evaluate their impact on co-optimized bulk power-gas infrastructure investments and operations under deep decarbonization scenarios. Applying the framework to study the U.S. New England region in 2050 across 20 weather scenarios, we find high electrification of the residential sector can increase sectoral peak and total electricity 
demands by up to {56-158\% and 41-59\%} respectively relative to business-as-usual projections. 
Employing demand-side measures like building envelope improvements under high electrification, however, can reduce the magnitude and weather sensitivity of peak load as well as induce overall efficiency gains, reducing the combined {residential sector energy demand for power and gas} by {28-30\%} relative to the present day. Notably, a combination of high electrification and envelope improvements yields the lowest bulk power-gas system cost outcomes. Accounting for \bt{midstream} methane emissions from gas supply chain increase the reliance on low-carbon fuels, which indirectly improves the cost-effectiveness of end-use electrification. Similarly, we find that demand flexibility programs can reduce the total system cost by up to 6.3\%. 
}


\maketitle

\section{Introduction}

Economy-wide decarbonization by mid-century requires accelerated CO$_2$ emissions reduction efforts across all sectors of the economy. Yet, to date, the building sector has seen limited progress in decarbonization efforts relative to other sectors. For example, in the U.S. between 2005 and 2021, power sector CO$_2$ emissions declined by 34\% while the total emissions in the building sector declined by only 18\%  \citep{EPA-sources-GHG}. Natural gas (NG) remains an important contributor to the sector's emissions.  As of 2021,  NG accounted for 80\% of {on-site} fossil fuel consumption in the residential sector, or 5\% of total U.S. GHG emissions, with heating being the dominant use case \citep{eia-emis-end-use}. This share of emissions can be higher in cold climate regions with more NG heating; for example, in New England, residential NG usage accounted for 8\%  \citep{EIA-residential-consumption} of economy-wide emissions in 2021. The electrification of space heating via efficient air-source heat pumps (ASHPs), combined with a low-carbon power sector, is one of the most widely recognized and viable pathways to considerably reduce buildings sector emissions \citep{EIA-net-zero-report, GaurEtal2021}. Accordingly, federal and state policies increasingly aim to facilitate the deployment of heat pumps \citep{IRAinfo, MA-climate-plan, ACEEE-building}. 

The electrification of space heating, however, introduces new challenges for long-term planning and operations of electricity and gas infrastructure under emissions constraints. First, electrification is anticipated to increase the magnitude and alter the pattern of electricity consumption by buildings while displacing NG consumption \citep{EFSETA2022,DeetjenEtal2021,VaishnavFatimah2020}. For example, prior studies of heat electrification suggest that cold climates like the U.S. Northeast will experience a 2.5-3.9 times increase in state-level peak electricity demands {for the building sector}, with the peak occurring in the winter rather than the present summer-peaking conditions \citep{WaiteModi2020}.
Importantly, the projected impacts of heating electrification on final energy demand for electricity and NG are strongly dependent on the building stock as well as technology deployment assumptions including the \textit{size} of the heat pumps. For example, under a \textit{whole-home electrification} approach, ASHPs are sized to serve heating needs throughout the year, including during the coldest periods. In contrast,  a  \textit{hybrid electrification} approach would use smaller ASHPs coupled with backup heating fuel, where the latter is used during especially cold periods to reduce peak electricity consumption and requisite power infrastructure investment but {may continue} to need the gas infrastructure \citep{WaiteModi2020}. While these strategies have each been studied individually \citep{DeetjenEtal2021, VaishnavFatimah2020, WaiteModi2020, WuEtal2022}, heterogeneous consumer behavior and policy will result in the simultaneous adoption of multiple sizing strategies across the housing stock. Furthermore, the bulk of the literature has overlooked the role of building envelope improvements in reducing the demand impacts of either electrification strategy.


A second challenge is that the altered electricity and gas consumption patterns due to heat electrification will also impact the utilization of bulk gas infrastructure by reducing its relevance for heating and potentially changing its role for the grid \citep{KhorramfarEtal2022, VonWaldEtal2022}. \bt{The economy-wide decarbonization studies overlook the impact of electrification on gas infrastructure and often rely on the coarse spatiotemporal representation of the power grid \citep{WilliamsEtal2021,WiseEtal2019}.}
To date, only a few studies have considered coordinated planning of regional power-gas infrastructure under high electrification and emissions constraints \citep{KhorramfarEtal2022, VonWaldEtal2022}. Third, heating electrification uniquely increases the sensitivity of electricity demand to weather \citep{DeakinEtal2021, Eggimann2020, Peacock2023}, which, along with the weather-dependence of variable renewable energy (VRE) electricity supply, complicates decarbonized grid infrastructure planning \citep{StaffellPfenninger2018, PereraEtal2020, DeakinEtal2021}. For example, resource adequacy assessments of decarbonized grids with high levels of heat electrification need to consider the extent to which cold-weather peak events correlate with VRE production, which requires considering inter-annual weather variations. 

Several previous studies have used planning models to evaluate power sector decarbonization pathways under scenarios of high heating electrification \citep{WhiteEtal2021, EFSETA2022}. However, these grid-centric studies overlook the resulting impacts of heating electrification on the gas infrastructure and the potential cost impacts from the reduced use of gas for heating.
Moreover, the impact of inter-annual weather variations and end-use technology adoption on final energy demand from buildings under high heating electrification and its effects on bulk energy infrastructure needs has not been extensively analyzed. This paper presents a quantitative framework to assess the bulk power and gas infrastructure implications of building heat electrification under deep decarbonization scenarios while considering inter-annual weather variations and a range of technological interventions on both demand and supply side. Our key contributions include: i) {conducting a spatio-temporally resolved evaluation of electrification's impact on the residential demand for power and gas under several end-use adoption scenarios}; ii) {modeling the} power and gas infrastructure interactions under emissions constraints, including quantifying the system impacts and costs of shifting demand from gas to electricity in the building sector; and iii) {considering} weather-induced temporal variations in demand and supply for both gas and electricity and implications for power system balancing. 

\begin{figure}
    \centering
    \includegraphics[width=1\textwidth]{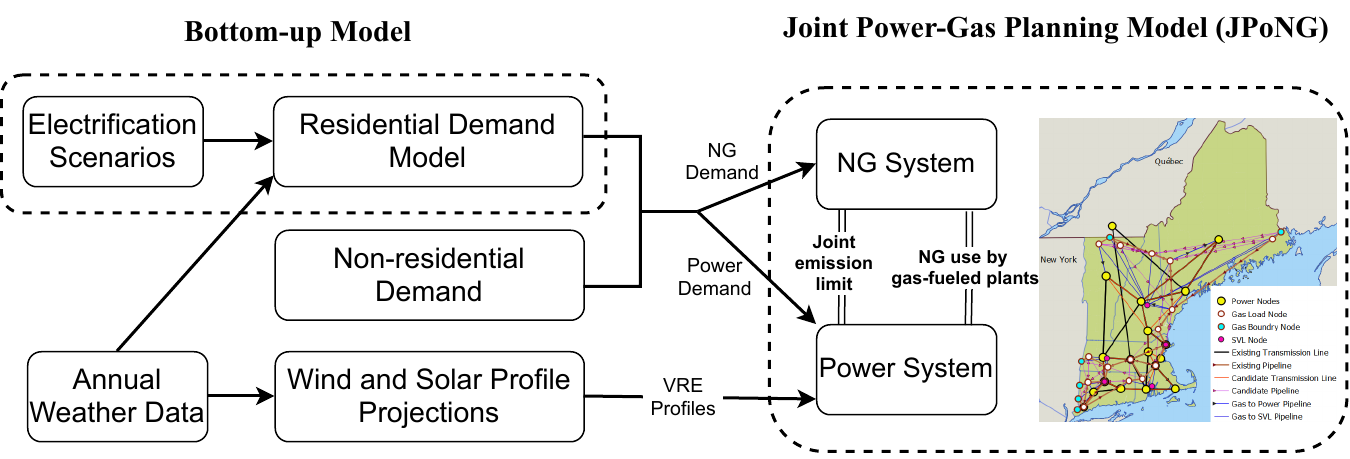}
    \caption{Overview of the modeling framework to evaluate the impact of residential heating electrification on \bt{energy infrastructure and operations planning outcomes}. Residential demand is explicitly modeled in the study to consider the impact of various demand-side technological interventions. Non-residential power and gas demand is held constant across all scenarios evaluated as per the projections for high electrification scenarios available from another study \citep{NREL2021ElecReport}.}
    \label{fig:bottomup-jpong-map}
\end{figure}

Our work also includes the following key methodological contributions. First, we develop a bottom-up model to project electricity and gas demand for residential buildings under several electrification pathways and weather scenarios that explicitly consider building stock heterogeneity, heat pump sizing, and inter-annual weather variations in the analysis. Second, {based on the initial model proposed in \citep{KhorramfarEtal2022}}, we formulate an optimization model for joint planning of electricity and gas infrastructure that is capable of exploring strategies for deep decarbonization of the joint system under various scenarios of electrification output from the bottom-up modeling. The updated model accounts for weather-induced temporal variations in demand and supply as well as the availability of alternative technologies including renewable generators (VRE and hydropower), short-duration battery storage (Li-ion), low-carbon fuels (LCF) that can substitute for NG, gas generation with and without carbon capture and storage (CCS), {demand-side flexibility} and transmission expansion (see Fig.~\ref{fig:bottomup-jpong-map}).

Applying our models to the case study of the New England region in the U.S. in 2050 reveals the following major observations. \textbf{First}, \bt{the aggressive electrification of heating in the residential housing stock (approximately 80\% of housing units) without envelope improvements results in up to a 56 to 158\% increase in peak sectoral electricity demand at the regional level relative to business-as-usual scenario. This peak increase can be mitigated by implementing envelope improvements or less aggressive electrification strategies. Envelope improvements also reduce the length of cold-weather peak events. } The aggregate peak sectoral demand change masks significant sub-regional variations - for instance, less populated regions (e.g. Vermont, Maine and rural New Hampshire) experience the greatest relative increases in peak load. \bt{Additionally, deploying electric resistance backup heating systems would likely create untenable increases in peak load.} \textbf{Second}, high electrification (i.e. electrifying $\sim$80\% of homes with an emphasis on whole-home electrification) coupled with envelope improvements results in the lowest combined {residential power and gas demand} that is up to {28-30\%} lower than values in the year 2020 and reduces inter-annual variation in peak electricity demand. \textbf{Third}, combining high electrification of the residential sector with building envelope improvements incurs the lowest cost of bulk energy infrastructure (i.e. not including housing stock retrofit or distribution infrastructure costs) needed to meet sectoral emissions reduction of 80-95\% relative to 1990 levels.


\textbf{Finally}, while cost-optimal power sector decarbonization strategies rely heavily on VRE supply (mostly wind), they are also accompanied by transmission expansion ({10-41\%} higher than the present), short-duration storage and low-carbon firm power resources such as gas-based generation with and without CCS, fueled by a combination of LCF and fossil gas (i.e., natural gas, NG). Notably, across the scenarios, the levels of total NG consumption are less sensitive to the extent of electrification than the level of {power-gas system} decarbonization, implying substitution effects between NG use in buildings and power. At the same time, the role of LCF is found to grow significantly under certain electrification and decarbonization scenarios, such as those with the inclusion of methane emissions associated with gas supply chain.

\section{Demand Impacts of Heating Electrification}
\begin{figure}[!htbp]
\centering
\begin{subfigure}[b]{.52\textwidth}
  \centering
  \includegraphics[width=1\textwidth]{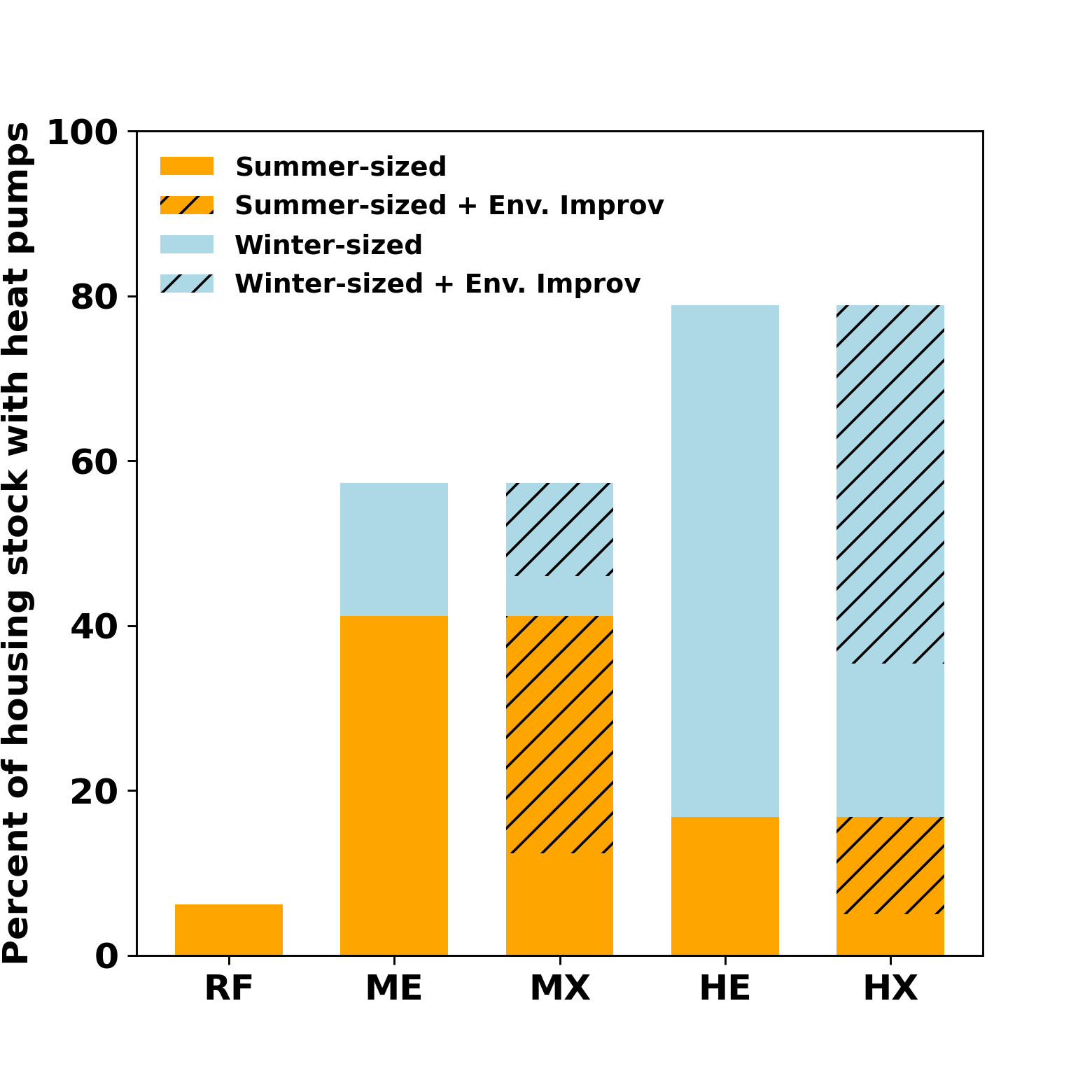}
  \caption{}
  \label{fig:techscenariossmall}
\end{subfigure}%
\begin{subfigure}[b]{.5\textwidth}
  \centering
  \includegraphics[width=1\textwidth]{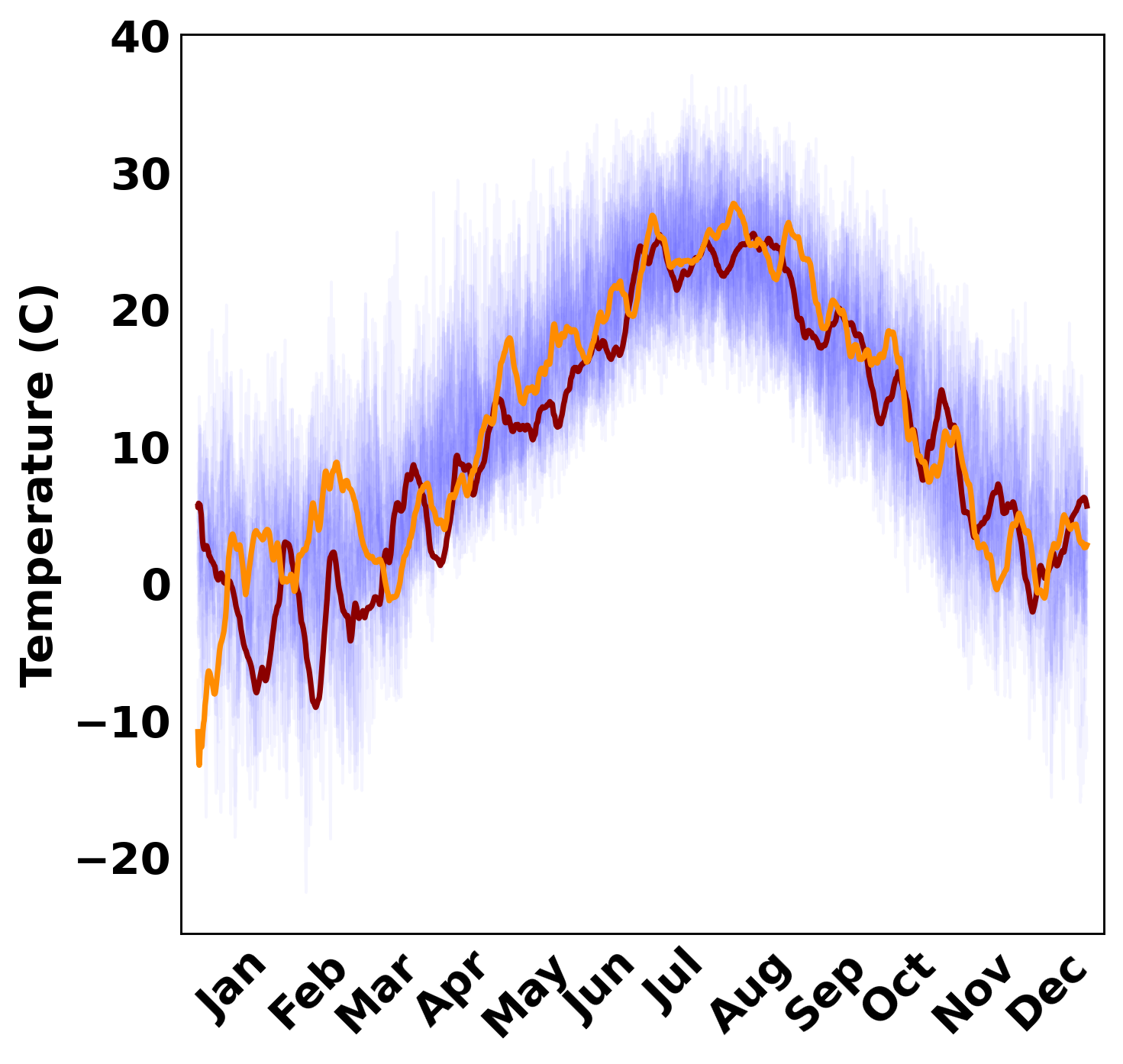}
  \caption{}
  \label{fig:weatherts}
\end{subfigure}
\caption{\textbf{(a)} Scenarios for electrification of residential building heating for 2050. ``Winter-sized'' homes have large whole-home heat pumps that are sized to provide heat through the winter and are used as the only source of heating and cooling. Smaller ``Summer-sized'' heat pumps are primarily sized to meet air conditioning needs in the summer, but also provide some heating in the winter, supplemented by a backup heating system. In our analysis, ``Summer-sized'' systems are almost all also hybrid systems because the existing heating systems in New England homes are typically gas or oil. \bt{We base our heat pump deployment scenarios on the Massachusetts Clean Energy and Climate Plan (CECP) for 2050, with our Medium Electrification scenarios corresponding to the CECP “Hybrid” scenario and the High Electrification scenarios corresponding to the CECP “High Electrification” scenario \citep{MA-climate-plan}}. In the MX and HX scenarios, we assume 70\% of all {homes given heat pumps} also receive envelope improvements. See further details in SI \ref{SIsec:deploymentscenarios}. 
     \textbf{(b)} Temperature profiles in Suffolk County, MA for the 20 weather years used in the analysis. Light blue lines in the background illustrate hourly temperature values for all 20 years overlaid on top of one another to illustrate the weather variations. Orange and bold red lines highlight weekly rolling averages of temperatures for two different years drawn from the dataset. {A more detailed depiction of the 20 years of temperature data can be found in Fig. \ref{SIfig:tempdists}.}} 
\end{figure}

We consider five residential electrification scenarios for 2050, each with different mixes of heat pump adoption, sizing, and envelope improvements that include basic retrofits to improve the thermal efficiency of the building exterior (see SI~\ref{SIsec:deploymentscenarios} and \ref{SIsec:envimprov}) as shown in Fig.~\ref{fig:techscenariossmall}. The first four scenarios include medium and high electrification levels (ME, HE) and variations of them with envelope improvements (MX, HX) that are parameterized to mimic 2050 electrification policy goals in the region. We also evaluate a reference scenario with lower electrification levels to reflect business-as-usual adoption rates (RF). We evaluate power-gas demand for these scenarios under 20 different projections of annual weather data that are based on historical weather patterns and adjusted for the effects of climate change through 2050 (Section \ref{ssec:Bottomup}), which we refer to as \textit{weather years}. As an example, Fig. \ref{fig:weatherts} depicts the diversity of the temperature profiles for the full set of weather years for the Boston area. We also simulate so-called ``present-day'' hourly demand estimates using our bottom-up method in order to compare the 2050 demand changes to current system conditions {approximately representing 2020} and to benchmark the accuracy of our modeling approach (see SI \ref{SIsec:presentdaymodel}).


\subsection{Substitution of NG Demand by Electricity Demand}

\begin{figure}[!htbp]
     \centering
     \begin{subfigure}[b]{0.85\textwidth}
             \centering    \includegraphics[width=\textwidth]{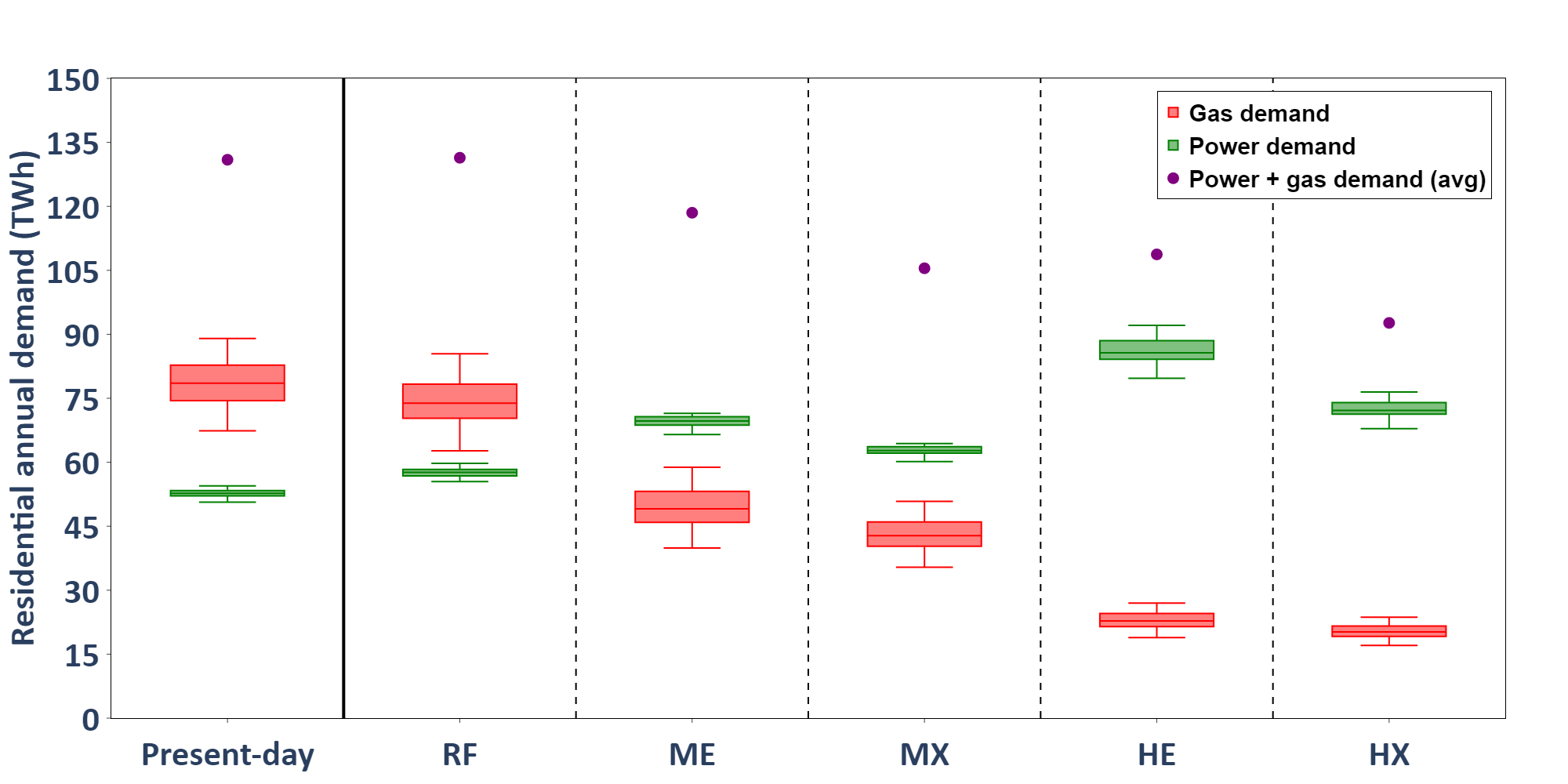}
    \caption{}
    \label{fig:annual}
     \end{subfigure}
     \hfill
    \begin{subfigure}[b]{0.85\textwidth}
             \centering    \includegraphics[width=\textwidth]{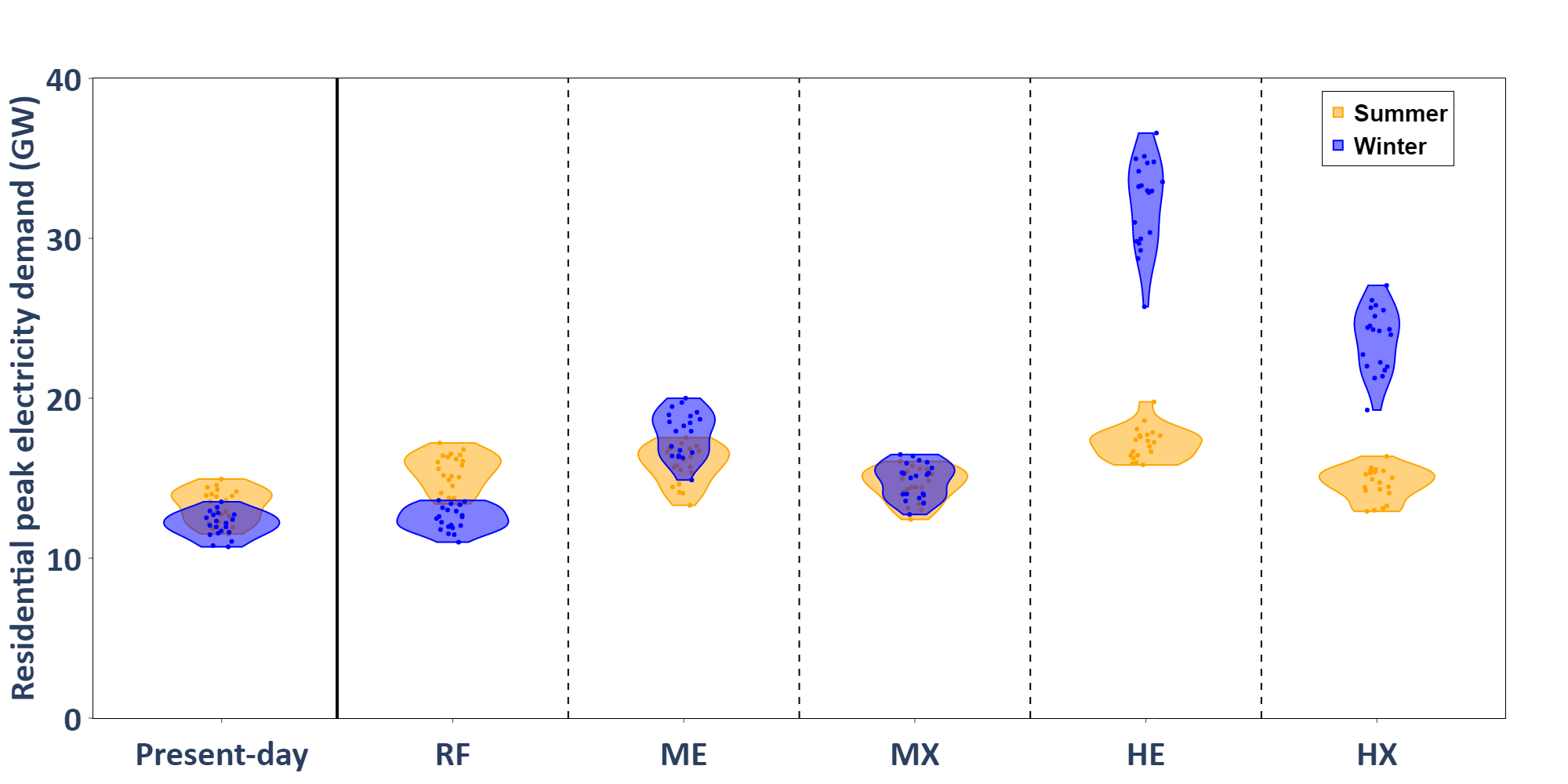}
        \caption{}
        \label{fig:violin}
     \end{subfigure}
     \caption{\textbf{(a)} Annual electricity and gas demands in residential sector under different electrification scenarios. Each box plot contains demand data for the 20 weather years. HE, HX, ME, MX, and RF are all 2050 scenarios. {The box edges correspond to interquartile range (IQR) and the whiskers indicate the most extreme points within the range Q1 - 1.5*IQR to Q3 + 1.5*IQR}. 
     \textbf{(b)} Summer and winter peak demands for the residential sector under different electrification scenarios. Each violin contains demand data simulated for 20 weather years. The width of plots increases where the density of data is higher, and its height represents sensitivity toward the weather variation. For New England, our "present-day" model and data workflow result in overestimation of residential gas consumption as compared to published outputs of ResStock runs by NREL \citep{ResstockComstock2022} as well as historical annual values available from EIA (shown in Figure~\ref{SIfig:histvspresent}). As discussed in SI \ref{SIsec:validation}, the difference between our present-day model results and EIA data is likely attributable to a combination of a) differences in weather data used in our analysis {(see Fig. \ref{SIfig:ERA5vsNREL})}, b) the limited number of building archetypes we consider to maintain computational tractability, and c) error within the ResStock model. The analysis presented in \ref{SIsec:validation} suggests that the error in the present-day model does not as strongly impact the demand results from our future electrification scenarios.}
        \label{fig:demandresults}
\end{figure}

With increasing amounts of heating electrification in the housing stock, Fig.~\ref{fig:demandresults}a shows that annual residential NG consumption declines while electricity consumption increases.  
For scenarios HE and ME, {residential} electricity demand increases over the present day by an average of {33.2 TWh (63\%) and 16.8 TWh (32\%)} across all weather years, respectively. The impact of these electrification scenarios on {site-level gas demand (excluding power generation)} is a reduction of {55.4 TWh (71\%) and 29.2 TWh (38\%)}. Due to the higher efficiency of heat pumps compared to the existing heating stock, the {total combined end-use} energy demand {for power and gas} also declines relative to the present day with increasing electrification, by {16-18\% and 28-30\%} for the HE and HX scenarios respectively.
At increasing levels of electrification, inter-annual variation in demand decreases for gas but increases for electricity because greater amounts of the heating load are being met by electricity. Deploying envelope improvements in parallel with heat pumps reduces the increase in electricity consumption; consumption of electricity is reduced by an average of {16\% between scenarios HE and HX, and by 10\%} between scenarios ME and MX. Envelope improvements also induce average gas demand savings of {11\% and 13\%}, respectively. 

Since our modeled electrification scenarios include some hybrid heating systems that leverage the existing gas heating system as backup, their operation alone is responsible for {2.8-8.8 TWh} of gas consumption across the weather years in the high electrification scenarios (HE, HX) and {6.9-21.6 TWh} for the medium electrification scenarios (ME, MX). 

 

\subsection{Changes in Residential Electric Peak Demand}



Similar to other studies on heat electrification, we find that peak electricity demands, which are generally an important metric for grid planning purposes, increase with electrification and become more weather-dependent, as highlighted in Fig.~\ref{fig:violin}. The weather dependence of peak demand is illustrated by the fact that the maximum and minimum annual peak demands across the 20 weather years differ by {3.0 GW for the present-day and 10.9 GW for HE}. {Because the peak demand generally coincides with extreme cold weather, this result suggests that higher levels of electrification will magnify the adverse peak demand impacts during periods of extreme cold.} Importantly, increases in peak electricity demand at high electrification are greater than the increases in annual demand - for example, peak electricity demand and annual electricity demand for the HE scenario, which has the highest demand, are {56-158\%} and {41-59\%} greater than the RF scenario %
across 20 weather years, respectively. 

Demand-side measures like envelope improvements can partly mitigate increases in peak demand with electrification as well as reduce its sensitivity to weather variations. Envelope improvements provide reductions in peak demand by an average of {8.5 GW} for HX compared to HE (resulting in a peak {17\% to 92\%} higher than RF) and {2.5 GW} for MX compared to ME (resulting in a peak {between 8\% lower and 21\%} higher than RF). They also reduce the duration of peak electricity demand events\bt{, which may be important to reducing the amount of firm resources needed on the system}. These events are characterized as the consecutive hours immediately preceding and following the seasonal peak, during which demand exceeds {75\%} of the demand recorded in the peak hour. The longest peak demand event \bt{across the 20 weather years considered} can last up to {60} hours in the case of the HE scenario but is reduced to {33} hours or lower in the HX scenario (Fig.~\ref{SIfig:Peaklength}). The supply-side results shown in the next section indicate that the reduction in the duration and magnitude of peak demand enabled by envelope improvements leads to reduced power capacity investment and lower total power-gas system costs under deep decarbonization scenarios.


Fig.~\ref{fig:violin} highlights that high electrification of the residential sector alone results in sectoral peak demands that are comparable to the present-day New England system-wide peak demands of 24.4-26.0 GW \citep{ISONE2023}.  While the residential sector's average summer peak in the present-day scenario currently exceeds the winter peak by {an average of 1.2 GW}, the residential winter peak exceeds summer peak by an average of {0.3, 2.0, 9.0, and 15.0 GW} under the MX, ME, HX, and HE scenarios, respectively. New England's system-wide summer peak currently exceeds the winter peak by an average of 5.4 GW \cite{ISONE2023}. Thus, {depending on peak timing in other sectors,} HE and HX can drive New England into a winter-peaking system due to the effects of residential electrification alone. 


Installing electric resistance heat as backup in lieu of hybrid systems with fuel backup can further reduce fuel consumption without needing a larger heat pump, \bt{which may come at reduced fixed costs for households}. However, using electric resistance backup dramatically increases peak electricity consumption. Figure~\ref{SIfig:ElecBackupPeaks} shows that using electric resistance backup instead of existing backup further increases peak loads by up to {184\%} across the various electrification scenarios.
This increase in peak electricity demand, which is particularly drastic for the ME and MX scenarios for their higher deployment of smaller heat pumps, may not be economical from a supply-side perspective as it may reduce utilization of dispatchable electricity generation, \bt{illustrating a potential misalignment between customer costs and system costs}.


\begin{figure}[!htbp]

\centering
\begin{minipage}{.5\textwidth}
  \centering
  \includegraphics[width=1\textwidth]{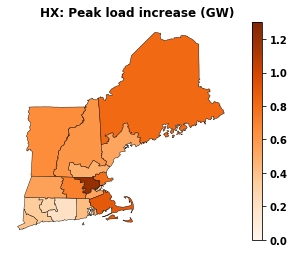}
  \captionof*{figure}{(a)}
\end{minipage}%
\begin{minipage}{.5\textwidth}
  \centering
  \includegraphics[width=1\textwidth]{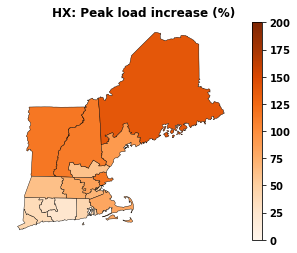}
  \captionof*{figure}{(b)}
\end{minipage}
\caption{Average \bt{(statistical mean)} residential peak load increase across the 20 weather years for scenario HX above present-day for 17 zones in New England. A summary of the zone geography is presented in SI \ref{SIsec:Powernodes}. Numerical results shown in Table \ref{SItab:spatialdemand}. \textbf{(a)} Results in GW and
     \textbf{(b)} on percent basis.}
     \label{fig:loadincmap}

\end{figure}

While regionally aggregate peak demand is an important metric for transmission planning,  the spatial distribution of peak loads is important for distribution network planning. Fig. \ref{fig:loadincmap} shows the average \bt{peak load} increase, both in absolute and percent terms, for scenario HX across the weather years for each load zone. We find that while \bt{absolute growths}, as shown in Fig. \ref{fig:loadincmap}a, are distributed diversely across the region, Fig. \ref{fig:loadincmap}b reveals that \bt{percentage growths} relative to the present day are greater in the more rural, colder regions. For example, the state of Vermont incurs an average {residential peak demand growth of 115\%} under the HX scenario with a population of 647,064 \cite{VTQuickFacts}, while Suffolk County, where Boston is located, has a much lower peak growth of {45\%} with a population of 766,381 \cite{SuffolkQuickFacts}. A key potential implication is that the demand increase in rural areas could require deeper distribution network buildout relative to the population of ratepayers who will likely incur the cost through retail tariffs. Conversely, the more densely populated southern regions of New England may not need as dramatic network upgrades, the costs of which can also be spread across the larger population.





\section{Supply-side Impacts of Heating Electrification}
We evaluate the impact of different electrification scenarios and inter-annual weather variation on bulk power-gas system outcomes using {a substantially enhanced} version of the \textit{\ModelName}  model \cite{KhorramfarEtal2022}. The model co-optimizes bulk power and gas infrastructure to minimize the sum of their annualized investment and operational costs while adhering to constraints related to technology operation, resource availability, and policy, namely emissions constraints and resource adequacy requirements (see details in Section~\ref{ssec:JPoNG} and SI~\ref{SIsec:model-formulation}). We evaluate model outcomes for an emissions goal of 80\% sectoral emissions reduction relative to 1990 levels, consistent with regional policy goals for 2050, as well as a more ambitious policy of 95\% emissions reduction \cite{E3Team2020, MAdecarbRoadMap2020}. Key model decision variables relate to investment and operation of VRE, short-duration battery storage, power, and gas transmission as well as gas generation with and without CCS. For the latter, we model two gas fuel options: NG imports as well as well as imports of LCF. The latter is presumed to be carbon-neutral and compatible with the gas infrastructure but is available {at cost premium relative to NG according to a supply curve (10-50 \$/MMBtu vs. \$5.45/MMBtu - see Table~\ref{tab:key-JPoNG-params} in Section~\ref{sec:methodology})}. LCF is meant to serve as a proxy for synthetic or biogenic methane that has been considered by other studies modeling regional and national decarbonization \citep{MAdecarbRoadMap2020,MITEI2022FES,ColeEtal2021}. {We do not model hydrogen blending owing to its uncertainty related to technological readiness, costs (e.g. to retrofit the pipelines) and emissions impacts (e.g. due to H$_2$, CH$_4$ leakages).}

\subsection{Cost-effective resource portfolios} \label{ssec:cost-effective-resource-portfolio}
\begin{figure}[!htbp]
    \centering
    \includegraphics[width=\textwidth]{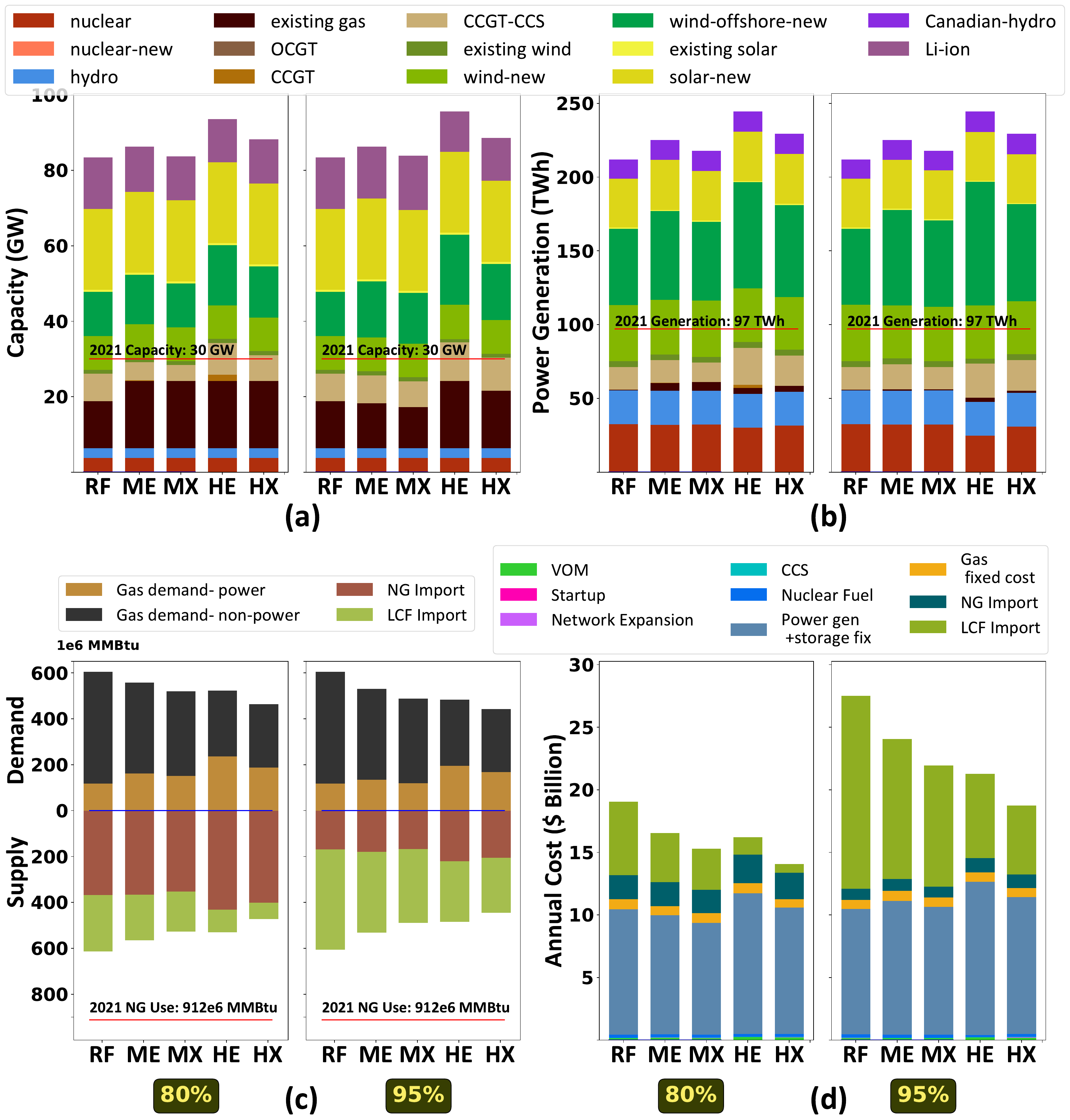}
    \caption{{System outcomes for the power and gas systems in 2050 where the load projection is based on an exemplary weather year.} (\textbf{a}) Power capacity, (\textbf{b}) power generation. {`OCGT' and `CCGT' are open and combined cycle generators, respectively. `CCGT-CCS' is a `CCGT' plant with carbon capture and storage technology.}  (\textbf{c}) gas consumption (top) and supply (bottom), and (\textbf{d}) annual {power-gas} system costs for exemplar weather year under different electrification scenarios, as described in Fig.~\ref{fig:demandresults}, and decarbonization targets. {The investment costs are annualized.}
    In the upper legend, ``Li-ion'' is short-duration battery storage.
    In the lower-right legend, ``VOM'' and ``FOM'' are variable and fixed operating and maintenance costs, respectively. ``CCS'' is the cost of establishing carbon capture and storage infrastructure. 'Power gen + storage fix' is the investment and FOM costs for storage and power plants. {`Network Expansion' is the cost of establishing new transmission lines.} }
    \label{fig:cap-gen-cost-2003}
\end{figure}

To illustrate the least-cost resource portfolios across various electrification and emissions scenarios, Fig.~\ref{fig:cap-gen-cost-2003} shows the model's outcomes for an exemplar weather year that has the highest combined energy demand for power and gas across the 20 years we evaluate. We include results across all weather years in SI \ref{SIsec:GTresults} (see Figure~\ref{SIfig:GTresults}).


Across electrification and decarbonization scenarios, Fig.~\ref{fig:cap-gen-cost-2003}a-b shows that {VRE resources make up the majority of capacity and generation. Despite being deployed in smaller amounts than solar in terms of capacity, onshore and offshore wind dominate the generation owing to their higher availability} (capacity factor of 50\% for offshore wind vs. {19}\% for solar PV {in the exemplar weather year}), especially during periods of increased demand in the winter months. For the same reason, as electrification increases, offshore wind is the primary resource used to meet the incremental demand - for example, offshore wind's share of total generation increases from \bt{26\%} to \bt{36\%} between the RF and HE scenarios under 95\% decarbonization goal. Across all scenarios evaluated in Fig.~\ref{fig:cap-gen-cost-2003}, solar and onshore wind are built to their potential capacity limit in the region \citep{E3Team2020}, implying that increasing land availability for these resources could further increase their deployment in lieu of other resources and reduce system costs. 

All electrification scenarios maintain approximately \bt{10.6-14.4} GW of short-duration battery storage (Li-ion) with a rated duration of \bt{3.6-4.7} hours, highlighting the fact that battery storage is required to compensate for the variable nature of VRE, particularly solar. The medium and high electrification scenarios drive transmission expansion by \bt{2.5-5\%} compared to current network capacity across the emissions scenarios (\ref{SIsectrans-lines}).

As can be seen by comparing HE to ME in Fig.~\ref{fig:cap-gen-cost-2003}c, increasing electrification also reduces overall gas consumption by reducing its use in the building sector which more than offsets any increased use of gas in the power sector. This outcome is likely partially driven by the aforementioned efficiency benefits of heat pumps. Moreover, the LCF consumption for HE and HX scenarios varies between \bt{70e6 to 98e6} MMBtu under 80\%, and between \bt{239e6 to 263e6} MMBtu under 95\% decarbonization target; this is almost within the production capacity of the region for the biogenic LCF (see Table~\ref{tab:key-JPoNG-params} in Section~\ref{sec:methodology}) and highlights the need for high electrification.
Implementation of envelope improvements coupled with electrification on the supply side generally reduces power capacity investments, which is most notable for high electrification scenarios (Fig.~\ref{fig:cap-gen-cost-2003}a),  as well as reduces gas consumption across power and building sectors (Fig.~\ref{fig:cap-gen-cost-2003}c).


The comparison between Fig.~\ref{fig:cap-gen-cost-2003}a and \ref{fig:cap-gen-cost-2003}b also reveals that the system retains most of the existing unabated gas-fired power plants but these plants are utilized in very limited amounts (with capacity factors of \bt{0.2-3.7\%}) and are primarily used to meet peak demand {during extreme cold-weather events} (as shown in SI \ref{SIsec:dispatch}). Fig.~\ref{fig:cap-gen-cost-2003}a also shows that CCS-based plants are highly utilized, with capacity factors ranging from \bt{24\% to 35\%}. At \bt{17.7-28} GW, the overall capacity of the fleet of gas power plants is roughly \bt{1-57\%} larger than the existing capacity but makes up only \bt{8.4-13.4}\% of overall generation as compared to 52\% for New England in 2022 \citep{ISONE-Mix2022}. 
Motivated by the low regional CO$_2$ storage availability and lack of regional policy interest in CCS, we evaluated a sensitivity case without CCS for the electrification and decarbonization scenarios. The results (Fig.~\ref{SIfig:cap-gen-cost-noCCS}) point to an increased reliance on offshore wind and LCF that is used via unabated gas generation to balance the system in the absence of CCS, which can be done at similar or slightly increased cost (see Fig.~\ref{SIfig:diff-no-CCS}).

Fig.~\ref{fig:cap-gen-cost-2003}c reveals that a more stringent decarbonization goal results in higher LCF deployment \bt{(239e6 to 437e6 MMBtu)}, nearly doubling across all electrification scenarios for the 95\% emissions reduction target, and also making a large impact on the combined power and NG system cost (see Fig.~\ref{fig:cap-gen-cost-2003}d). The expensive LCF supply comprises a large portion of overall cost, ranging from on average \bt{18\%} in the 80\% decarbonization case to \bt{41\%} under the 95\% decarbonization target. This results in a system-wide cost increase of on average \bt{39\%} across the electrification scenarios when moving from an 80\% emissions goal to 95\%.


\subsection{Impact of Electrification Pathways on Power-Gas System Costs}

\begin{figure}[!htbp]
\centering
    \begin{minipage}{\textwidth}
        \includegraphics[width=\textwidth]{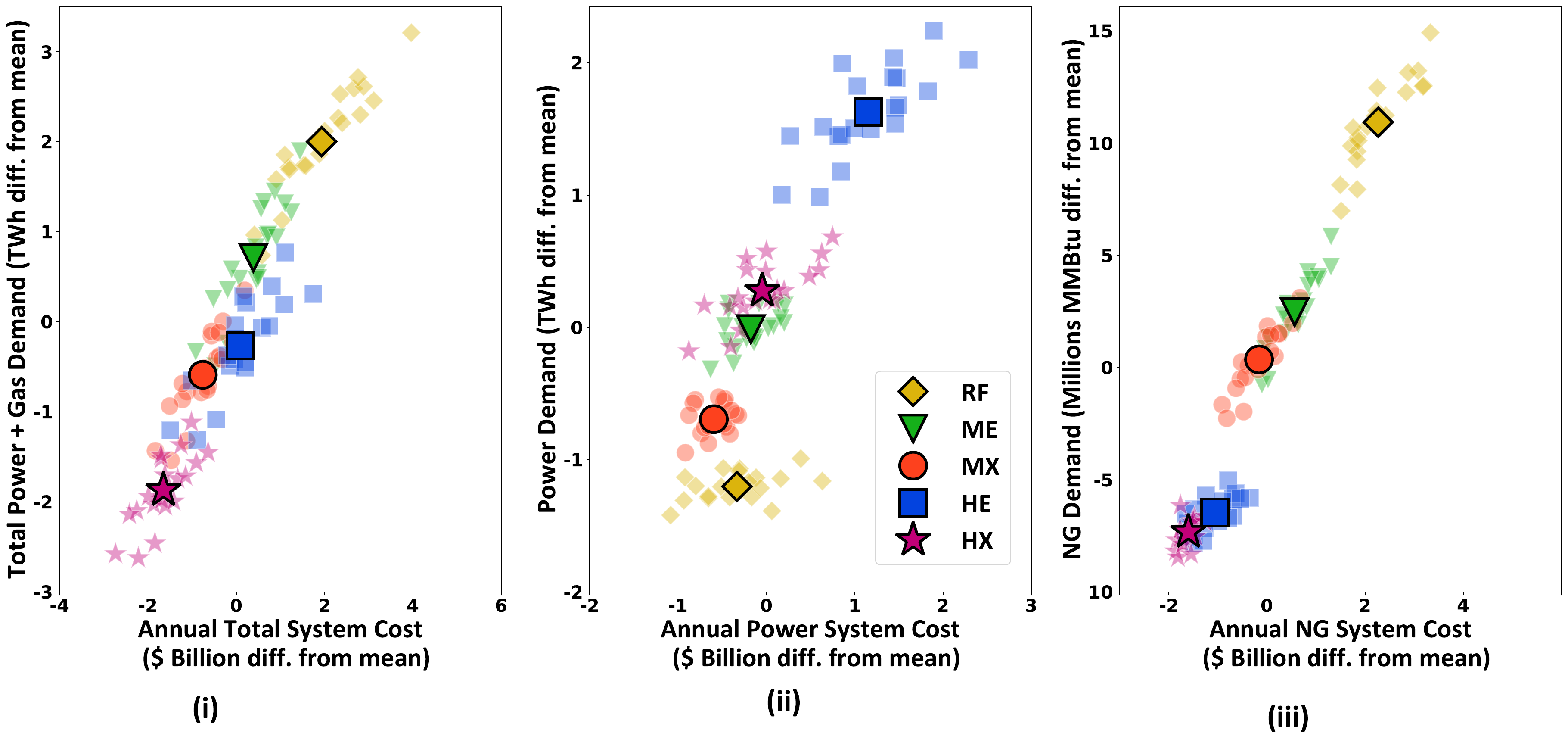}
    \captionof*{figure}{(a) Decarbonization Target = 80\%}
        \label{fig:load-vs-cost-80}
    \end{minipage} 
    
    \begin{minipage}{\textwidth}
        \includegraphics[width=\textwidth]{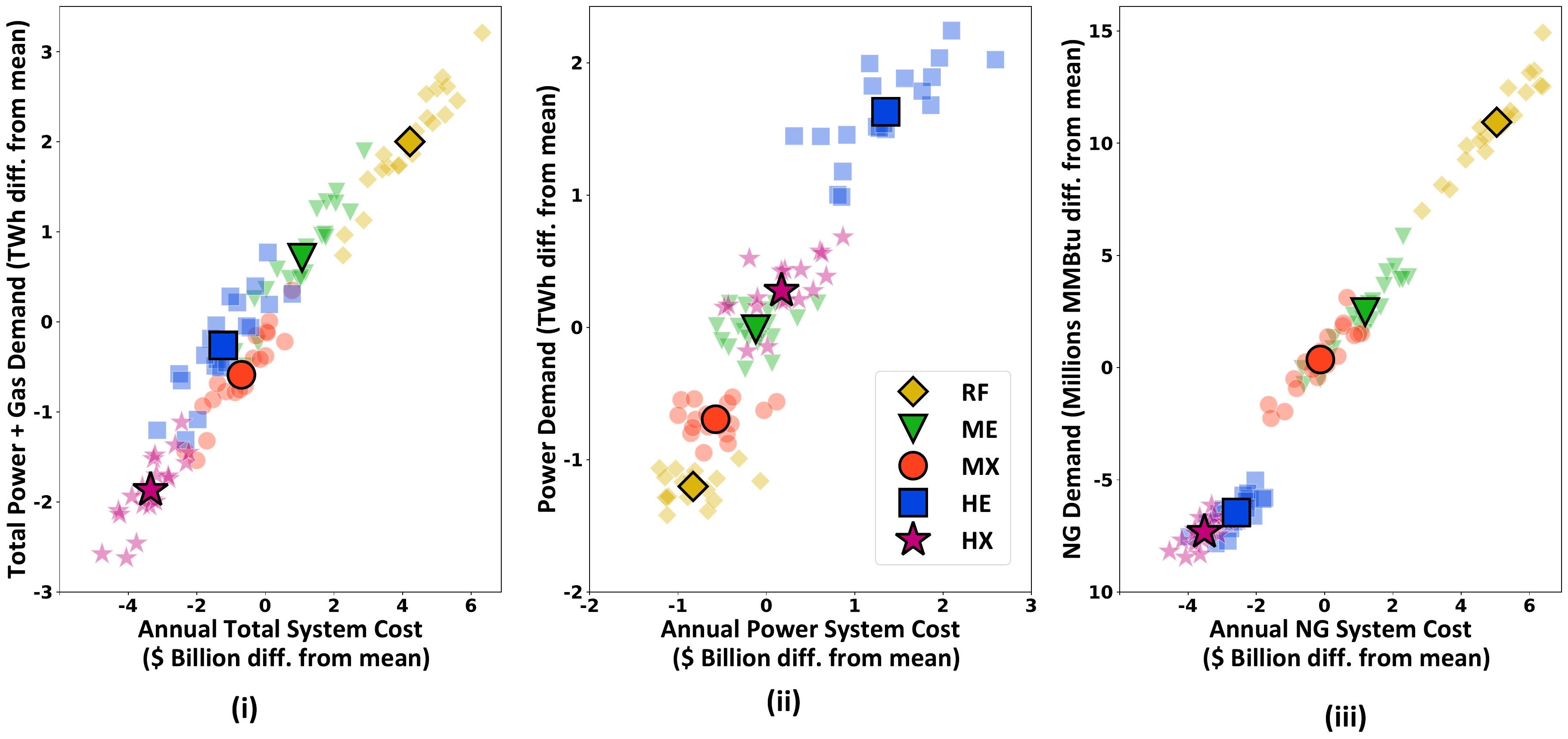}
        \captionof*{figure}{(b) Decarbonization Target = 95\%}
        \label{fig:load-vs-cost-95}
    \end{minipage}
     \caption{Total bulk power-gas system cost versus energy consumption for (\textbf{a}) 80\% decarbonization and (\textbf{b}) 95\% decarbonization. Subfigures (i) compare the combined system cost with the combined energy demand for power and NG. Subfigures (ii) and (iii) have the same individual comparisons for power and gas with their corresponding demands. Axes are normalized according to their difference with the average value across all points. Translucent points show all 20 weather years for each electrification scenario, while bold points indicate the averages. Note that bulk power-gas system costs do not include demand-side costs of electrification, such as heat pump and envelope improvement retrofits as well as electricity and gas distribution infrastructure costs.
     }
        \label{fig:load-vs-cost}
\end{figure}

Across the weather years and decarbonization scenarios, Fig.~\ref{fig:load-vs-cost} highlights that the average bulk power-gas system costs (shown by the emphasized points) are lowest for high levels of electrification that entail envelope improvements. 
While a full assessment of the cost of envelope improvement retrofits is outside the scope of this work, a simple net-present-value analysis demonstrates that the direct bulk system savings alone offset roughly {35-57\%} of the cost of envelope improvements under high electrification (see SI \ref{SIsec:NPVEnvImprov}). This analysis does not account for other co-benefits of envelope improvements (see Section~\ref{sec:discussion}). 

It is instructive to note that the RF scenario, corresponding to the least electrification, has the highest average power-gas system costs across both decarbonization targets in Fig.~\ref{fig:load-vs-cost}, which reinforces the importance of electrification as a strategy for economy-wide decarbonization; notably, the costs for RF are even higher than the scenarios with more electrification that lack any envelope improvements. On average relative to the ME, MX, HE, and HX scenarios, RF yields \bt{9\%, 16\%, 11\% and 21\%} higher costs under the 80\% decarbonization goal and \bt{12\%, 19\%, 21\%, and 29\%} higher costs under 95\% emissions reduction goal, respectively. While higher levels of electrification {can lead to increased costs} in the power system as illustrated in Fig.~\ref{fig:load-vs-cost}a(ii) and 
\ref{fig:load-vs-cost}b(ii), they lead to substantial savings in the NG system (Fig.~\ref{fig:load-vs-cost}a(iii) and 
~\ref{fig:load-vs-cost}b(iii)) that partially or entirely offset power system cost increases. This highlights the importance of joint energy system planning and indicates that evaluating the gas or power systems in isolation from each other can lead to misleading results.

\subsection{Impact of Methane Leakage}\label{sec:methane_leakage}

\bt{Our analysis in this paper is based on decarbonization goals which primarily target the emissions of carbon dioxide. However, methane emissions associated with NG supply chain is another factor driving NG's climate impact \citep{AlvarezEtal2018}.}
\bt{Table~\ref{tab:methane_leak} section highlights the impact of accounting for methane emissions associated with the supply chain of LCF and NG fuels on the supply-side outcomes. This analysis is based on accounting for methane emissions, in (CO$_2$ equivalent terms), associated with the NG supply chain (production, gathering, processing, transportation and storage) and LCF supply chain (transportation and storage). While emissions from NG distribution and end-use are also prevalent, they remain a relatively small share of the overall leaks \citep{AlvarezEtal2018}. Moreover, as aforementioned, many regions including the state of Massachusetts\citep{MassGSEPProgram} have implemented pipe replacement programs which when completed would further reduce these leaks.  Therefore, considering a 2050 scenario, we did not include distribution-related CH$_4$ emissions. Section~\ref{SIsec:methaneleakage} provides the details of our estimation method for methane leakage. }

\bt{Table~\ref{tab:methane_leak} shows the changes in total system cost, LCF imports and NG consumption between the instances with and without methane leakage for an exemplar weather year across scenarios of interest evaluated in Figure 5. Across all scenarios, accounting methane emissions in the modeled decarbonization target leads to reductions in NG and increasing use of lower emissions-intensive, but expensive, LCF, thus increasing the total system cost. The reference scenario demonstrates the largest total cost increase (3.3-11.7\% (\$0.9-\$2.2B/year), whereas higher electrification scenarios are more robust to methane leakage, especially under 95\% reduction target. 
Overall, we find that accounting for methane emissions in the decarbonization target increases the cost-effectiveness of high electrification; for example, the difference between HE/HX vs. RF is 6.4-9.2 \$B/year with methane emissions included vs. 5-8.8 \$B/year without accounting for methane emissions.}



\begin{table}[htbp]
\caption{\bt{Changes in total system cost, LCF, and NG consumption as a result of including estimated methane emissions associated with NG and LCF supply chains in the modeled emissions constraint. Results correspond to exemplar weather and reported as a percent difference vs. the model results without accounting for methane emissions in the emissions constraints (Fig.~\ref{fig:cap-gen-cost-2003})}}
\label{tab:methane_leak}
\bt{\begin{tabular}{l|ccc|ccc}
\toprule
   & \multicolumn{3}{c}{80\% Reduction Goal}                                     & \multicolumn{3}{c}{95\% Reduction Goal}                                     \\
   \midrule 
   & \begin{tabular}[c]{@{}l@{}}Total \\      system cost\end{tabular} & \begin{tabular}[c]{@{}l@{}}LCF \\      import\end{tabular} & \begin{tabular}[c]{@{}l@{}}NG \\      import\end{tabular} & \begin{tabular}[c]{@{}l@{}}Total \\      system cost\end{tabular} & \begin{tabular}[c]{@{}l@{}}LCF \\      import\end{tabular} & \begin{tabular}[c]{@{}l@{}}NG\\      import\end{tabular} \\
   \midrule 
RF & 11.7  & 21.9& -14.6   & 3.3   & 5.2 & -12.7   \\
ME & 6.8   & 7.7 & -10.9    & 2.8   & 5.2 & -9.9   \\
MX & 4.8   & 23.3& -11.8    & 3.5   & 5.7 & -9.2    \\
HE & 3.8   & 42.2& -9.7     & 2.4   & 5   & -6.6    \\
HX & 5.7   & 62.4& -10.6   & 2.7   & 7.2 & -1.8  \\ 
\bottomrule
\end{tabular}}
\end{table}

\subsection{LCF Availability}\label{sec:LCF_availability}
\begin{figure}[htbp]
 \centering
        \includegraphics[width=.7\linewidth]{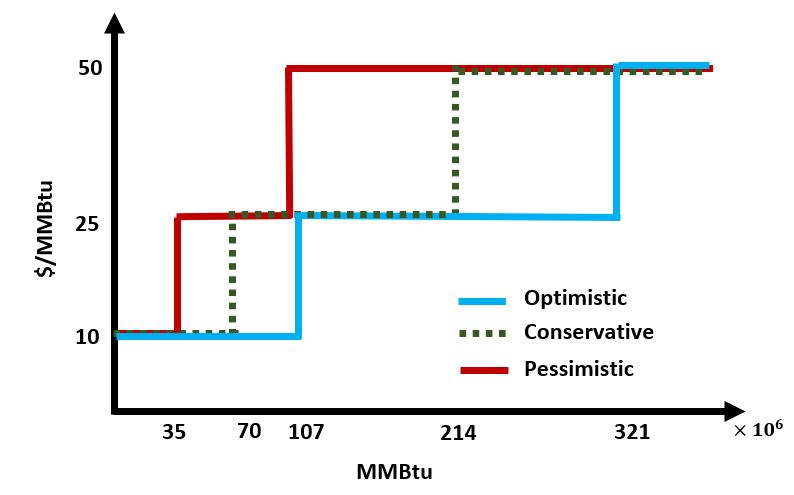}
        \caption{Supply curve for optimistic, conservative and pessimistic estimate of LCF availability. }
        \label{fig:LCF_supply_curve}    
  \end{figure}

\begin{figure}[htbp]
      \centering
      \includegraphics[width=0.8\linewidth]{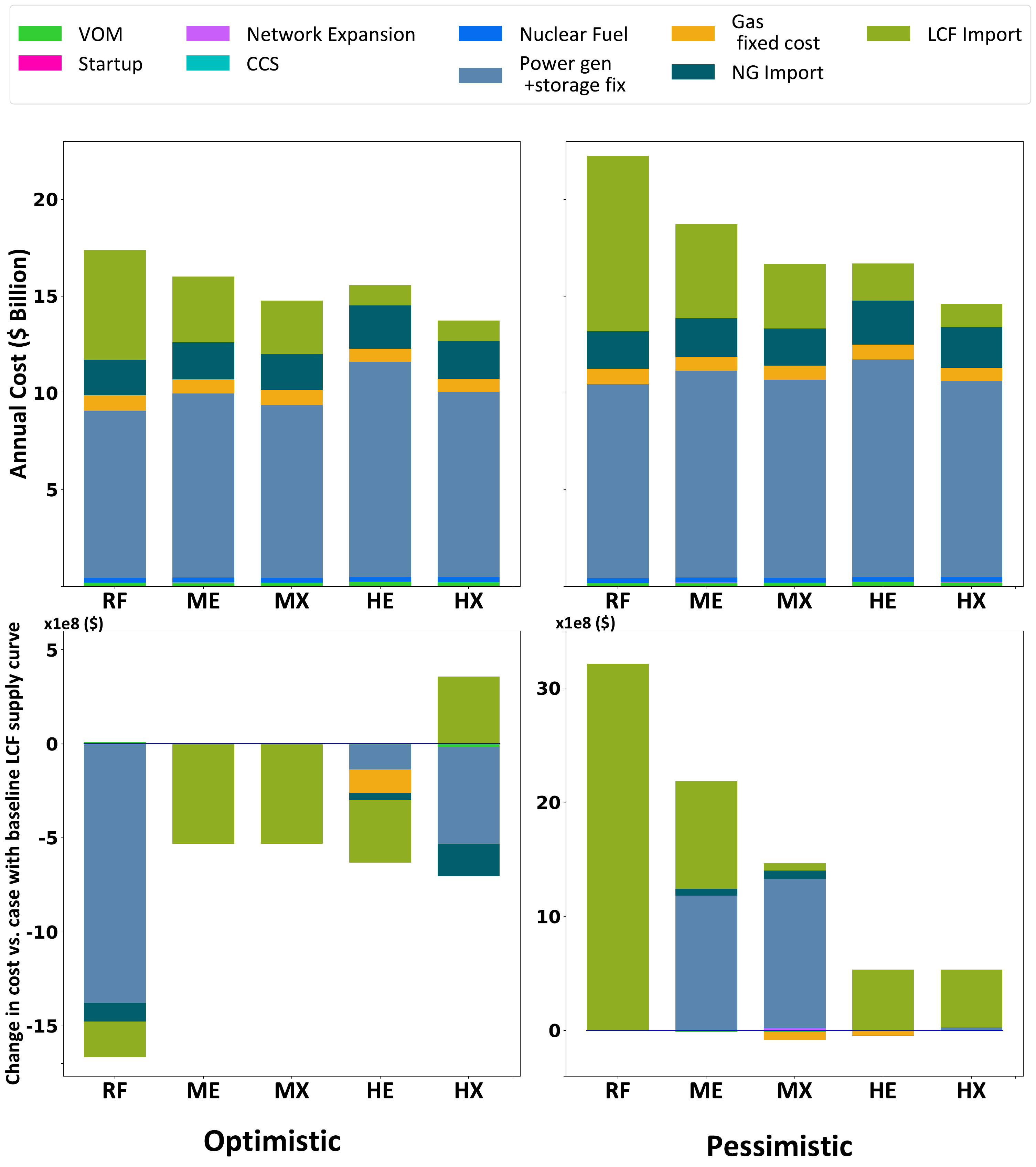}
      \caption{Changes in system cost and gas supply as function of LCF supply curve assumptions (top panel) compared to the model outcomes with the baseline LCF supply curve scenario. Results shown for optimistic and pessimistic LCF supply curves defined in Fig. \ref{fig:LCF_supply_curve}, for exemplar weather year under 80\% decarbonization scenario. Baseline LCF supply curve results summarized in Figure \ref{fig:cap-gen-cost-2003}}
      \label{fig:LCF_availability_comparison}
  \end{figure}

\bt{Even though many power and gas system decarbonization studies consider the role for LCF for system balancing in deep decarbonization \citep{RicksEtal2024_NE,ColeEtal2021,BaikJenkinsEtal2021}, often with much lower costs (e.g. \$20/MMBtu in \citep{ColeEtal2021} and \$15/MMBtu in \citep{BaikJenkinsEtal2021}), there is considerable uncertainty on their availability, cost and embodied emissions. We tested the sensitivity of our results to LCF availability and cost, by evaluating the system outcomes based on alternative \textit{optimistic} and \textit{pessimistic} LCF supply curves, shown in Fig.~\ref{fig:LCF_supply_curve}. The optimistic and pessimistic supply curves assume 50\% higher and lower than the baseline supply curve, described further in Section~\ref{SI:LCF_avail}.  Fig.~\ref{fig:LCF_availability_comparison} shows the resulting absolute system cost breakdown (upper plots) for optimistic and pessimistic estimates and their cost difference from the baseline case (lower plots) under 80\% reduction goal.}

\bt{Similar to a conservative estimate, the HX remains the most cost-effective scenario in new estimate levels. The RF scenario is most vulnerable to LCF estimations with 8.8\% (\$1.7B/year) system cost decrease in optimistic, and 16.8\% (\$3.2B/year) increase in pessimistic estimate. Unlike conservative and optimistic estimates, the second cost-effective scenario is HE in the pessimistic estimate, emphasizing the importance of high electrification as a hedge against LCF uncertainty. }

\section{Discussion}\label{sec:discussion}
Our study illustrates the importance of taking an integrated planning approach to heating electrification and applying methods that recognize the importance of weather variations. We show that, without such mitigating measures as envelope improvements, peak electricity demand is likely to considerably increase under deep electrification, necessitating further investment in grid supply resources that may be under-utilized.  Regardless, we find that high electrification, even without envelope improvements, can result in reduced bulk power-gas system costs vs. the business-as-usual scenario with modest electrification (where less than 10\% of the housing stock is electrified, see scenario RF in Fig.~\ref{fig:techscenariossmall}) under {both} decarbonization scenarios considered. This is in contrast to some suggestions that more aggressive electrification pathways are more expensive due to the capital investment needed on the supply side \cite{DPUSubmission, AGA2018, Strbac2018}. Although overall bulk power-gas system costs are lower at high levels of electrification, future analysis should evaluate the allocation of these costs, particularly as it pertains to low-income customers who could be stranded on legacy gas infrastructure. In addition, our analysis does not model the cost associated with power-gas distribution as well as end-use technology adoption, which would need to be factored in when considering the cost-benefit analysis of different levels of electrification. For example, hybrid systems are often promoted (and may be favored by consumers) for their potential to reduce the upfront capital costs of ASHP installations and also could reduce the extent of power distribution network upgrades needed to accommodate high electrification levels. 





Although they furnish cost savings on the bulk system, envelope improvements also entail retrofit costs at the household level that have not been considered in this analysis. While preliminary calculations suggest that the bulk system cost savings from envelope improvements could partially offset the costs of retrofit implementation (see SI \ref{SIsec:NPVEnvImprov}), there are other co-benefits that are worth noting. These include: avoided distribution infrastructure investment that will be needed to support higher loads, household benefits such as energy bills savings \cite{Tonn2018}, and health benefits \cite{Tonn2018}. However, these benefits can be undermined by the evidence that traditional energy efficiency measures, including envelope improvements, may not deliver on the full technical potential of their estimated savings \cite{SorrellEtal2009}. The analysis presented here can be coupled with other studies focused on the energy bill impacts of heat pump adoption to provide a holistic view \citep{WilsonEtal2024_Joule,WaiteModi2020}.


Our results highlight the value of integrated power-gas planning in future decarbonized energy systems, particularly in regions like New England where NG is prevalent as a heating fuel and also currently plays a major role in electricity supply.  Our modeling results suggest that electricity demand under decarbonization scenarios will largely be met by VRE including high levels of wind deployment and a complementary mix of solar and storage. However, even under high levels of heating electrification, there may be a need for gas fuel for balancing the power system as well as servicing residual gas demand in the heating stock.
These interactions necessitate coordinated planning of the power and gas sectors going forward in order to explicitly evaluate the tradeoffs and explore the full spectrum of potential solutions.

A theme of our findings is that a flexible low-carbon electricity resource and/or low-carbon energy carrier will likely be needed to supply the energy demands of electrification in a cost-effective manner while meeting decarbonization objectives in cold climates. This finding is robust to sensitivities regarding the level of electrification or the availability of CCS. Although the literature recognizes a need for flexible resources more generally \cite{SepulvedaEtal2018}, this study highlights their role under different levels of heating electrification and interactions with the gas system. While CCS and LCF-based power generation are examples of flexible, low-carbon resources, competing technologies like nuclear-based power generation and long-duration energy storage could also be equally important to consider as highlighted by other studies \citep{SepulvedaEtal2018,KhorramfarEtal2022}. 

In the case of CCS-based power generation and LCF, further scrutiny is needed to understand their role in deeply decarbonized energy systems as both technologies face future cost uncertainties due to the lack of their deployment at scale currently. 
In the context of uncertain LCF supply and cost, high electrification serves as an important hedge as it leads to the lowest LCF consumption across the modeled demand scenarios. Moreover, the LCF consumption in these scenarios (HE-HX: \bt{70e6-263e6} MMBtu across both decarbonization targets) is similar to the maximum technical potential of biogenic methane assumed for the region (\bt{see Fig.\ref{fig:LCF_supply_curve}}) under conservative estimates.
In addition, the GHG emissions benefit of using NG in conjunction with CCS and using LCF with or without CCS could be reduced when factoring methane emissions associated with the supply chains of these fuels. Unsurprisingly, accounting for the supply chain methane emissions for these fuels in the emissions constraints raises the system cost of decarbonization \bt{by up to \$2.2 billion}, through increased substitution of NG with less-carbon intensive LCF, as highlighted in SI \ref{SIsec:methaneleakage}. Moreover, when methane emissions are included as part of the 80\% and 95\% emissions targets, the cost savings of the HE and HX cases vs. the RF scenario becomes even greater, saving the region an additional \bt{\$0.39-1.6B/year}  (\ref{SIfig:diff-methane-leak}).  At the same time, the continued reliance on these fuels even when accounting for supply-chain methane emissions, highlights the need for policy mandates to measure and mitigate supply-chain emissions, such as those being undertaken in the Methane Emissions Reductions Program implemented under the Inflation Reduction Act \citep{EPAEmisReducProgram}.  

Our analysis also highlights that the weather-induced demand variability experienced under heating electrification leads to notable variations in the optimal investment outcomes on the supply side, such that key capacity investments can vary by up to \bt{18\%}. Similar to the focus on the impact of inter-annual weather variation on VRE generation, planners will have to consider the impact on demand, particularly in a future with high heating electrification. This sensitivity calls for principled planning approaches that account for weather-induced demand variation, or alternatively, the implementation of demand-side solutions to mitigate the variability and enable convergence to a single ``best'' system plan.

Future work could serve to consider other key dimensions of the modeling problem. Evaluating the hourly impacts of commercial building sector heat electrification would be valuable, but so far has proven to be a difficult bottom-up modeling problem \cite{ResstockComstock2022}. Future analysis should also consider the role of demand flexibility, which could be instrumental in reducing peak demand and power system costs under electrification, similar to the envelope improvement scenarios modeled here. As an initial investigation into the potential role of demand flexibility under heat electrification, we test the model's sensitivity to flexible transportation loads and find that activating flexibility in this sector could result in large system-wide cost savings \bt{of up to 5.2\% (\$1B) amounting to around \$1.2 billion} (see SI \ref{SIsec:transportflexibility}). \bt{Therefore, more investigation will be needed to determine how behavioral change in response to electrification and/or efficiency improvements may impact load profiles}. Furthermore, given that our model results call for network expansion and power import from neighboring regions (e.g., Canada), the evaluation of resiliency-related issues such as network outages and coordination of supply and demand with neighboring regions during extreme cold weather events will need to be studied further.



\section{Methodology} \label{sec:methodology}
Our framework consists of bottom-up and planning models as depicted in Fig.~\ref{fig:bottomup-jpong-map}. The bottom-up model projects the sectoral demand of power and gas for the year 2050 under various electrification and weather years. Using the projected values, the planning model provides investment and operational outcomes for each unique instance of weather year and electrification scenario. 


\subsection{Bottom-up Engineering Model for Energy Demand}\label{ssec:Bottomup}

\begin{figure}[!htbp]
    \centering
    \includegraphics[width=0.9\textwidth]{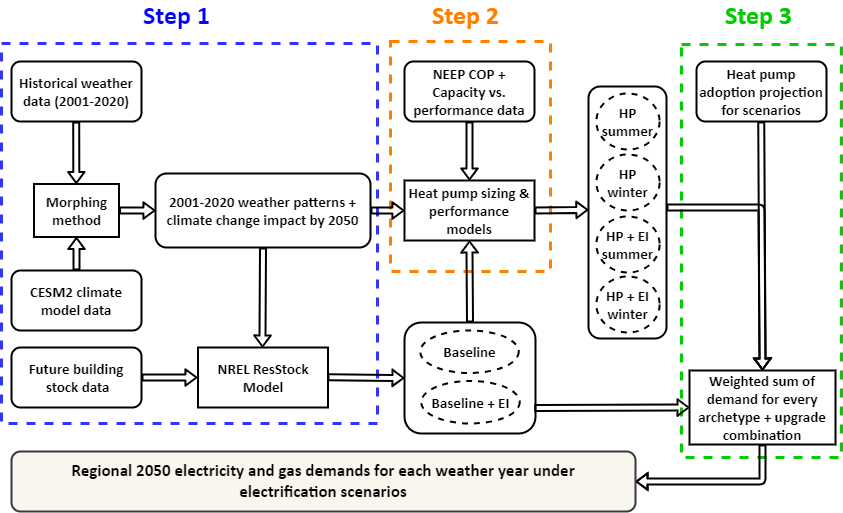}
    \caption{Diagram of bottom-up {residential} model. ``Summer'' and ``winter'' correspond to the two heat pump sizing methods applied in this study and capture both the whole-home and hybrid approaches to heat pump retrofits (see Figure \ref{fig:techscenariossmall}). ``Baseline'' corresponds to the remaining homes that do not receive heat pumps. ``EI'' indicates the subset of homes that receive envelope improvements. See SI  \ref{SIsec:bottomup} for more details. Step 1 (see \ref{SIsec:ResStock}) determines heating and gas fuel demands for the original archetypes using future weather and building stock projections. Step 2 determines the electricity and gas demands for the archetypes when they are electrified under different mixes of heat pump sizes and envelope upgrades (see \ref{SIsec:HPmodel}). Step 3 aggregates the demand profiles of all of the archetypes to determine the hourly regional loads for the residential sector while accounting for the mix of possible heat pump sizes and envelope improvements applied to the stock (see \ref{SIsec:aggmethod}).
    Historical weather data is obtained from ERA5 \cite{era5} via \cite{OikoLab} {(see Fig. \ref{SIfig:weathercomparison} for more details on rationale for using the ERA5 dataset)}, morphing method from \cite{BelcherEtal2005,DickinsonBrannon2016,JentschEtal2008}, CESM2 climate model data from \cite{DanabasogluEtal2020}, and future building stock data from \cite{MA-climate-plan}.
    }
    \label{fig:bottom-up-approach}
\end{figure}

We develop a bottom-up model to project the electricity and gas demands in the residential sector of the U.S. New England region in the year 2050 using 20 years of historical weather data, climate change projections, and other publicly available data and open-source resources. SI \ref{SIsec:bottomup} provides a detailed description of the model. Because our focus is on the bulk system, we do not evaluate the impacts of electrification on demand for other heating fuels such as fuel oil or propane, \bt{although the input scenarios assume homes with these heating systems electrify at equal rates to those with natural gas heating}. Our modeling framework allows us to leverage existing heat-pump adoption rate projections (see \ref{SIsec:deploymentscenarios}) to produce demand profiles for different levels of electrification. We divide New England into 17 zones each representing a population of between 0.6 and 1.6 million people (see SI \ref{SIsec:Powernodes}). The zones serve as the nodes of the power system network and the regions for which demand data is generated. Our process consists of three modeling steps as shown in Fig.~\ref{fig:bottom-up-approach}.

 In the first step of the demand modeling, we leverage NREL’s ResStock model \citep{ResstockComstock2022,WilsonEtal2017}. Given the housing stock characteristics and weather, ResStock generates a sample of housing archetypes and simulates their hourly energy consumption for key end-uses. We modify ResStock’s housing stock data inputs to resemble the projected housing stock of New England in 2050 (see SI \ref{SIsec:stockprojections}). As the basis for our 20 years of weather inputs, we procure hourly historical weather data for weather years 2001-2020 for 44 locations in New England corresponding to the DOE’s typical meteorological year (TMY) locations \citep{BianchiFontanini2021} (see SI \ref{SIsec:weatherdata}). We simulate each archetype with the weather data of the closest location. We account for the climate change impact between the present-day and 2050 by applying the \textit{morphing method}, a common tool in the building energy modeling literature \citep{BelcherEtal2005,JentschEtal2008,DickinsonBrannon2016} to modify the hourly weather data according to monthly climate model projections that characterize the change of key weather metrics over time. More information on the morphing method is included in SI \ref{SIsec:morphingmethod}. ResStock’s hourly outputs exhibit a smoothing effect with increasing numbers of archetypes \citep{ResstockComstock2022}. We run simulations for approximately 400 archetypes in each of the 17 demand zones in order to balance the benefits of the smoothing effect with the computational burden of more simulations. We discuss the implications of sample size on simulation outputs and model error in \ref{SIsec:validation} and note that this value has been used previously in similar work \citep{DeetjenEtal2021}. In addition to the baseline stock, we use ResStock to model envelope improvements that may occur in parallel with electrification (see SI \ref{SIsec:envimprov}). 



In the second step, we determine the impacts of electrification for each archetype. We use the heating and cooling demands from ResStock as inputs to a heat pump model. The efficiency of ASHPs to provide cooling or heating decreases as the difference between indoor and outdoor temperatures increases. Maximum heating capacity also degrades at lower temperatures. Therefore, we calculate capacity and efficiency at each hour by fitting a regression model to a large dataset of temperature-dependent heat pump performance. {The performance curves are such that the heat pumps have a heating efficiency of 273\% (COP of 2.73) at 0 C and 183\% at -20 C.} Further details about these models are included in SI \ref{SIsec:NEEPdata}-\ref{SIsec:capmodel}.

We model the heat pump’s energy demand under two possible forms of sizing to capture distinct electrification strategies: 1) whole-home or winter sizing and 2) summer sizing, where the heat pump is supplemented by the existing heating (backup) system, most often in a hybrid configuration because most existing heating systems in the region use fuels rather than electricity. 

\bt{We use industry-standard heat pump sizing methods to set each archetype's heat pump size based on modeled heating and cooling demands using TMY data (see SI \ref{SIsec:sizing})}. For summer-sized systems, we assume the heat pump and backup system operate based on the characteristics and limitations of the existing system in each home as described in SI \ref{SIsec:Auxheating}. 
End-use demands excluding heating and cooling are taken from the ResStock model. For the electrified homes, we also assume the water heater is converted to a heat pump water heater (see SI \ref{SIsec:waterheater}). 

The third step produces the zone-level demands for each combination of weather year and electrification scenario.
To yield the zone-level hourly gas demands for each scenario, we compute a weighted sum of the demand profiles for all of the archetypes in the zone. The weights are equal to the number of homes represented by each archetype and take into account the heterogeneous mix of sizing and envelope improvements that may apply to different portions of the stock represented by each archetype (more detail in SI \ref{SIsec:aggmethod}) as well as population change through 2050 (SI \ref{SIsec:popdata}).
Once the residential load is obtained, we use existing sectoral {energy demand projections} from the literature for 2050 for all non-residential sectors (see SIs \ref{SIsec:powerloaddata} and \ref{SIsec:NGloaddata}). 

\subsection{Joint Power-Gas Planning Model}\label{ssec:JPoNG}
We use an extended version of the \ModelName{} \citep{KhorramfarEtal2022} model to analyze the power-gas system planning in this paper, with a complete mathematical formulation provided in SI~\ref{SIsec:model-formulation}. {These extensions include modeling decommissioning of legacy gas assets, planning reserve constraints, hydro imports from Canada, supply curve for LCF, electricity demand flexibility and accounting for methane emissions associated with gas supply chain.}  
The investment decisions include establishing power plants, transmission lines, dedicated pipelines for carbon sequestration, gas pipelines, and power storage technologies. The model also considers decommissioning legacy power assets as well as gas pipelines. Key model constraints include i) power and gas system supply-demand balancing at hourly and daily time resolution, respectively; ii) network constraints for electricity (DC-OPF) and gas (pipe and bubble formulation);  iii) capacity reserve margin for electricity system; iv) resource availability limits for VRE, Canadian hydropower imports and LCF; and v) system-wide emissions reduction goals.


The operations of both systems are coupled through two sets of constraints. The first set ensures gas flow (either NG or LCF) to the power system to fuel gas-fired power plants. The second coupling constraint limits the CO$_2$ emissions caused by the consumption of NG in both systems to a specified target. 
The model captures an entire year of daily variability of gas demand and hourly operations of the power system over 24 representative days that are selected by the method proposed in \citep{BrennerEtal2023}. 



The system is modeled as a semi-isolated energy system with no power or NG export but with the possibility of power, NG, and LCF import. We construct the network for each system using publicly available data detailed in \ref{SIsec:input-data}. The resulting power and gas networks have 18 (17 nodes inside New England, one node to represent an import point from Quebec, Canada) and 23 nodes, respectively. This level of spatial granularity is sufficient to facilitate detailed analysis and adequately capture the network congestion effects. Overall, the power system has 39 existing and 34 candidate transmission lines and the gas system includes 36 existing and 46 candidate pipelines. SI~\ref{SIsec:input-data} provides details of assumptions and technical and economic parameters in the power system. Table~\ref{tab:key-JPoNG-params} shows key system parameter values used in this paper. Most of these parameters are sourced from \citep{ATB2021, SepulvedaEtal2021,MA_RNG_report2022, ReimersEtal2019}. The detailed gas system data and parameters are provided in SI~\ref{SIsec:input-data}. {Although we use \ModelName{} for the case study of New England, the model is broadly applicable to other regions and countries for both local and national level planning. In particular, the data sets we have used are also available for other U.S. regions (e.g. ResStock, ERA5) and thus the workflows developed could in principle be extended to study other regions in the U.S. readily.}

\bt{The resulting JPoNG model is a mixed-binary linear program and is developed in Python using the Gurobi solver. The details of the code and implementation are given in the subsequent section. The code along with all the data used in this study are available in the GitHub repository \citep{JPoNG-GitHub}. All instances are run on the MIT Supercloud system which uses an Intel Xeon Platinum 8260 processor with up to 48 cores and 192 GB of RAM \citep{Supercloud2018}. The system currently uses Gurobi 10.1. We solved all instances with 30 representative days which we select based on the method presented in \citep{BrennerEtal2022}. All instances are solved within 12 hours with an optimality gap of less than 1\%.} 


\begin{table}[htbp]
    \centering
    \begin{tabular}{l|r}
      Parameter   &  Value\\
      \midrule
        NG price ($\$$/MMBtu) & 5.45\\
        LCF availability levels (MMBtu) & 70.8e6, 214.5e6, 1000e6\\
        LCF availability price ($\$$/MMBtu)  & 10, 25, 50 \\
        Power load shedding cost (\$/MWh) & 10,000\\
        Gas load shedding cost (\$/MMBtu) & 10,000\\
        Reserve margin requirement & 15\% \\
        \bottomrule
    \end{tabular}
    \caption{Key system parameter assumptions in \ModelName. The reserve margin is enforced for all modeled operational periods.}
    \label{tab:key-JPoNG-params}
\end{table}

\section*{Acknowledgments}
We thank John Parsons of MIT for our hydropower import data and for helpful discussion, suggestions and comments. {We additionally thank Zack Schmitz of the MIT Energy Initative for his assistance with model runs and troubleshooting.} We also thank NREL's ResStock team for assisting with the usage of the ResStock model. All authors acknowledge support from the MIT Energy Initiative Future Energy Systems Center. R.K. and S.A. acknowledge funding from MIT Climate Grand Challenges ``Preparing for a new world of weather and climate extremes,'' and MIT Energy Initiative's project Hurricane Resilient Smart Grids. The responsibility for the contents lies with the authors.

\newpage
\section*{Supplementary Results}




\subsection{Impacts of electrification on length of peak event}\label{SIsec:peaklength}
\begin{figure}[htbp]
    \centering
    \includegraphics[width=\textwidth]{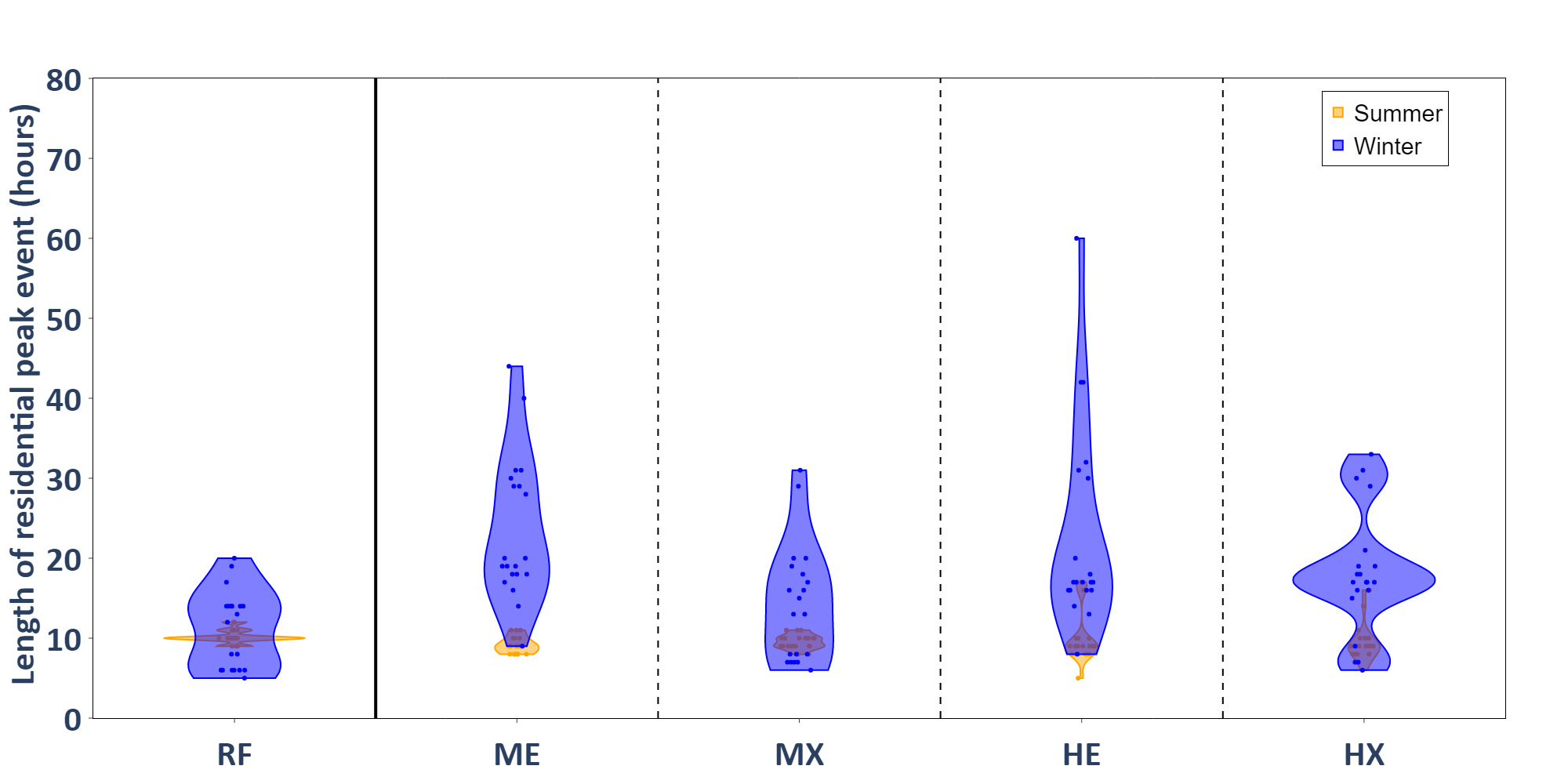}
    \caption{Length of the residential peak electricity demand event for summer and winter across all 20 weather years for each electrification scenario. Length is defined as the number of continuous hours in the seasonal peak event for which the demand is greater than {75\%} of the peak seasonal demand. The width of the violins is proportional to the density of the data.}
    \label{SIfig:Peaklength}
\end{figure}

While peak electricity demand is one of the metrics commonly used in grid capacity planning, the length of the peak event is instrumental to determine the required mix of resources. Fig.~\ref{SIfig:Peaklength} depicts the length of peak events during the summer and winter seasons for different electrification scenarios. We define a ``peak event'' as the continuous period for which the electricity demand is within {25\%} of the seasonal peak demand, including the periods directly before and after the peak hour.

For certain weather years under certain electrification scenarios, the peak event can span multiple days, especially in the winter. When electrification is implemented in the absence of envelope improvements, peak events can become longer in duration. Envelope improvements can reduce the length of peak events. 

\subsection{Demand implications of electric resistance backup}\label{SIsec:ElecBackup}

\begin{figure}[htbp]
    \centering
    \includegraphics[width=0.9\textwidth]{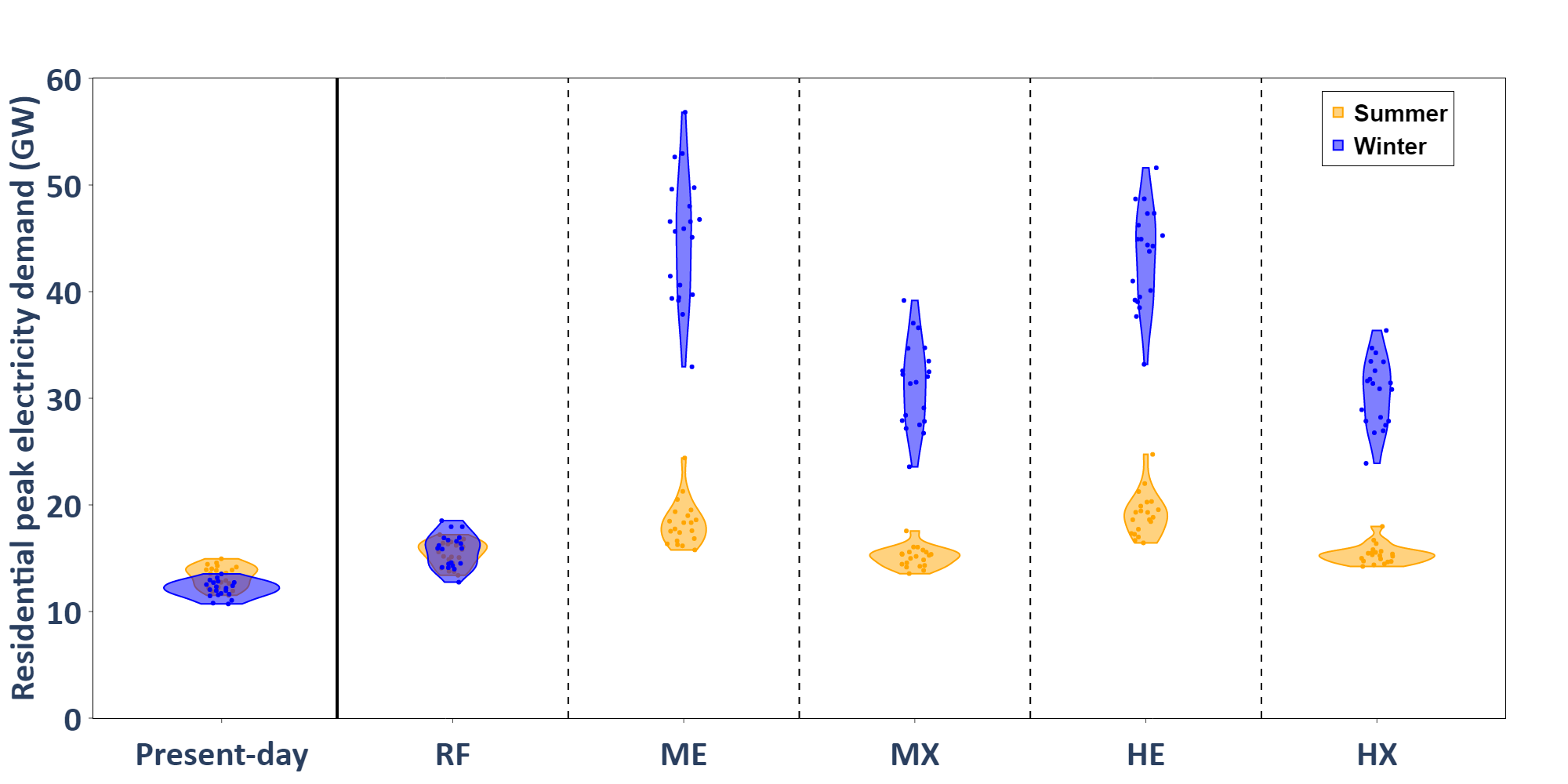}
    \caption{Summer and winter peak loads under the electrification scenarios presented in the main text, but for which summer-sized systems use electric resistance backup rather than the existing system. Peak loads significantly increase.}
    \label{SIfig:ElecBackupPeaks}
\end{figure}

Our main analysis assumes that summer-sized systems will maintain their existing systems for backup heating. However, it is also possible that households will adopt electric resistance systems as their backup systems, for example, when their existing system eventually malfunctions and needs to be replaced. With a COP of nearly 1, electric resistance backup is inefficient in comparison to heat pumps. In Fig.~\ref{SIfig:ElecBackupPeaks}, we highlight the peak demand impacts of using summer-sized heat pumps with electric resistance backup. 

Electric resistance instead of existing backup can further increase peak load between {0-184\%} for 2050 across scenarios, with large increases across the board for winter-peaking scenarios. Scenario ME, which emphasizes the deployment of summer-sized systems, now has the highest median peak of {45.7 GW, 25\% higher than even the highest peak of 36.6 GW} presented under the HE scenario in the main text. The sensitivity of peak load to interannual weather variation also increases drastically across the scenarios. For example, the range of peak demand under the ME scenario with electric resistance vs. existing backup is {33.0-56.8 GW} vs. {16.3-20.0 GW}, respectively. Therefore, the adoption of electric resistance may not be economical from a power system planning perspective.


\subsection{Validation of model against historical data}\label{SIsec:validation}

\begin{figure}[htbp]
    \centering
    \includegraphics[width=\textwidth]{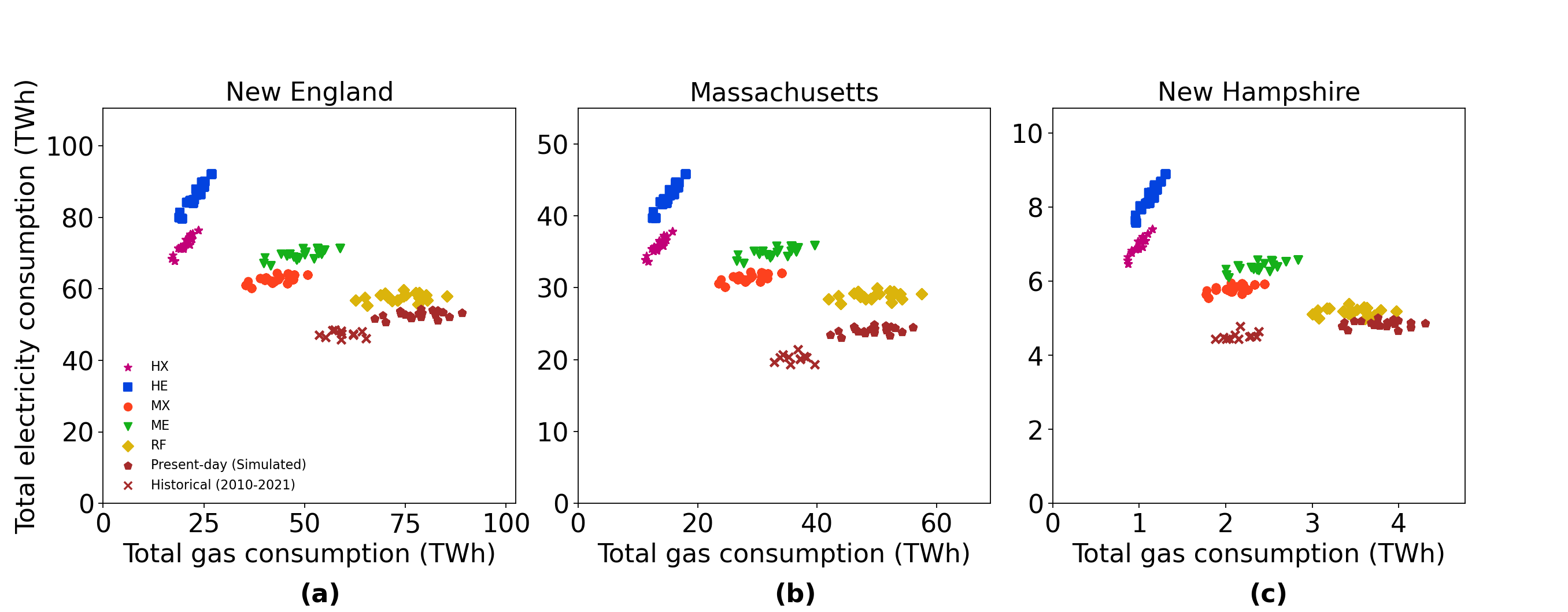}
    \caption{Annual demand results including the historical EIA data and our present-day simulated results, indicated by the points in brown. All results shown use 400 archetypes per load zone. Results for Massachusetts and New Hampshire are highlighted to show how Massachusetts contributes an outsized portion of the error relative to other states. Note the difference in scale on the axes, preserved in order to enable visibility of the data clusters.}
    \label{SIfig:histvspresent}
\end{figure}

Comparing our simulated present-day results to recent historical consumption values illustrates our model's performance. We include a comparison for the New England region in Fig.~\ref{SIfig:histvspresent}, which shows that our present-day model estimates the historical power demand similar to EIA data \citep{eiaWebsite2022_NG_pipe, eiaWebsite2022_ElectricPowerData}. However, it generally overestimates residential gas demand, by {18-37\% (average of 28\%) or 10.4 - 21.3 TWh (average of 16.7 TWh)}. Massachusetts contributes the majority of this demand error, by an average of {11.5 TWh}, due to its large population and high proportion of homes with gas heating. The error for electricity demand is significantly smaller, with an error of {9-15\% (average of 12\%)} across New England. 

We identify three potential sources of error to explain the overestimation of gas demand relative to historical data from EIA \cite{EIAStateProfiles} shown in Fig.~\ref{SIfig:histvspresent}:
\begin{enumerate}
    \item Systematic errors intrinsic to the ResStock model.
    \item Insufficient number of building archetypes in each zone.
    \item Errors in weather data.
\end{enumerate}

{We also note that the EIA data itself may have errors relative to actual historical consumption -- for example, some utilities report demand data to EIA in the form of billing periods rather than calendar months, which may misalign some of the demand data temporally \cite{ResstockComstock2022}. However, because we are unable to quantify this error and others, we treat the EIA data as a satisfactory baseline for actual consumption.}\\

\noindent\textbf{Errors in ResStock model:}

The calibration and validation of the ResStock model have been well-recorded by NREL \citep{ResstockComstock2022}, which used 550,000 archetypes nationally to analyze hourly building energy demands for a single weather year (2018). The subset of model outputs for the New England region compares reasonably well against EIA state-level monthly consumption data for gas and electricity (see Appendix F in the report \cite{ResstockComstock2022}). For example, for the most populous state of Massachusetts, gas demand from NREL's ResStock results are 10\% higher than the data from EIA (42.1 TWh vs. 38.2 TWh). Nonetheless, this deviation between ResStock model outputs vs. EIA data also propogates into our own analysis using ResStock with fewer archetypes per state (see next section) and multiple weather years. {In general, we opt not to attempt to calibrate ResStock to better fit historical demand. First, NREL already undertook a considerable effort to calibrate the model using immense amounts of proprietary and spatio-temporally resolved demand data \citep{ResstockComstock2022}, which we lack. Additionally, there would be no certainty that this would result in a more ''accurate'' result once the model is used to evaluate future levels of electrification that will fundamentally alter demand patterns from the historical baseline.}

\vspace{\baselineskip}\noindent\textbf{Number of building archetypes:}

As observed in ResStock's performance, higher numbers of archetypes lead to smoothing and convergence in the demand outputs of ResStock \cite{ResstockComstock2022}. Therefore, another source of the error can be attributed to the number of archetypes we consider in our bottom-up model, which in the main text is 400 archetypes per zone. 
Table~\ref{SItab:samplesizes-presentday} presents the impact of archetype count on Massachusetts, the most populous state in the region with the highest contribution to the gas demand overestimation.

\begin{table}[htbp]
\centering
\caption{Impact of archetype count on annual present-day model error for 2018 in Massachusetts. Differences calculated relative to EIA empirical data and results from NREL's ResStock runs in the End Use Load Profiles Project \cite{ResstockComstock2022}. Our demand analysis for all scenarios in the main text uses 400 archetypes per zone.}
\label{SItab:samplesizes-presentday}
{\begin{tabular}{lcccccc}
\toprule
Archetypes & Total & \%diff. from & \%diff. from& \%diff. from& \%diff. from \\
per zone & archetypes&NREL (gas)& EIA (gas)& NREL (power) &EIA (power)\\
\midrule
400    & 2,800  & +17.5 & +29.7 & +0.92 & +22.7 \\
1,000  & 7,000  & +12.1 & +23.7 & +3.46 & +25.8 \\
1,700  & 11,900 & +10.3 & +21.7 & +4.26 & +26.7 \\
2,500  & 17,500 & +7.5 & +18.5 & +4.84 & +27.4 \\
\bottomrule
\end{tabular}}
\end{table}

The results show that increasing the number of archetypes consistently reduces the gas demand error relative to EIA data, from an overestimate of {29.7\%} at the 400-archetypes-per-zone resolution used in our main analysis, down to {18.5\%} with 2,500 archetypes per zone. We expect the trend to continue with even higher archetype sizes as the quota sampling algorithm used in ResStock selects combinations of archetype features proportional to their probability distributions \cite{ResstockComstock2022} and thus better captures outliers with higher archetype counts. We note that this observation is not aligned with NREL's report in which 1000 archetypes per zone of interest is recommended to obtain convergence in demand behavior \cite{ResstockComstock2022}. However, the emphasis in that effort was calibrating electricity demand rather than gas demand; our findings indicate it may take greater numbers of samples to approach convergent gas demand results in ResStock.

{As for the power demand, it can be seen that for Massachusetts, the model experiences a 22-28\% deviation from EIA, which is outsized relative to the rest of New England. However, this deviation is evidently due to intrinsic error in the ResStock model - our present-day model result for power demand never deviates further than 5\% from NREL's estimate. We do not expect that we could improve the power demand results, given that this would require calibrating ResStock itself, as aforementioned.}


\vspace{\baselineskip}\noindent\textbf{Weather data:}

\begin{figure}[htbp]
    \centering
    \includegraphics[width=\textwidth]{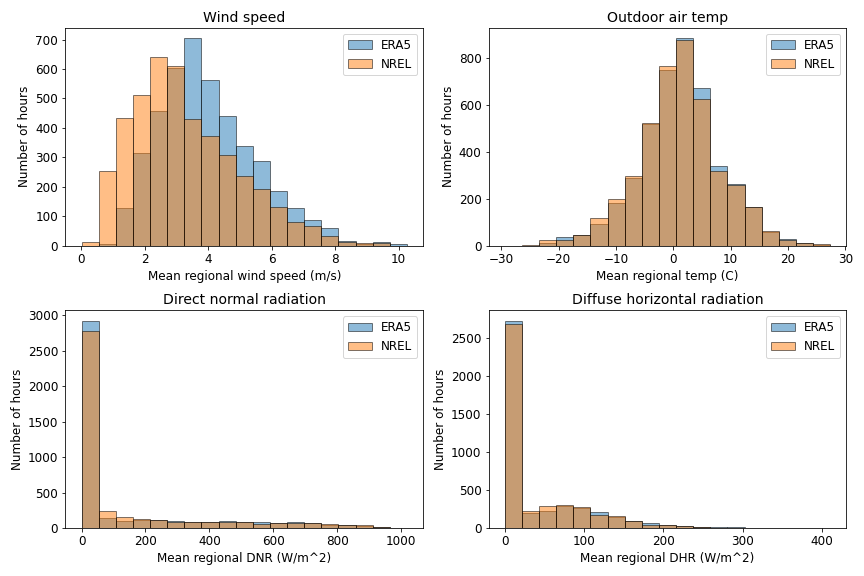}
    \caption{Comparison between key weather features for the ERA5 and NREL weather datasets during the heating season (October - March). The hourly values are averaged across all of the 44 weather locations used in our models.}
    \label{SIfig:ERA5vsNREL}
\end{figure}

Because weather is also a key input to the bottom-up framework and significantly impacts the demand output, we sought to quantify its impact. {NREL weather data encompasses the same parameters and geographic scope as our own weather data while producing gas demand outputs much closer to the EIA baseline; therefore, we compare our 2018 weather data to it in an attempt to determine what might be causing our overestimate. Furthermore, as mentioned in Section \ref{SIsec:weatherdata}, the NREL weather largely is constructed from NOAA observational data for 2018, making it closer to a ``ground truth'' dataset. In particular, we compare key weather features in the heating season (defined here as October through March), which would be more likely to adversely affect gas demand because the majority of gas demand is due to space heating. We show comparison histograms for key weather features in Fig.~\ref{SIfig:ERA5vsNREL}. It should be noted that the comparison of all weather-related parameters is not possible as NREL's published weather data purposefully excludes key variables like cloud cover and infrared radiation.}


{The figures reveal that while there is alignment between our ERA5 weather data and NREL's among most key variables, the wind speed has some discrepancies. When we weigh each location's weather data by its population to capture the likely impact on residential energy demand, we find that the ERA5 data generally overestimates winter wind speeds, by an average of 0.55 m/s (22\%) annually. This value is perhaps substantial enough to appreciably change the demand outputs of the bottom-up model by increasing the demand for heating, as higher wind speeds increase building heat loss on cold days. We note that discrepancies with wind speeds in ERA5 have been noted by other researchers in this domain \citep{Staffell2023}. However, the complexity of the weather inputs coupled with the sophisticated impact of weather on heating demand in the bottom-up approach makes it difficult to concretely determine the key parameters that contribute most to the differences between the results of the two datasets.}

\subsection{Impact of historical gas demand overestimation on future demands}
In the previous section, we identify a handful of potential input-related sources of the gas demand overestimation. Although increasing the archetype count can potentially reduce the error, it will greatly increase computational and data storage requirements, especially given the wide input parameter space of weather and electrification scenarios fundamental to this study. 
{As explained in Section \ref{SIsec:weatherdata}, we also opt to maintain our usage of the ERA5 data because it meets the needs of the bottom-up approach while performing well in our study region relative to other candidate datasets.}

Importantly, the errors in the present-day analysis may not necessarily propagate to the future electrified scenarios that underlie the bulk of our findings. To see how the sources of gas demand overestimation impact future demand, we compare our results for New England with those of an alternatively parameterized model - an ``NREL-parameterized'' model - for a handful of our scenarios. The premise is to generate an instance that would reasonably approach the current ``best'' published output of the model by mirroring the setup NREL used in its extensive calibration efforts. The ''NREL-parameterized'' model setup has two differences from the model setup used to generate the results described in the main text. First, the number of archetypes in the ``NREL-parameterized'' case is the same as what NREL used in its calibration of ResStock for New England (26,701 total, divided among the states roughly proportional to the number of homes). Second, the weather data is the same as NREL's 2018 weather data, except we use TMY data to fill in the cloud cover data which was purposefully redacted, enabling ResStock to interpolate the remaining missing weather features. 
The rest of the model inputs, including building stock evolution, population change, climate change, and electrification scenarios for 2050 are the same for both model setups.


\begin{table}[htbp]
\centering
\caption{Percentage difference between electrification scenario results in the main analysis and ``NREL-parameterized'' results}
\label{SItab:NRELparameterized}
{\begin{tabular}{lccc}
\toprule
Scenario& gas demand& power demand& power peak load\\
\midrule
HX & +0.64 & -4.95 & -2.15 \\
MX & +0.38 & -5.79 & -7.92 \\
RF & +0.77 & -7.52 & -4.82 \\
\bottomrule
\end{tabular}}
\end{table}

The results shown in Table~\ref{SItab:NRELparameterized} reveal that our future projections are more robust to the sources of overestimation impacting the present-day simulations. Across the HX, MX, and RF scenarios, annual gas demand is overestimated relative to the ``NREL-parameterized'' model setup by approximately just {0.38-0.77\%} as opposed to {17.5\%} for the present-day results. Annual electricity demand is also quite robust at only {4.95\% to 7.52\%} lower than the ``NREL-parameterized'' results. Importantly, the difference in peak demand between the two parameterizations is small, with less than an {8\%} difference. The results indicate that using a larger archetype size or an alternative weather dataset would likely not confer a significant impact on our findings with respect to future supply-side investments. Furthermore, all the results in this discussion have pertained solely to the residential sector, which only makes up a portion of the economy-wide demand data fed through to our supply-side \ModelName{} model, therefore tempering the sensitivity of our model outputs to the residential demand alone.


\subsection{Spatial Distribution of Peak Load Increases}\label{SIsec:spatialdemand}

In addition to the map in the main text, we include a tabular summary of the spatially diverse increases in peak load in Table \ref{SItab:spatialdemand} below.

\begin{table*}[ht]
\caption{Average peak electricity demand increases by zone. Zone topology presented in Section \ref{SIsec:Powernodes}.}
\centering
\scriptsize
{\begin{tabular}{l|rr}
\toprule
Zone & Average Peak Electricity & Average Peak Electricity \\
     & Demand Increase (\%) &  Demand Increase (GW) \\
\midrule
1  & 115.77 & 0.79 \\
2  & 84.06 & 1.20 \\
3  & 44.99 & 0.34 \\
4  & 70.03 & 0.53 \\
5  & 81.24 & 0.89 \\
6  & 77.04 & 0.71 \\
7  & 62.43 & 0.55 \\
8  & 138.96 & 0.81 \\
9  & 90.23 & 0.53 \\
10 & 115.46 & 0.65 \\
11 & 60.18 & 0.48 \\
12 & 112.06 & 0.61 \\
13 & 41.72 & 0.41 \\
14 & 42.28 & 0.39 \\
15 & 37.05 & 0.34 \\
16 & 31.02 & 0.27 \\
17 & 25.25 & 0.20 \\
\bottomrule
\end{tabular}}
\label{SItab:spatialdemand}
\end{table*}

\subsection{Spatial Distribution of New Investment}\label{SIsec:CapacityMap}
\begin{figure}
     \centering
     \begin{subfigure}[b]{0.5\textwidth}
             \centering    \includegraphics[width=\textwidth]{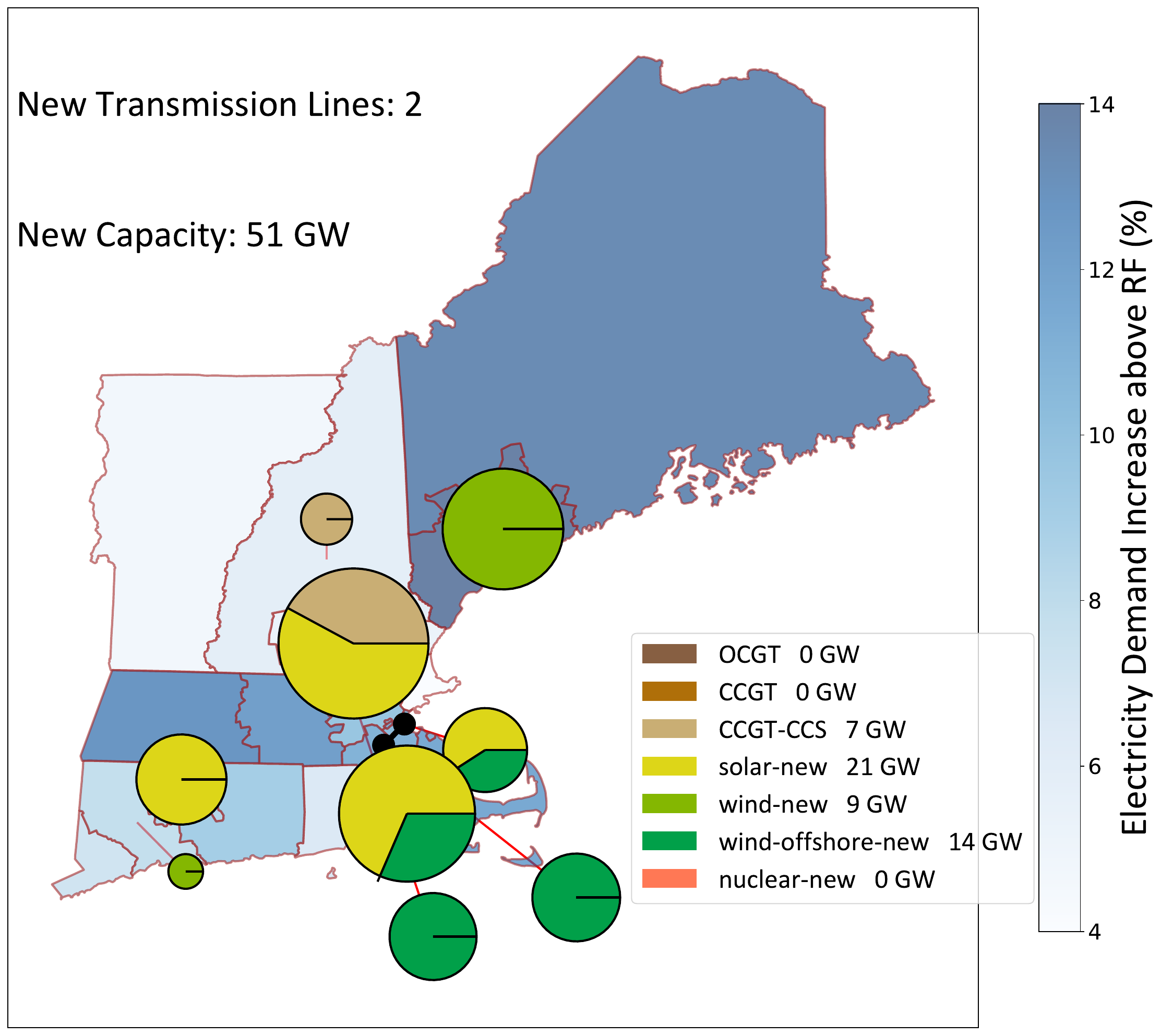}
    \caption*{(a) 80\% decarbonization target}
    \label{SIfig:spatial-cap-80}
     \end{subfigure}%
     \hfill
    \begin{subfigure}[b]{0.5\textwidth}
             \centering    \includegraphics[width=\textwidth]{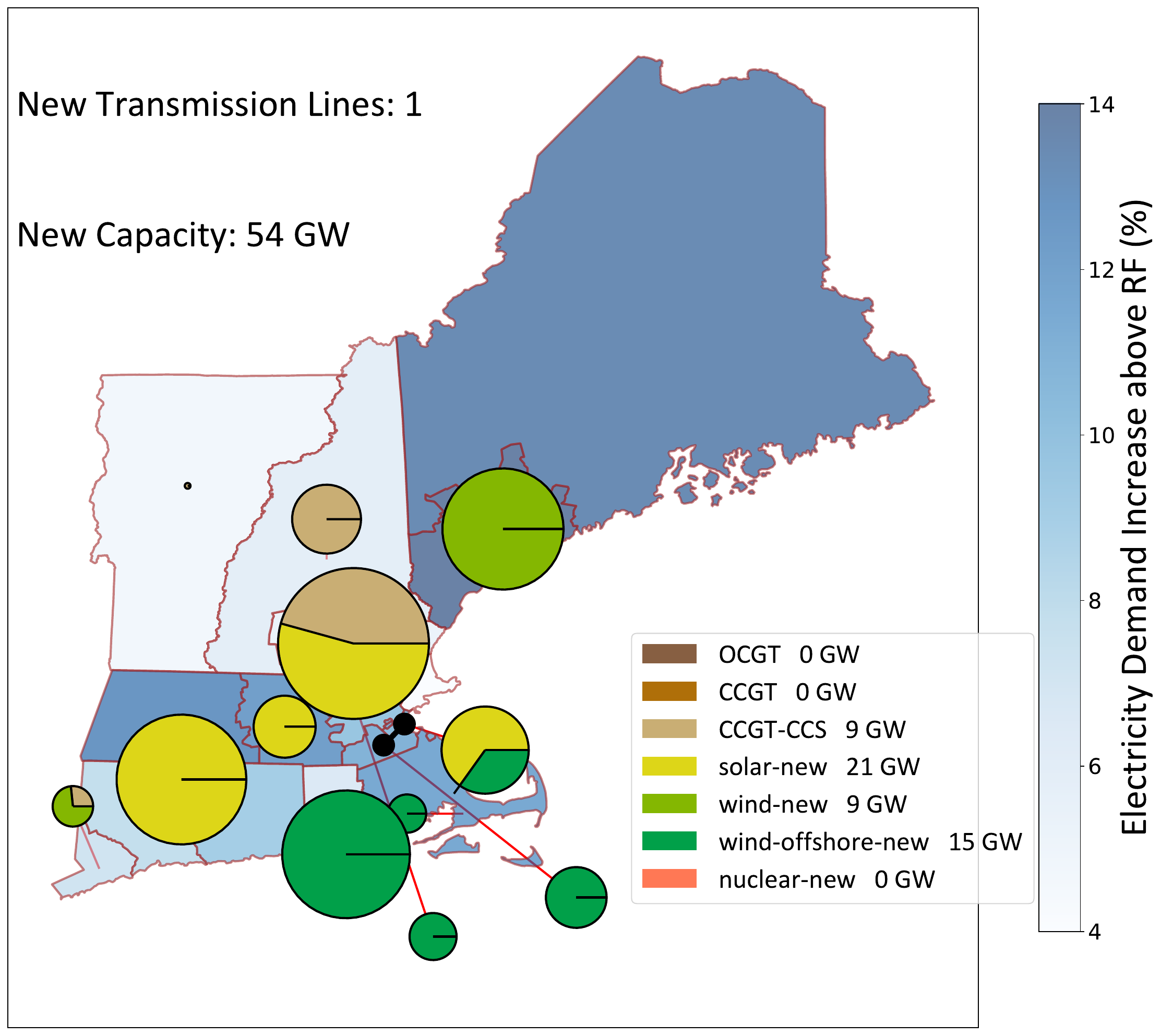}
    \caption*{(b) 95\% decarbonization target}
    \label{SIfig:spatial-cap-95}
     \end{subfigure}
     \caption{New transmission lines and capacity addition in the region under HX scenario and exemplar weather year of 2003. The size of the pie charts is proportional to total capacity it represents. The color intensity of each zone indicates the percentage of electricity load increase between HX and RF scenarios.}
     \label{SIfig:spatial-cap}
\end{figure}

The spatial distribution of new generation and transmission capacity for the HX scenario  under 80\% and 95\% decarbonization targets is highlighted in Fig.~\ref{SIfig:spatial-cap}.  As pointed out by previous studies \citep{Brattle2019,E3Team2020}, the region needs substantial expansion of its current transmission line infrastructure to support a generation mix dominated by variable renewable energy (VRE) and achieve decarbonization targets. For the exemplar 2003 weather year, Fig.~\ref{SIfig:spatial-cap} shows substantial investment for both decarbonization targets in new offshore wind and solar generation in coastal regions of Massachusetts. 

The decarbonization target's stringency necessitates investment in new power plants. However, a mild expansion of transmission lines is also required. As explained in Section~\ref{SIsectrans-lines}, all candidate transmission lines are assumed to have the same maximum flow capacity. 
Across the two scenarios, the spatial distribution in capacity does not strongly correlate with the spatial distribution in incremental load increases which can be due to sufficient transmission capacity that connects the states with higher load increase, such as Maine, to states with higher capacity expansion, such as Massachusetts.  

\subsection{Example Winter Day Operations}\label{SIsec:dispatch}
\begin{figure}
     \centering
     \begin{subfigure}[b]{\textwidth}
             \centering    \includegraphics[width=\textwidth]{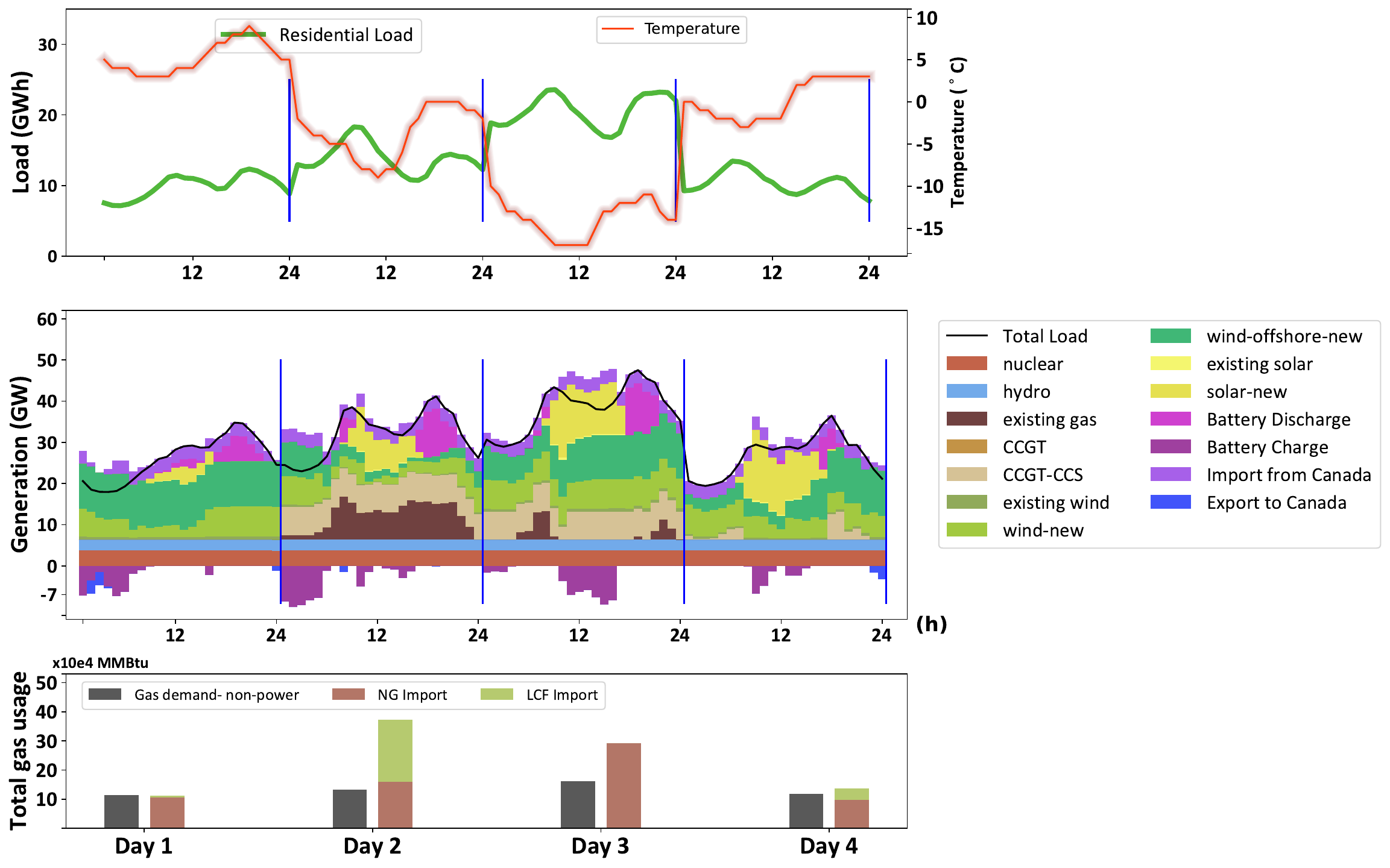}
    \caption*{(a) 80\% decarbonization target}
    \label{SIfig:temp-power-gas-80}
     \end{subfigure}
     \hfill
    \begin{subfigure}[b]{\textwidth}
             \centering    \includegraphics[width=\textwidth]{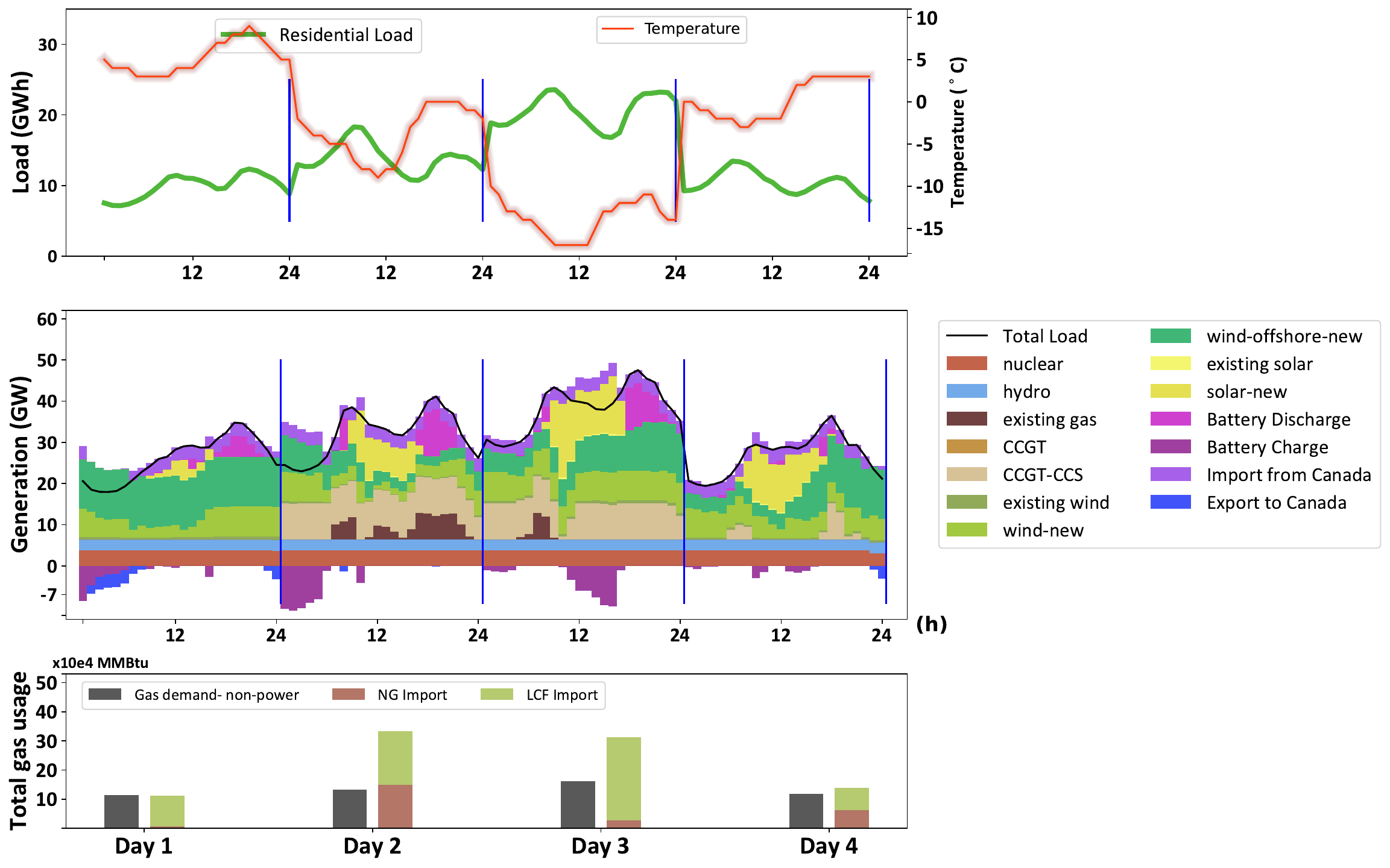}
    \caption*{(b) 95\% decarbonization target}
    \label{SIfig:temp-power-gas-95}
     \end{subfigure}
     \caption{Model inputs and outcomes for four representative winter days with exemplar 2003 weather year under HX scenario. The days are not necessarily sequential. The top plot shows the hourly residential load and ambient temperature on the left and right x-axes, respectively. The middle plot is the hourly dispatch decisions, battery charge and discharge, import from and export to Quebec, Canada, and the total hourly load of the region. The bottom plot depicts the daily gas load as well as daily NG and LCF imports.}
     \label{SIfig:temp-power-gas}
\end{figure}

As an example of system operation under high electrification and decarbonization scenarios, Fig.~\ref{SIfig:temp-power-gas} shows the various input and output indicators for four different winter days with the exemplar 2003 weather year and HX scenario under 80\% and 95\% decarbonization targets. 

The total load generally follows the temperature; the total load is higher on {colder ``Day 3'', and much lower on warmer ``Day 1'', ``Day 2'', and ``Day 4.''} On the supply side, nuclear and domestic hydropower are operating close to nameplate capacity for all days, while CCGT-CCS is utilized in ``Day 2'' and ``Day 3'' in relatively high capacity factors.
 In all four days, we see that
battery storage and hydro imports from Canada (details in Section~\ref{SIsec:import-node}) play a critical role in meeting peak demand. 

The generation mix depends on the total load and availability of VRE resources. ``Day 1'' has a low load, low solar CF but high wind CFs. {Accordingly, the system fully deploys wind generation assets and compensates for peak load hours with battery storage. ``Day 1'' has negligible gas generation. Conversely,  ``Day 2'' has considerable generation from CCGT-CCS and CCGT owing to relatively low wind CFs. A combination of high load {due to the extreme cold temperatures} and low solar CFs in the morning requires the dispatch of existing unabated gas generation.
The lower load of the day coupled with mild solar CF, battery storage and import from Quebec is sufficient to meet the demand. ``Day 3'' has a significantly higher load, and solar and wind CFs. The load magnitude of ``Day 4'' is similar to that of ``Day 1.'' However, the solar CFs are higher, but wind CFs are lower for the day, resulting in some deployment of abated gas generation under both decarbonization levels.}
 
{The imports of NG and LCF highly depend on gas generation plants in the power system. The total import and non-power demand for ``Day 1'' and ``Day 4'' are almost equal under both decarbonization targets. The discrepancy between the non-power gas demand and gas import (i.e., NG and LCF) increases on two other days (indicating gas is needed for power generation), culminating in ``Day 3'' where the total import is almost twice as the non-power gas load. LCF is a more expensive gas fuel than NG, hence its import surges only when the model needs substantial gas generation but the decarbonization targets prevent the import of carbon-intensive NG beyond a certain level.}


\subsection{The Impact of CCS Technology}\label{SIsec:CCS}
\begin{figure}[htbp]
    \centering
    \includegraphics[width=0.8\textwidth]{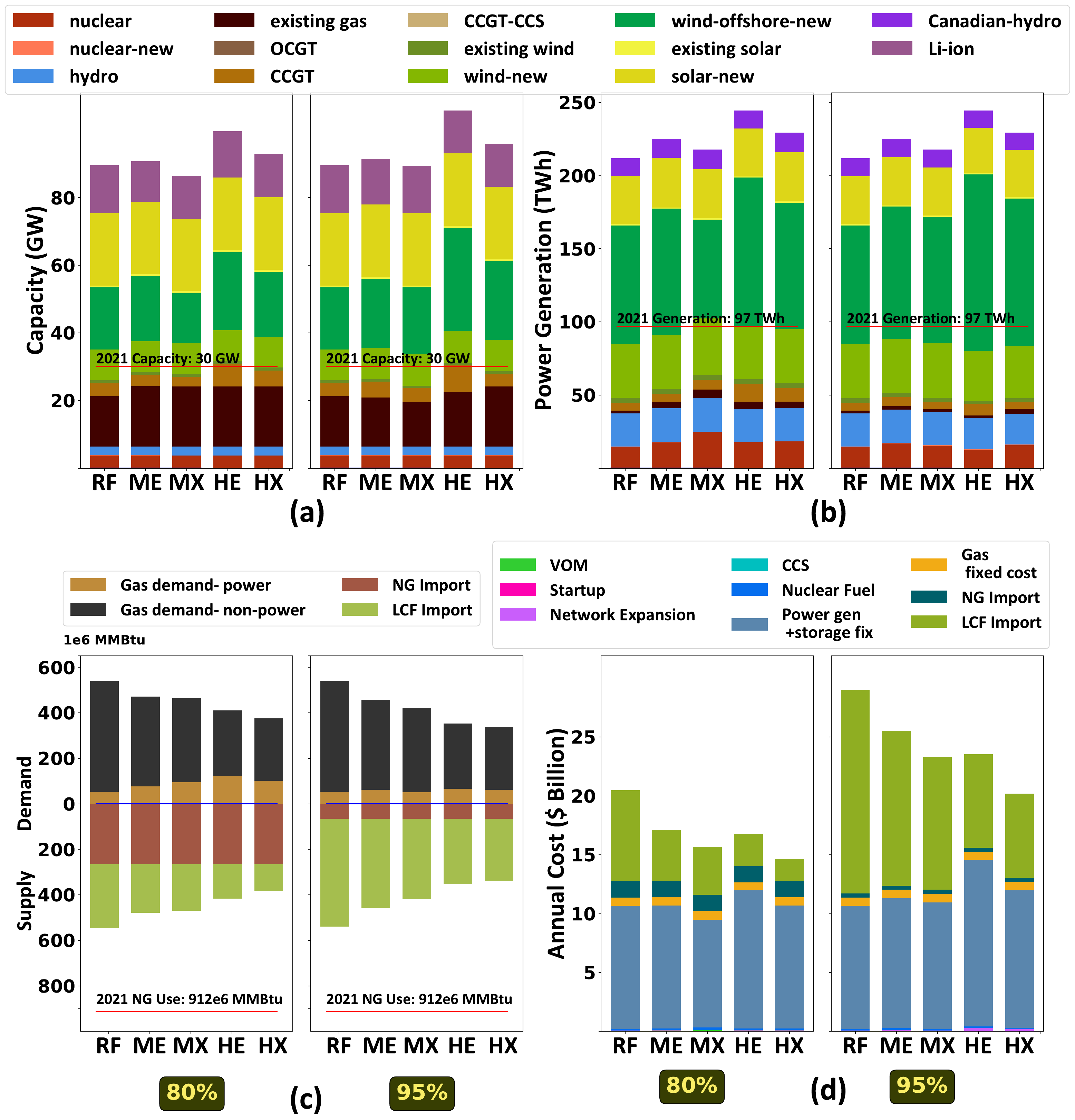}
    \caption{Capacity, generation, gas consumption characteristics, and cost breakdown for the no-CCS case. In the 80\% emissions cases, costs are comparable to the result where CCS is allowed. In the 95\% case, costs are higher. Both emissions cases require higher amounts of offshore wind, unabated thermal, and LCF deployment than the corresponding case where CCS is allowed.}
    \label{SIfig:cap-gen-cost-noCCS}
\end{figure}

\begin{figure}[htbp]
  \begin{subfigure}{1\textwidth}
    \centering
    \includegraphics[width=0.7\textwidth]{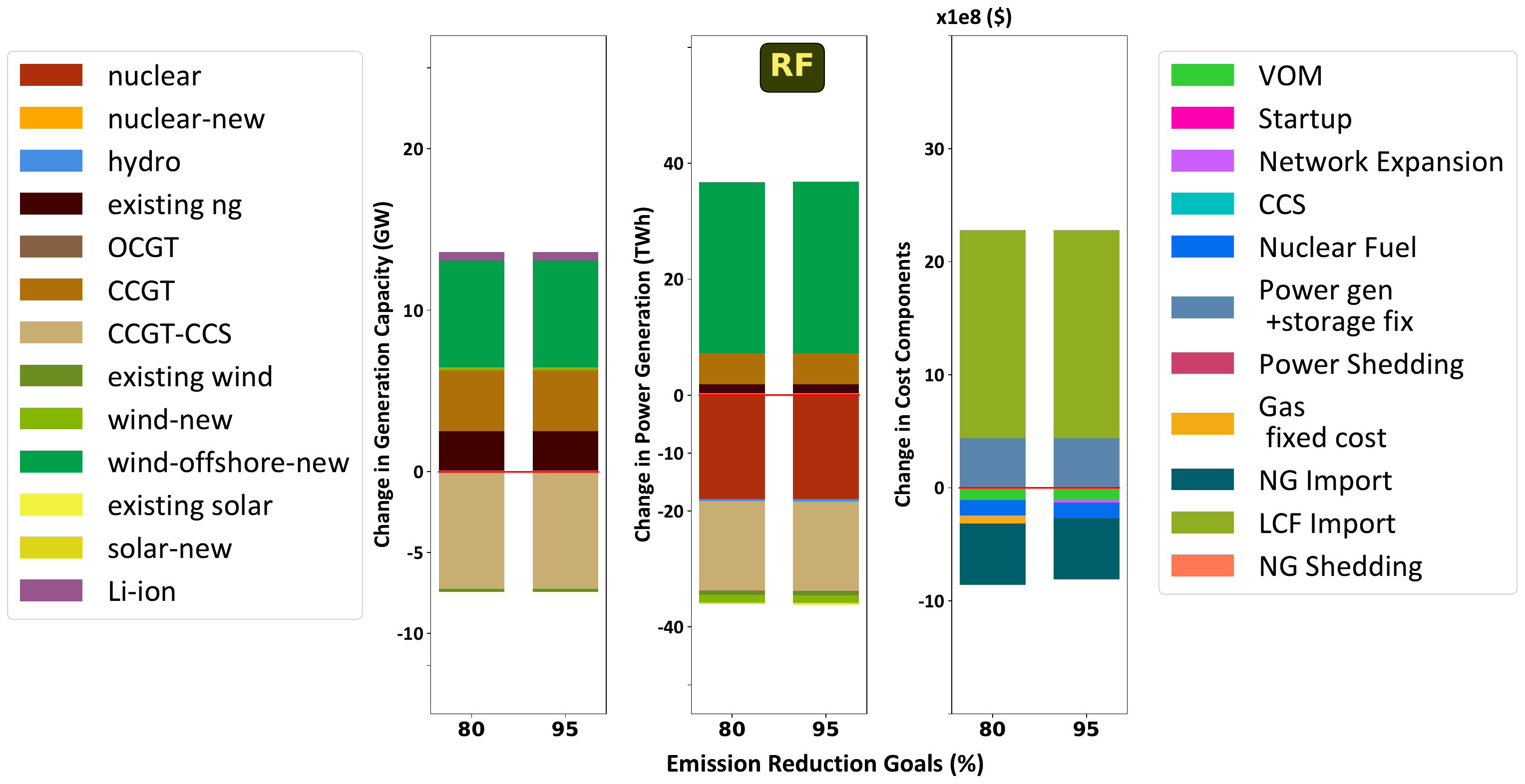}
  \end{subfigure}%
  \medskip
  \vspace{-0.5cm}
  \begin{subfigure}[t]{0.5\textwidth}
        \centering
        \includegraphics[width=0.7\textwidth]{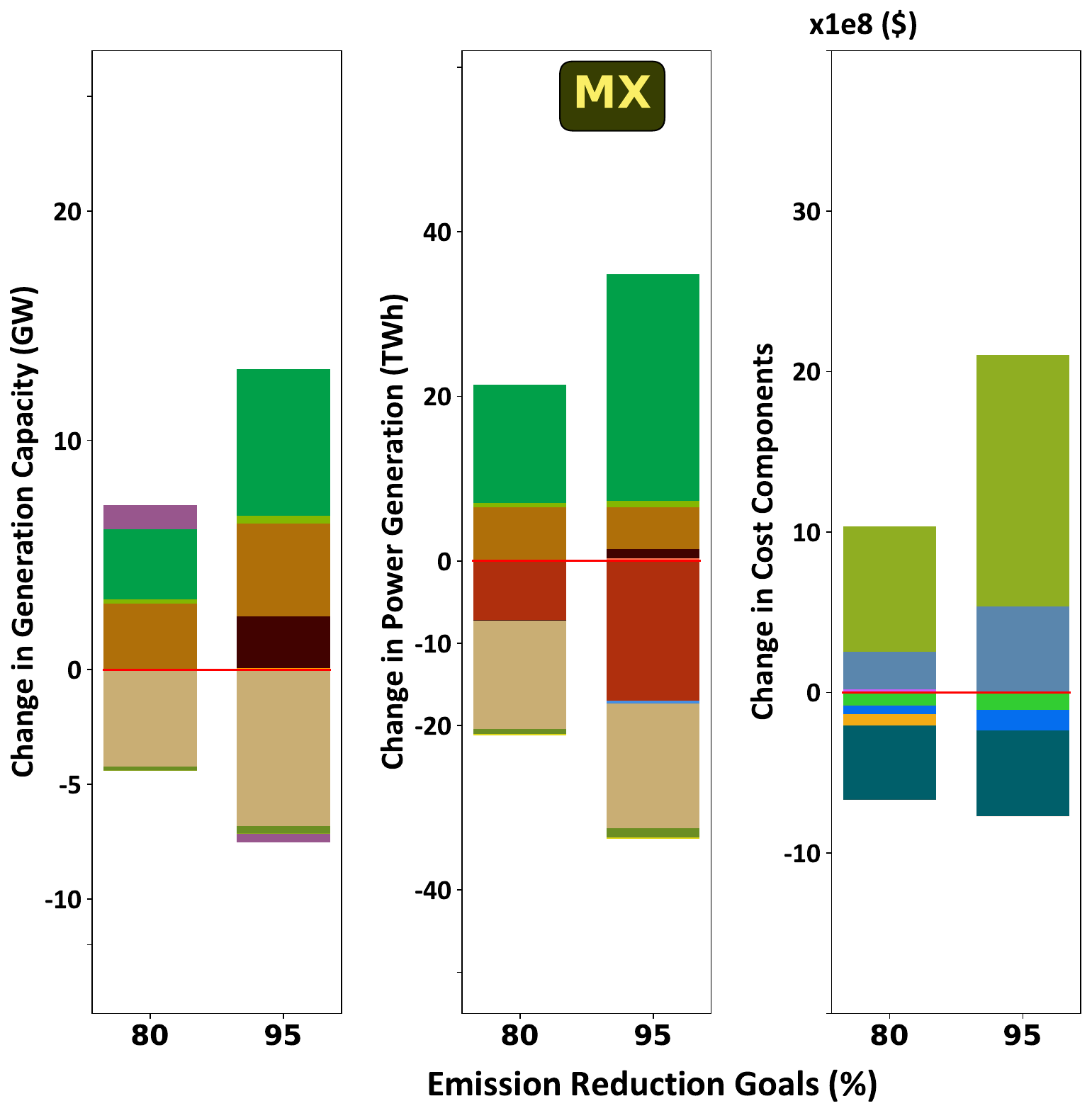}
    \end{subfigure}%
    \begin{subfigure}[t]{0.5\textwidth}
        \centering
        \includegraphics[width=0.7\textwidth]{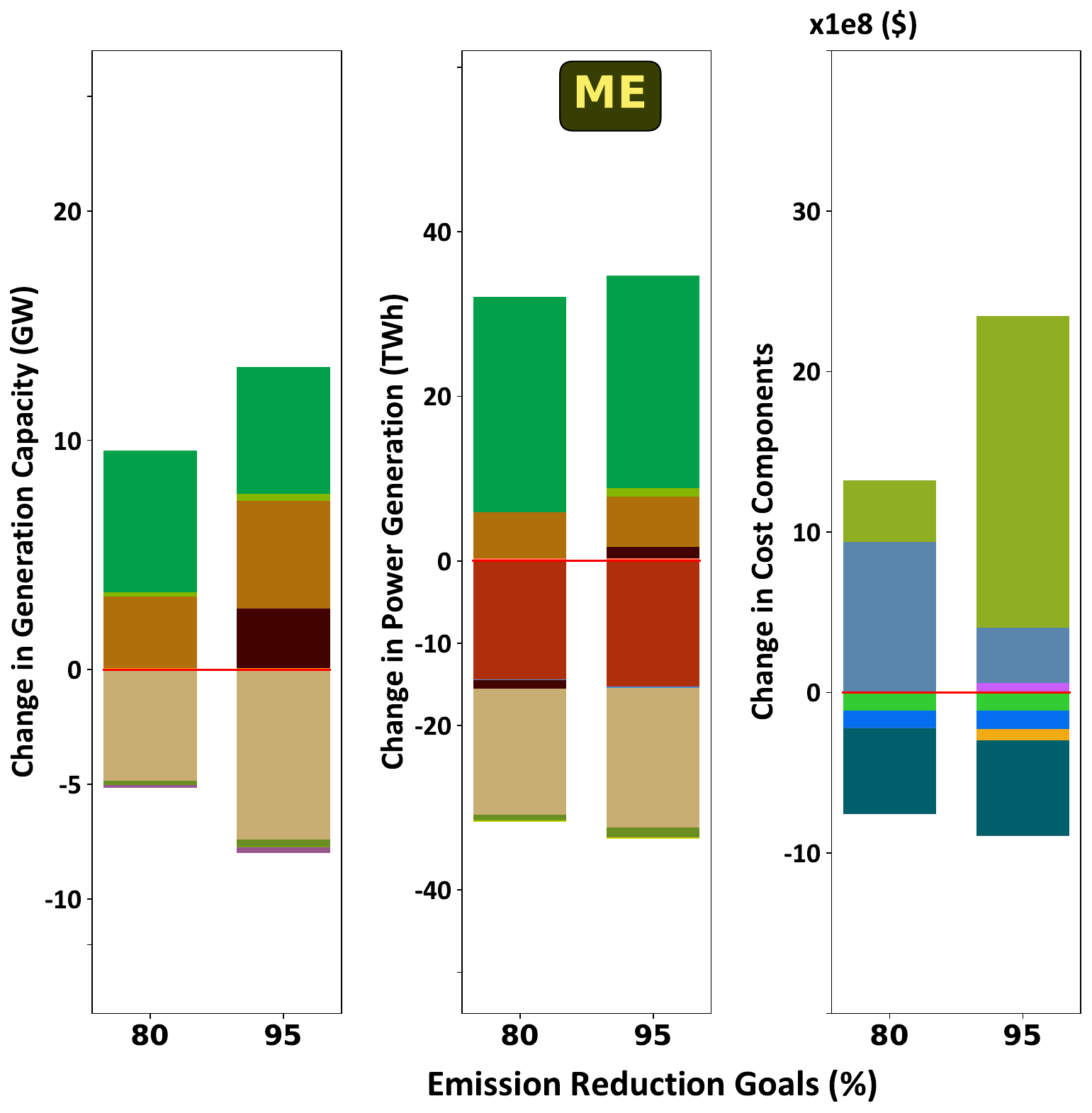}
    \end{subfigure}
    \medskip

    \vspace{-0.5cm}
      \begin{subfigure}[t]{0.5\textwidth}
        \centering
        \includegraphics[width=0.7\textwidth]{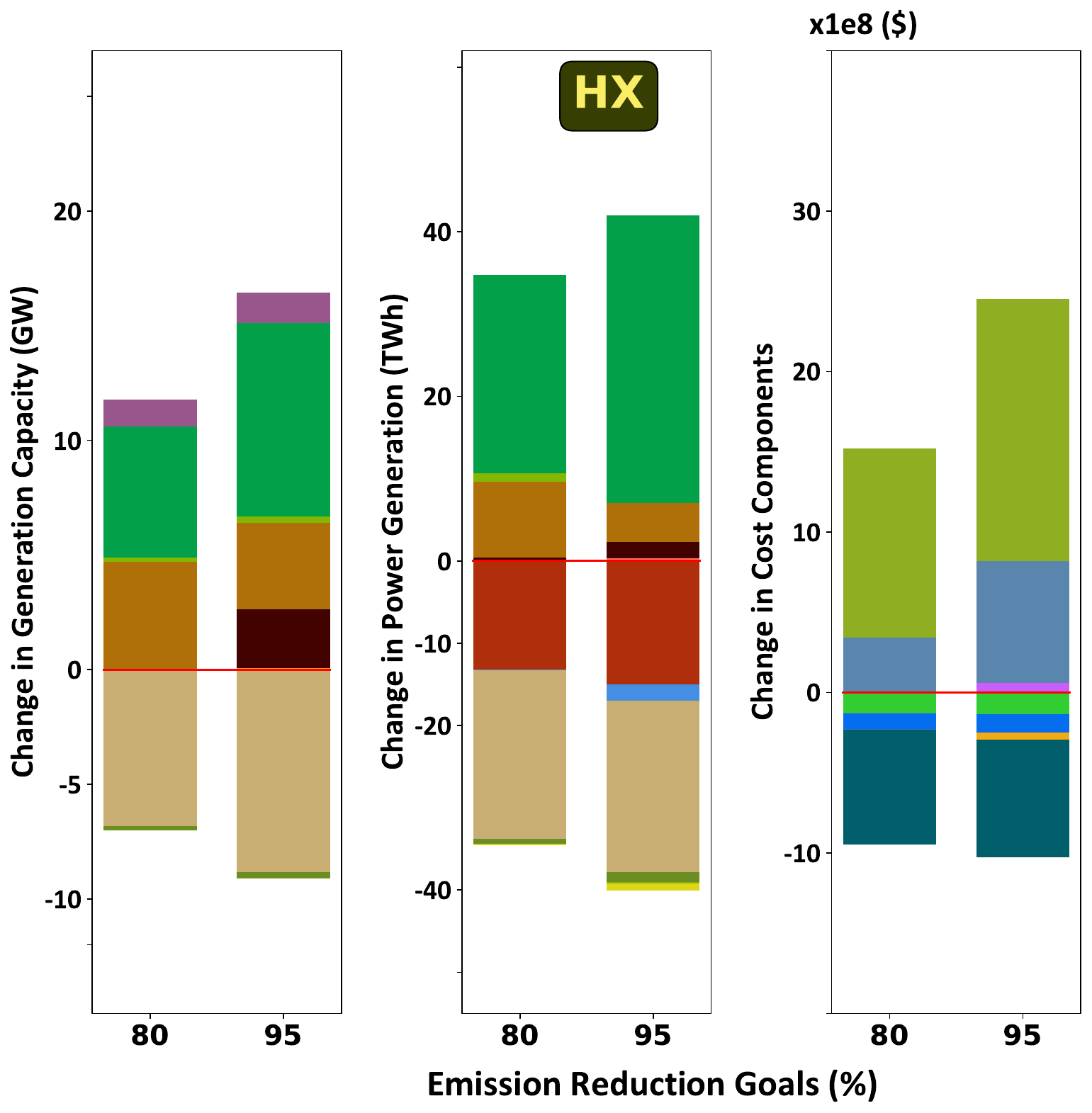}
    \end{subfigure}%
    \begin{subfigure}[t]{0.5\textwidth}
        \centering
        \includegraphics[width=0.7\textwidth]{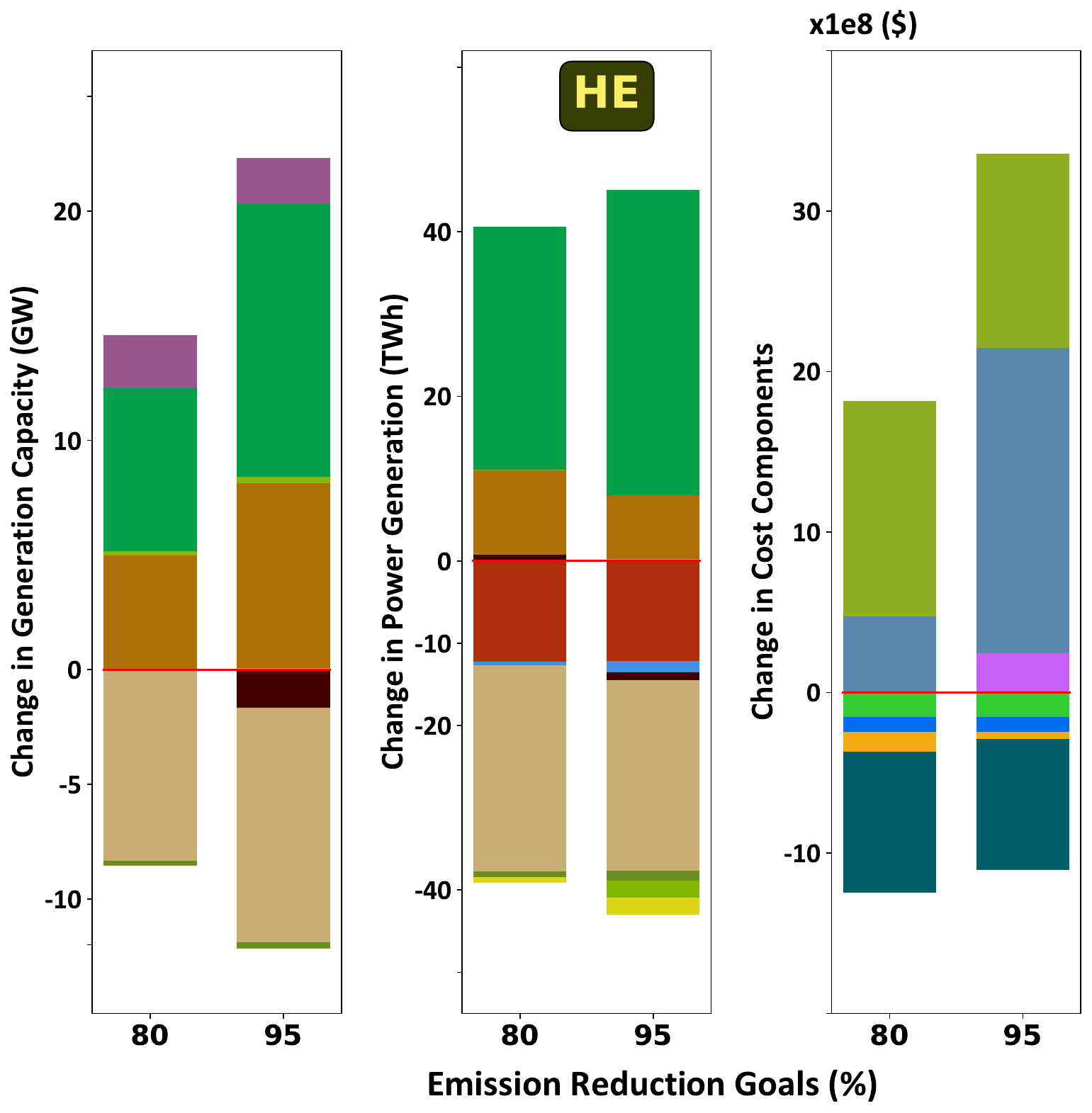}
    \end{subfigure}
    \caption{Difference in capacity, generation, and cost components between with and without CCS instances for exemplar weather year 2003 under different decarbonization targets and electrification scenarios.}
    \label{SIfig:diff-no-CCS}
\end{figure}

We explore the sensitivity of supply-side outcomes to the availability of CCS-based power generation, with results summarized Fig.~\ref{SIfig:cap-gen-cost-noCCS}, Fig.~\ref{SIfig:diff-no-CCS} and Table \ref{SItab:CCS_differences}  Without CCS, the system cost increases by up to \bt{7.3\%} in the 80\% emissions case and by up to \bt{5.2\%} in the 95\% emissions reduction cases. This highlights the increasing value of low-carbon firm dispatchable generation sources like CCS-based power generation with increasing stringency of emissions constraint. Without CCS availability, the cost-optimal power capacity portfolio has a greater reliance on offshore wind, unabated gas power capacity which requires using more LCF for power generation (Table \ref{SItab:CCS_differences}), and also battery energy storage capacity (primarily in the 95\% emissions reduction case).




\begin{table*}[ht]
\caption{Comparison of CCS and no-CCS cases for exemplar weather year 2003. Percent differences for each category span the five electrification scenarios presented in the main analysis. A positive percentage indicates a result that is higher for the no-CCS case.}
    \centering
    \scriptsize
    {\begin{tabular}{l|rrrrr}
    \toprule
  Emissions  &  Total &  Offshore & Unabated  & LCF  & Storage\\
 reduction & cost& capacity& thermal cap. &consumption&capacity\\
\midrule
80\% & \bt{2.3 to 7.3} \% &\bt{26 to 56\%} & \bt{-12 to -5\%} & \bt{7 to 66\%} & \bt{0 to 20\%} \\
95\% & \bt{0.5 to 5.2\%} & \bt{37 to 64\%} & \bt{-13 to 0\%} & \bt{8 to 13\%} & \bt{-2 to 18\%} \\
\bottomrule
    \end{tabular}}
    \label{SItab:CCS_differences}
\end{table*}


\subsection{The impact of methane leakage}\label{SIsec:methaneleakage} 

This section elaborates on our estimation of methane leakage. Recent estimates suggest that about 2.3\% of gross gas production in the US is lost due to methane emissions from the production points to end-use \citep{AlvarezEtal2018}, of which the upstream and midstream stages collectively are responsible for 96\% of all emissions. For LCF, whose supply and upstream emissions are quite uncertain, we only account for midstream emissions, assuming that they will be similar to those for natural gas.

\begin{figure}[htbp]
  \begin{subfigure}{1\textwidth}
    \centering
    \includegraphics[width=0.8\textwidth]{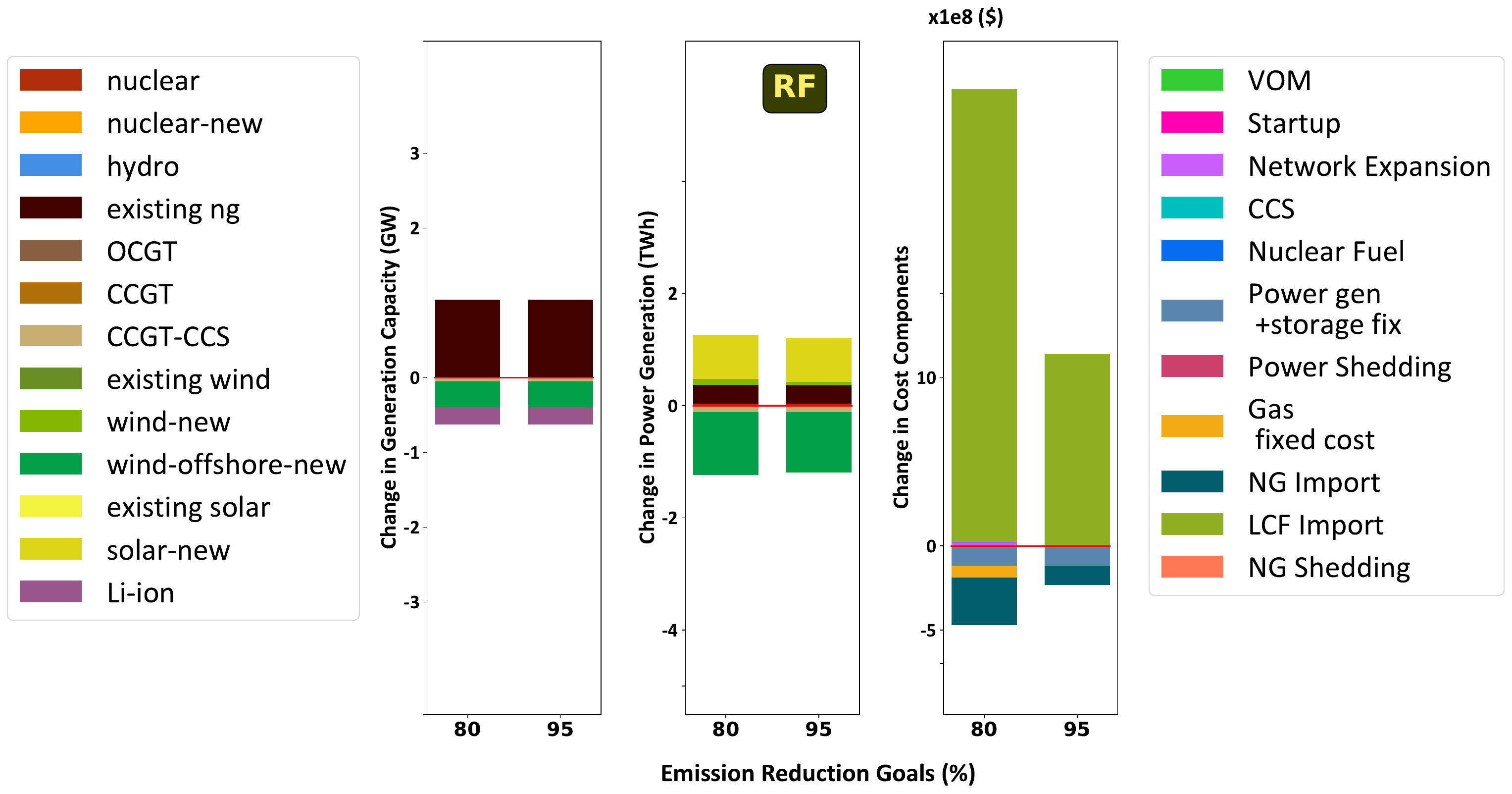}
  \end{subfigure}%
  \medskip
  
  \begin{subfigure}[t]{0.5\textwidth}
        \centering
        \includegraphics[width=0.8\textwidth]{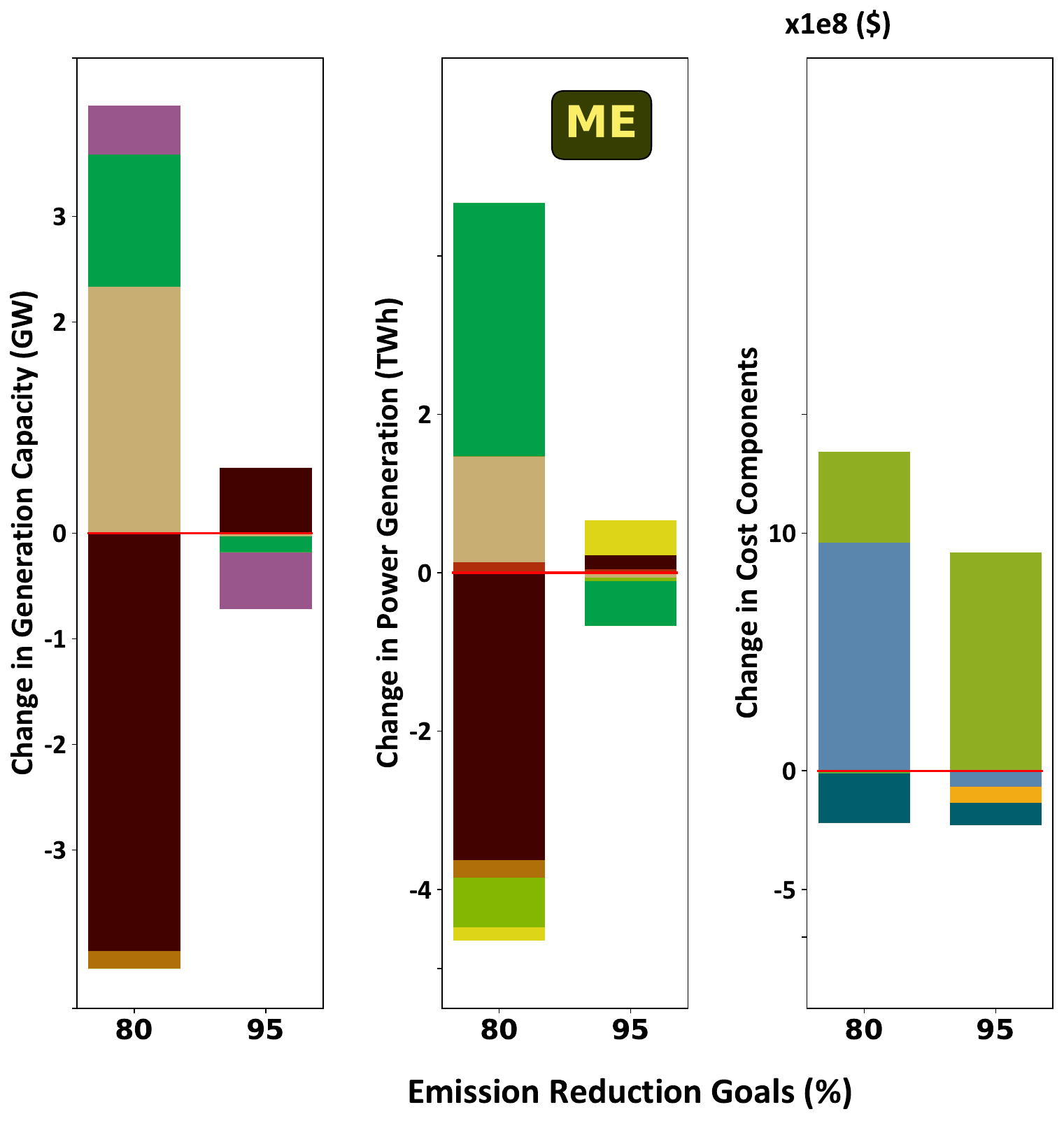}
    \end{subfigure}%
    \begin{subfigure}[t]{0.5\textwidth}
        \centering
        \includegraphics[width=0.8\textwidth]{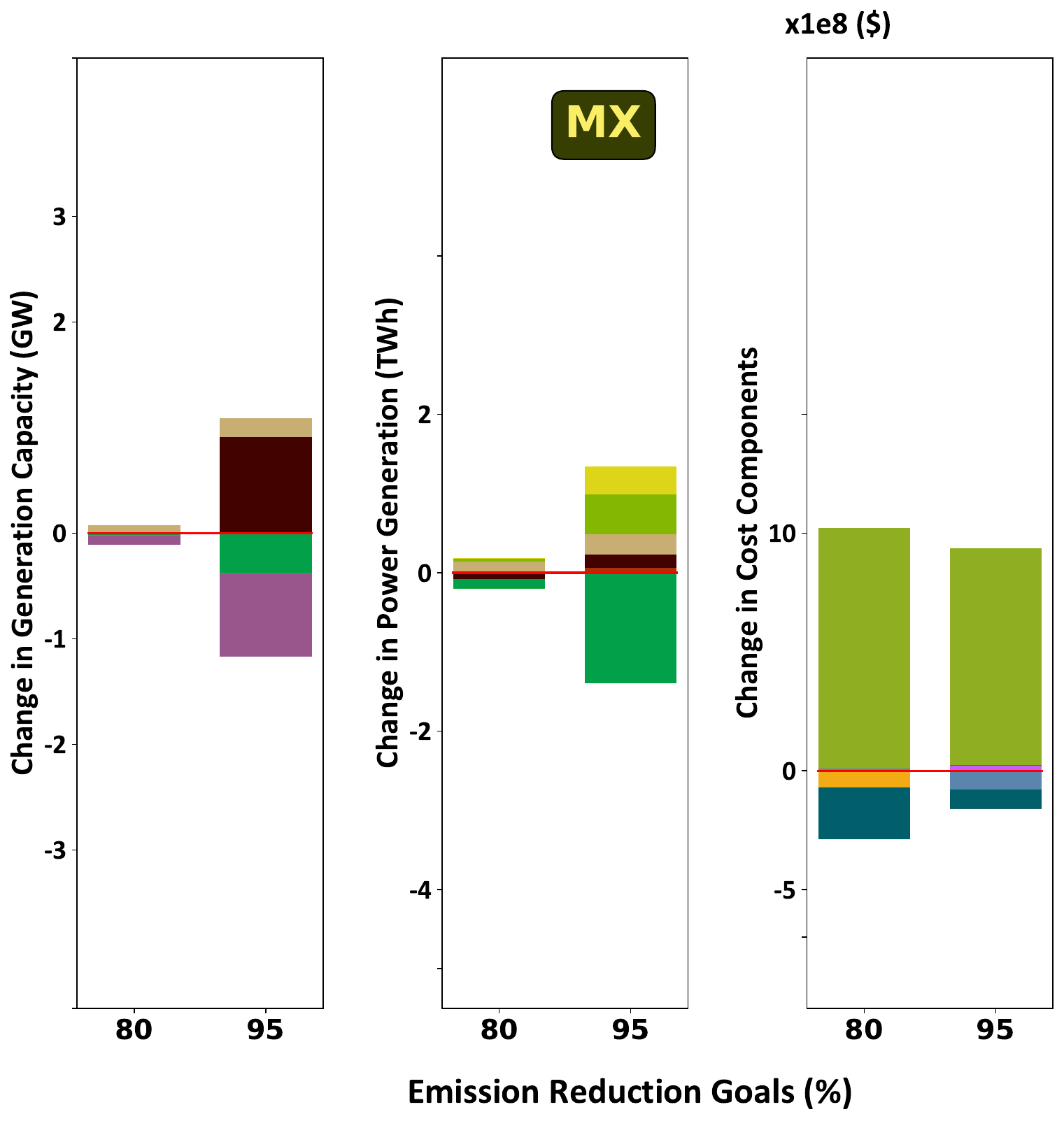}
    \end{subfigure}
    \medskip

      \begin{subfigure}[t]{0.5\textwidth}
        \centering
        \includegraphics[width=0.8\textwidth]{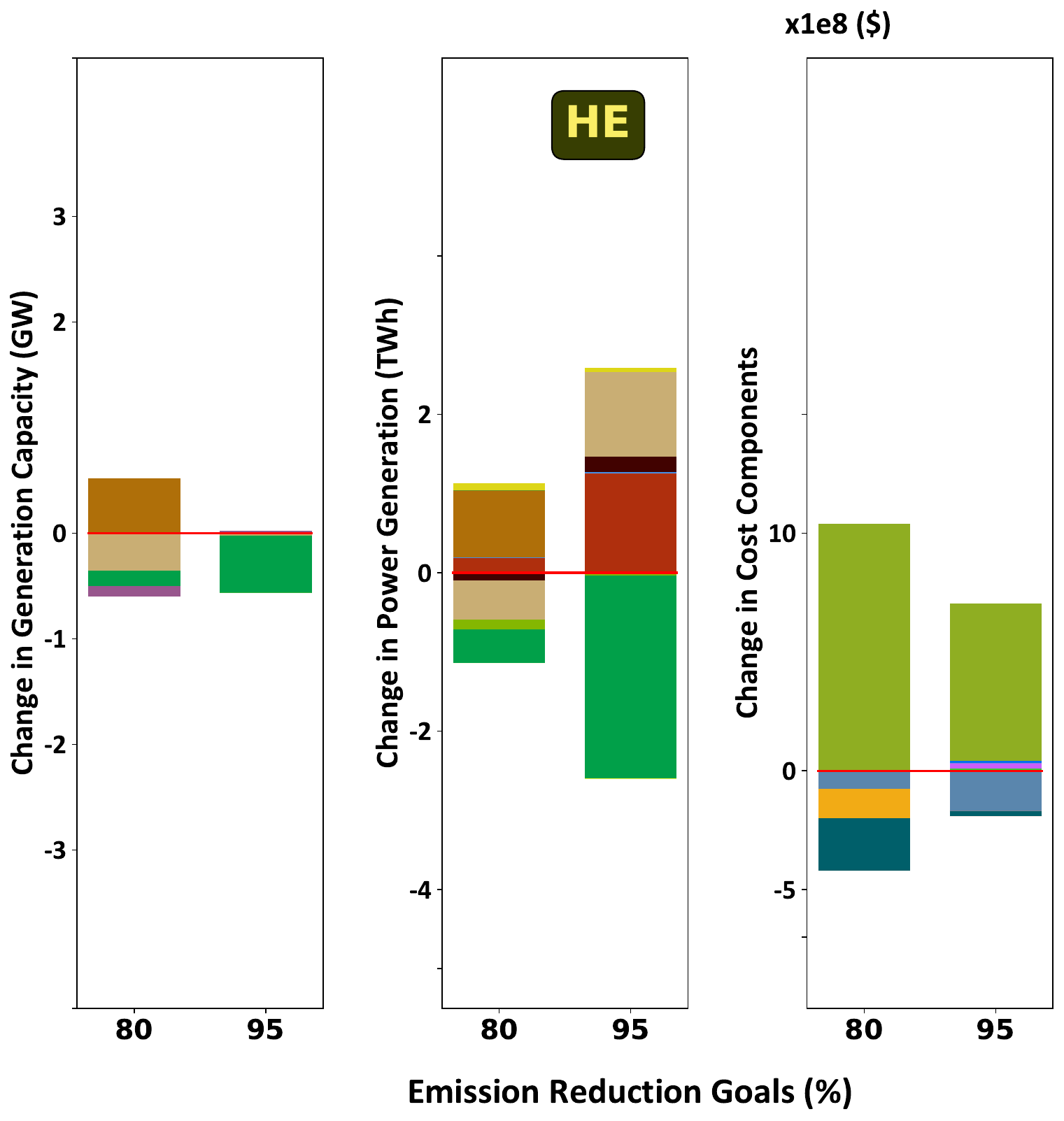}
    \end{subfigure}%
    \begin{subfigure}[t]{0.5\textwidth}
        \centering
        \includegraphics[width=0.8\textwidth]{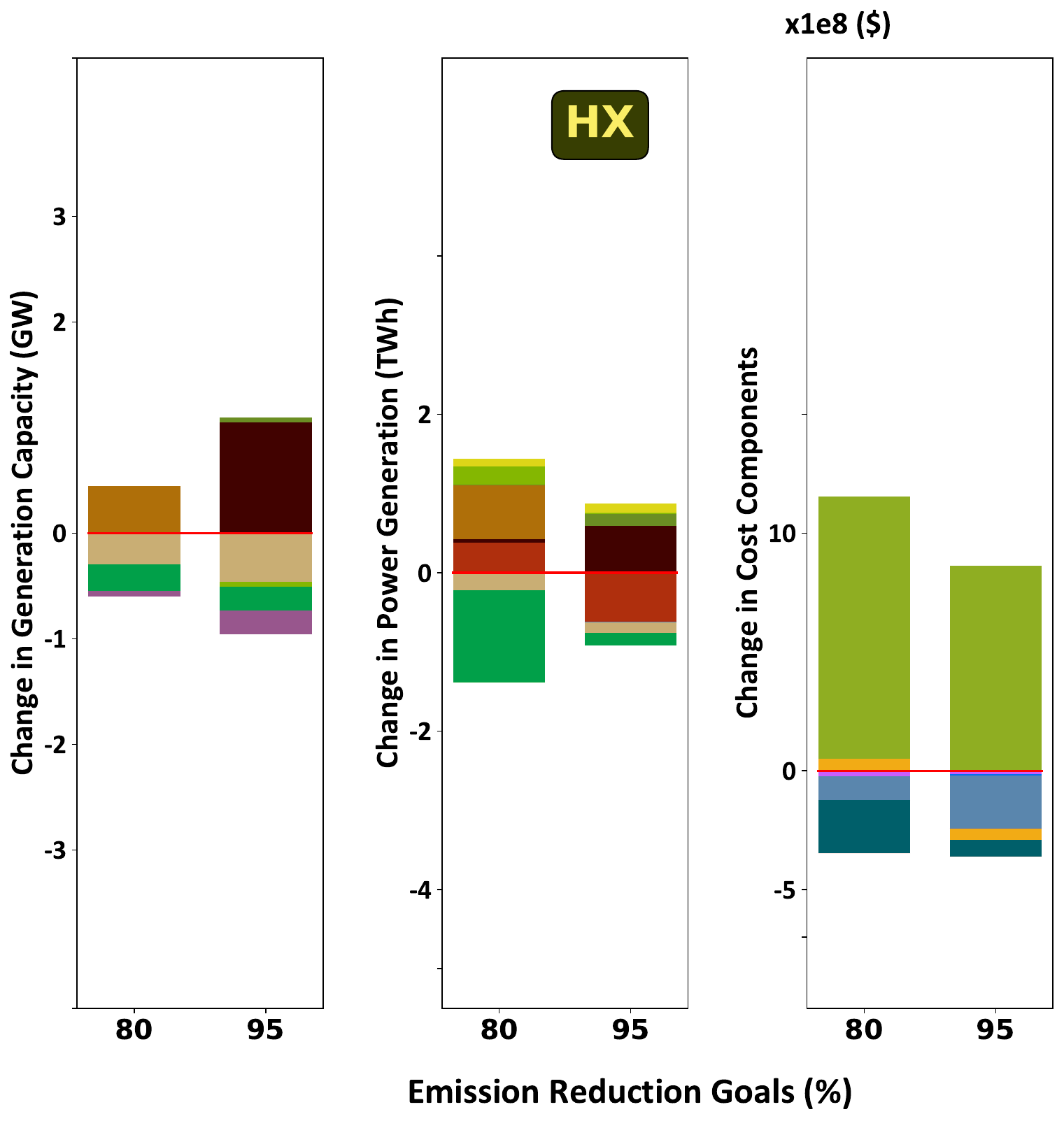}
    \end{subfigure}
    \caption{Difference in capacity, generation, and cost components between with and without methane leakage instances (instances with methane leakage - without methane leakage) for exemplar weather year 2003 under different decarbonization targets and electrification scenarios.}
    \label{SIfig:diff-methane-leak}
\end{figure}

Table~\ref{SItab:methane_emis} summarizes the methane emissions intensity and associated kgCO2 equivalent emissions intensity for natural gas and LCF considered in the evaluated scenarios. The difference in outcome per each electrification scenario, with and without methane leakage, is shown in Fig.~\ref{SIfig:diff-methane-leak}.

{\scriptsize
\begin{table}[htbp]
    \centering
    \begin{tabular}{l|lll }
    \toprule
    Property & Value & Unit & Source \\
    \midrule
    Volumetric energy density (HHV)&	1089	& btu/cf	&  \citep{GREET_2021}\\
    Gravimetric density (32F and 1 atm)$^1$	&22	& g/cf	&  \citep{GREET_2021} \\
  Carbon content &	72.40\%	&wt\%& \citep{GREET_2021}\\
  CO2 emissions intensity	&0.0536&	tCO2/MMBtu&\\
\midrule 
&&&\\
\textbf{Pure methane properties}&&&\\
Volumetric energy density (HHV)&	1068&	btu/cf& \citep{GREET_2021}\\
Gravimetric density (32F and 1 atm)$^1$&	20.3&	g/cf& \citep{GREET_2021}\\
Carbon content &75.00\%&	wt\%&\citep{GREET_2021}\\
CO2 emissions intensity	&0.0523& tCO2/MMBtu&\\
Methane content& 90\%&	vol\%&\\
\vspace{0.1cm}
Methane leakage rate&2.30\%&	gCH$_4$/gCH$_4$ in NG &\\
Upstream emissions (production,&	96\% &&\\
\vspace{0.1cm}
gathering, processing, trans. + storage)$^2$&&\\
Upstream emissions &	14\% &&\\
\vspace{0.1cm}
(transport + storage) of LCF$^2$&&\\
100-year Global warming potential CH$_4$& 29.80&	kgCO2eq/kgCH$_4$&\citep{IPCC_AR6_report2021}\\
Upstream methane leakage rate of NG&	0.022&	gCH$_4$/gCH$_4$&\\
\vspace{0.1cm}
Midstream methane leakage rate of LCF&	0.003&gCH$_4$/gCH$_4$&\\
Emissions leakage & 0.0226&	gCH$_4$/gCH$_4$&\\
per unit of methane consumed&&&\\
\midrule
{Total CO2eq emiss. intensity of NG}&0.0649&tCO2eq/MMBtu&\\
{Total CO2eq emiss. intensity of LCF}&0.00161&tCO2eq/MMBtu&\\
\bottomrule
    \end{tabular}
    \caption{Methane emissions intensity for natural gas and LCF\\
    {\footnotesize $^1$ NG density at 60F and 1 atm $\quad$ $^2$ of total CH$_4$ emissions $\quad$ $^3$ The capture rate for CCGT-CCS plants is reported for combustion and does not account for the upstream emissions. The inclusion of methane emissions increases the carbon intensity of NG by 22\%. Therefore, we reduce the capture rate of these plants by 22\%, to 69.3\%.} }
    \label{SItab:methane_emis}
\end{table}}

\subsection{The impact of transportation flexibility}\label{SIsec:transportflexibility}
{This section evaluates the impact of flexibility on transportation load. Although our analysis is primarily focused on the impacts of heat electrification in the residential sector under varying future scenarios, transportation is another sector with substantial potential for increased load in coming decades. Furthermore, transportation load is oft-considered as the best candidate sector for demand flexibility interventions, enabled by various demand response programs and tariff designs. }

{We used the modeling framework used in the GenX model \citep{GenX2017} to formulate the representation of transportation flexibility. Let $s^{\text{trRem}}_{nt}$ be the deferred demand remaining to be served during a flexibility event. Let $s^{\text{ser}}_{nt}, s^{\text{def}}_{nt}$ be served and deferred demand, respectively. We denote the number of hours the flexible transportation load can be advanced or delayed as $T^{\text{adv}}$ and $T^{\text{def}}$. Let $\rho^{\text{tr}}_{nt}$ be the share of the transportation load in the total load.  The following constraints are added to the formulation:}
{
\begin{subequations}
    \begin{align*}
        &s^{\text{trRem}}_{nt} = s^{\text{trRem}}_{n,t-1}+s^{\text{def}}_{nt}-s^{\text{ser}}_{nt} &n\in \mathcal{N}^{\text{e}}, t\in \mathcal{T}^{\text{e}}\backslash \{t^{\text{start}}_\tau \lvert \  \tau \in \mathfrak{R} \}\\
        & s^{\text{def}}_{nt} \leq \rho^{\text{tr}}_{nt} D^e_{n\phi^e_t}&n\in \mathcal{N}^{\text{e}}, t\in \mathcal{T}^{\text{e}}\\
        &\sum_{t'=t+1}^{t+T^{\text{def}}}s^{\text{ser}}_{nt'} \geq s^{\text{trRem}}_{nt}&n\in \mathcal{N}^{\text{e}}, t\in \mathcal{T}^{\text{e}}\\
        &\sum_{t'=t+1}^{t+T^{\text{adv}}}s^{\text{def}}_{nt'} \geq -s^{\text{trRem}}_{nt}&n\in \mathcal{N}^{\text{e}}, t\in \mathcal{T}^{\text{e}}\\
        &s^{\text{trRem}}_{nt} \in \mathbb{R},s^{\text{def}}_{nt}, s^{\text{ser}}_{nt} \in \mathbb{R}^+&n\in \mathcal{N}^{\text{e}}, t\in \mathcal{T}^{\text{e}}
    \end{align*}
\end{subequations}}

{Consequently, we change the hourly power supply-demand constraints (7) to account for flexible demand from the transportation sector:}
{\begin{align}
        &\sum_{i \in \mathcal{P}}p_{nti} +\sum_{m\in \mathcal{N}^{\text{e}}}\sum_{l\in \mathcal{L}^{\text{e}}_{nm}}\sign(n-m) f^{\text{e}}_{\ell t}+
        \sum_{r\in \mathcal{S}^{\text{e}}_n} (s^{\text{eDis}}_{ntr}-s^{\text{eCh}}_{ntr}) \notag \\ 
        &+s^{\text{def}}_{nt}-s^{\text{ser}}_{nt}+ a^{\text{e}}_{nt}=D^{\text{e}}_{n \phi^{\text{e}}_t}+d_n E^{\text{pipe}}\kappa^{\text{pipe}}_{n}+ E^{\text{cprs}} E^{\text{pump}}\kappa^{\text{capt}}_{nt}&n\in \mathcal{N}^{\text{e}}, t\in \mathcal{T}^{\text{e}}
\end{align}}

\begin{figure}[htbp]
  \begin{subfigure}{1\textwidth}
    \centering
    \includegraphics[width=0.8\textwidth]{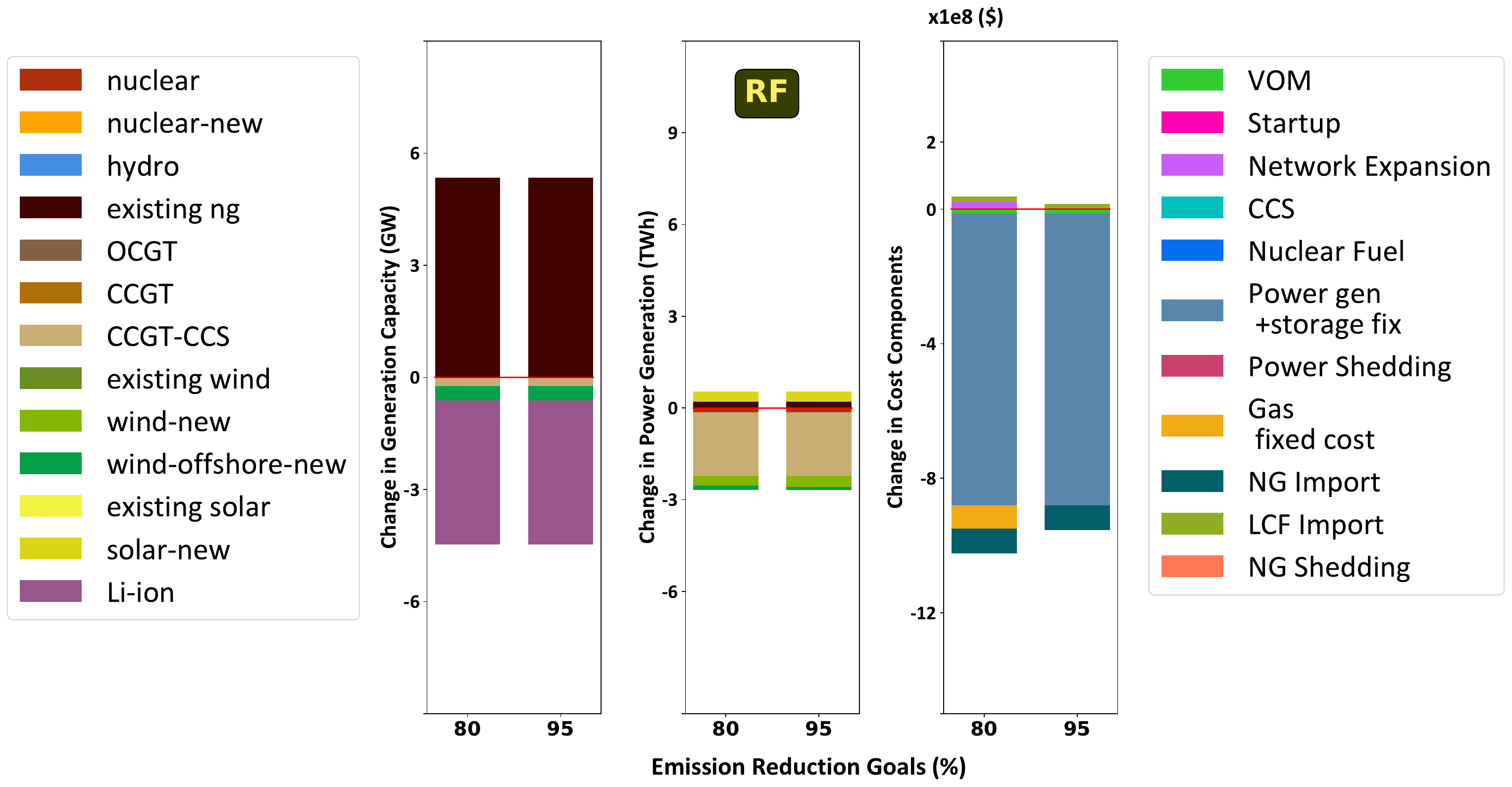}
  \end{subfigure}%
  \medskip
  
  \begin{subfigure}[t]{0.5\textwidth}
        \centering
        \includegraphics[width=0.8\textwidth]{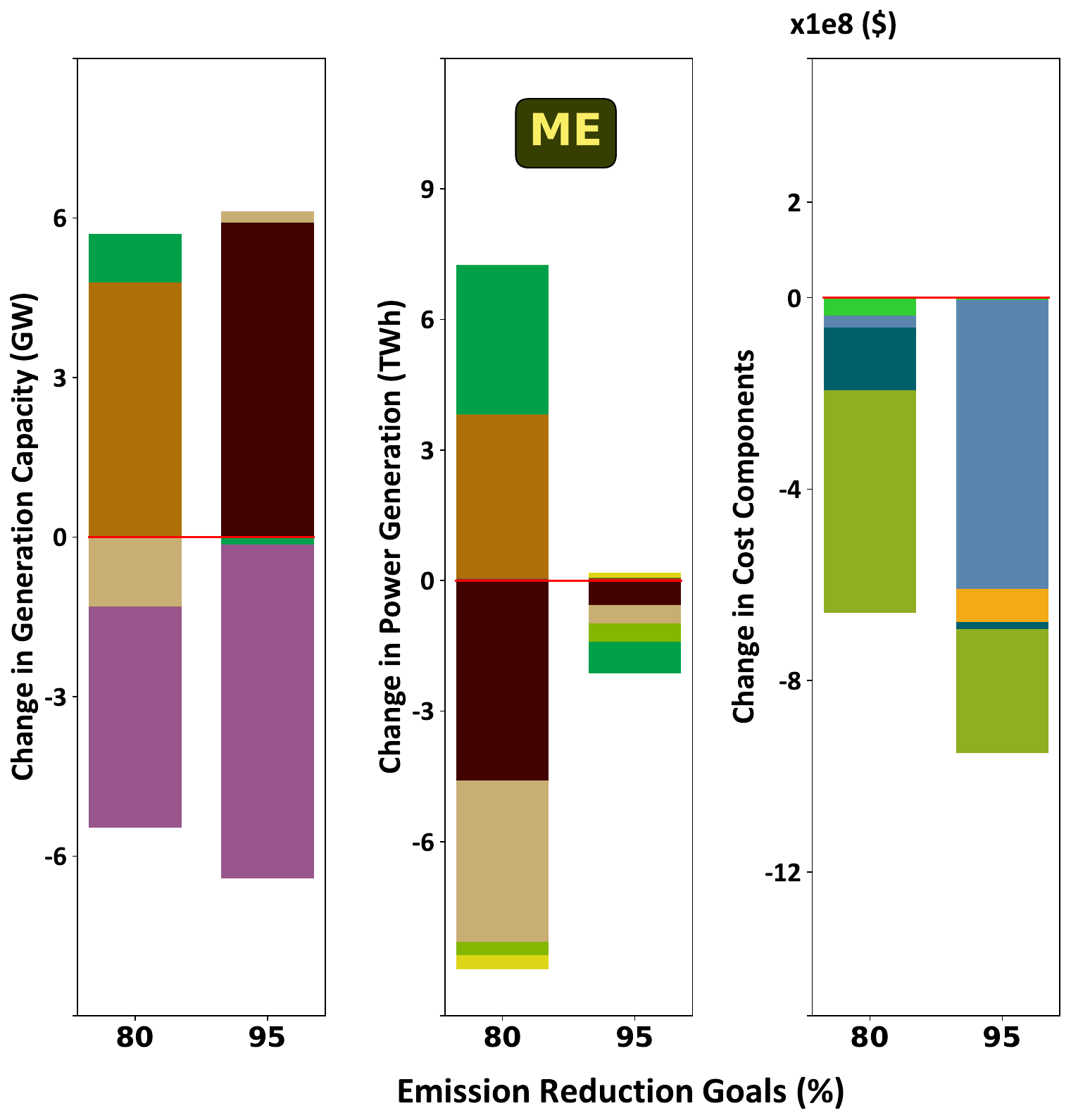}
    \end{subfigure}%
    \begin{subfigure}[t]{0.5\textwidth}
        \centering
        \includegraphics[width=0.8\textwidth]{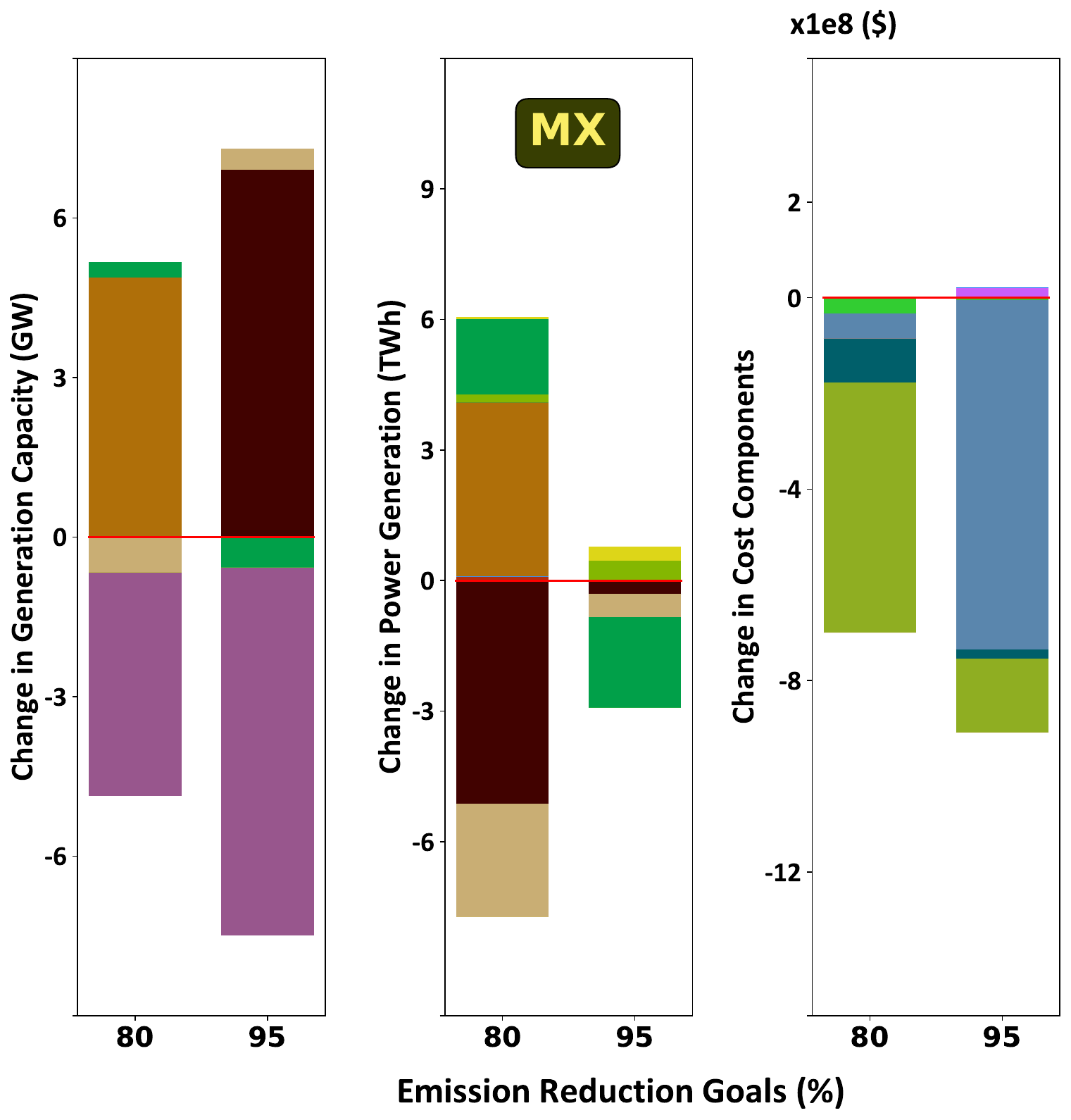}
    \end{subfigure}
    \medskip
    
      \begin{subfigure}[t]{0.5\textwidth}
        \centering
        \includegraphics[width=0.8\textwidth]{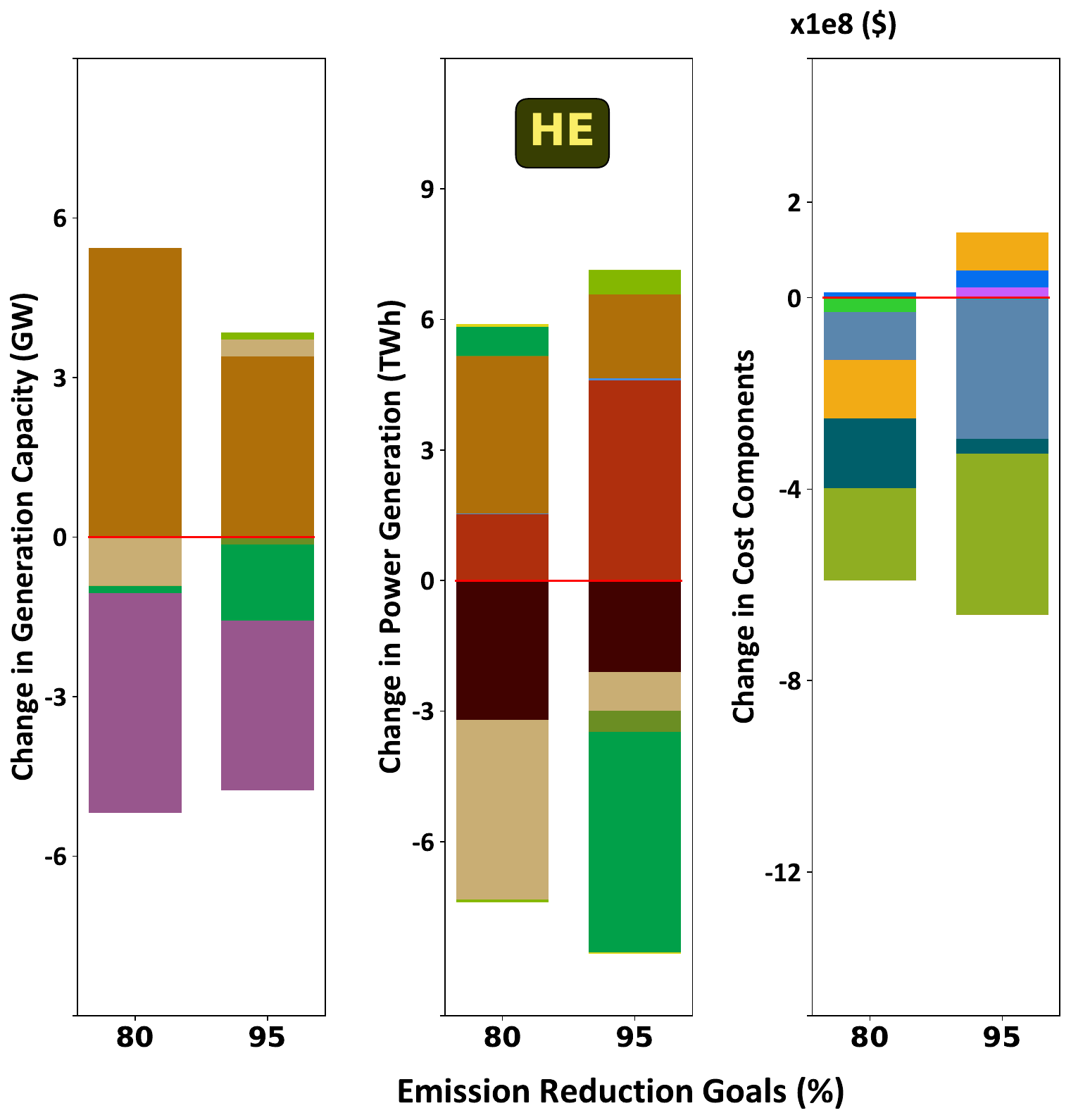}
    \end{subfigure}%
    \begin{subfigure}[t]{0.5\textwidth}
        \centering
        \includegraphics[width=0.8\textwidth]{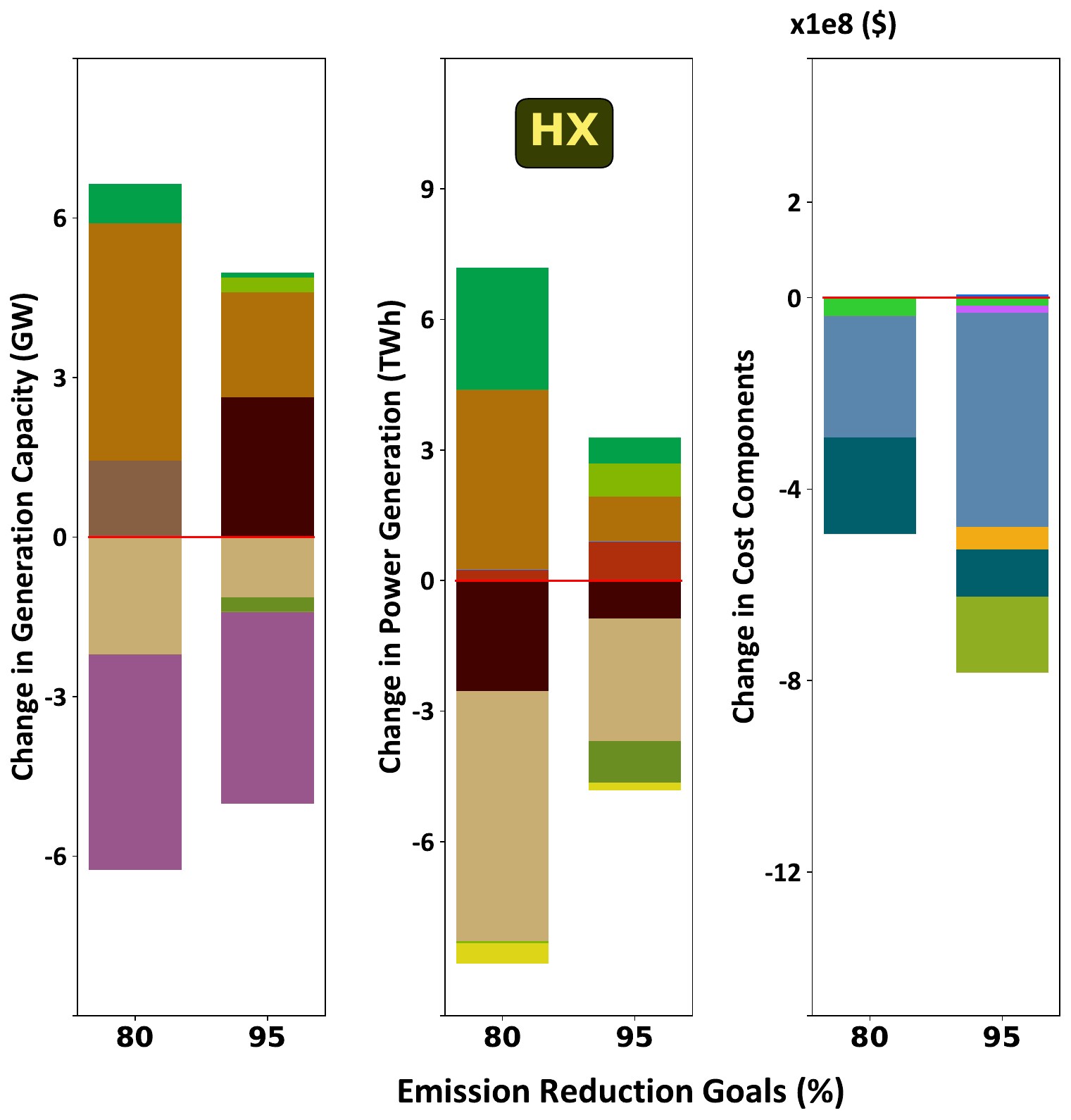}
    \end{subfigure}
    \caption{Difference in capacity, generation, and cost components between instances with and without transportation flexibility for exemplar weather year 2003 under different decarbonization targets and electrification scenarios.}
    \label{SIfig:diff-transp-flex}
\end{figure}

As indicated by other studies \cite{GenX2017, MITEI2022FES}, allowing the transportation load to be satisfied flexibly on an hour-to-hour basis allows for load-shifting within a day and thus reduces investment in Li-ion battery storage. We see a similar effect in our analysis, as shown in Fig.~\ref{SIfig:diff-transp-flex}, which illustrates the difference in outcomes for instances with and without transportation flexibility for the case in which transportation load for light-duty vehicles can be shifted ahead or forward by five hours.  \\
Across all residential sector demand and emissions scenarios, demand flexibility reduces investment in battery storage (by \bt{44-61\%}) and reduces overall gas consumption (see reduced biogas and NG import in cost charts). The reduction of battery storage as well as gas generation leads to total cost reduction of \bt{2.5-5.2\% (\$0.5- \$1 B)} across all instances. This reduction is more pronounced under more stringent decarbonization targets. Overall gas and LCF consumption decreased for all instances. However, the capacity of gas-fired plants increases in some scenarios to make up for the capacity contribution of battery storage resources to meet the capacity reserve margin constraint (Eq. 10a). As we did not consider the potential for demand flexibility to contribute to the resource adequacy (also called reserve margin), gas capacity has to make up for the reduced battery storage investment. Allowing demand flexibility to contribute to the capacity reserve margin constraint would presumably reduce gas capacity and further increase the cost savings from demand flexibility.

\subsection{Supply-side outcomes for all 20 weather years}\label{SIsec:GTresults}
\begin{figure}
     \centering
     \begin{subfigure}[b]{\textwidth}
             \centering    \includegraphics[width=\textwidth]{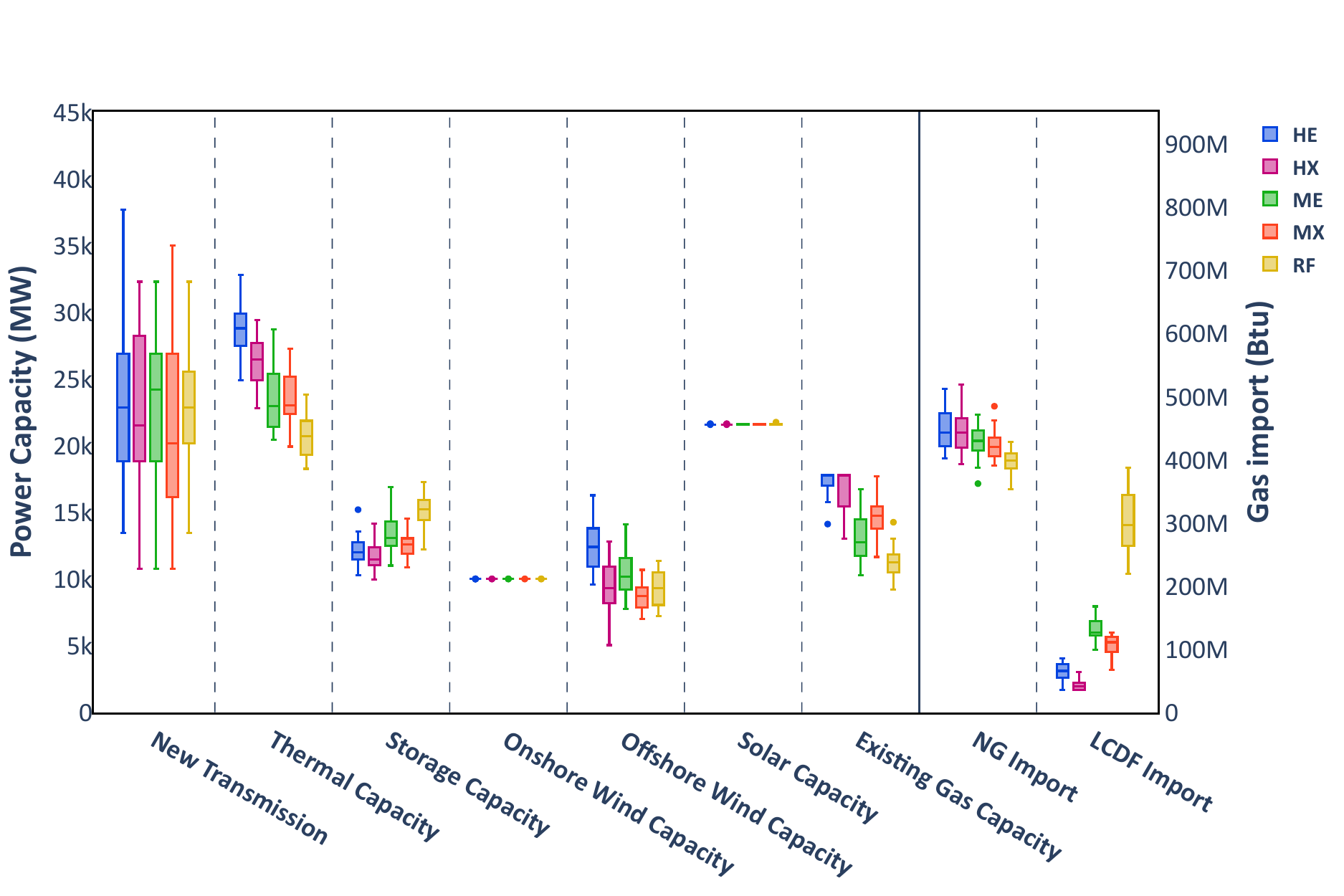}
    \caption*{{(a) 80\% decarbonization target}}
     \end{subfigure}
     \hfill
    \begin{subfigure}[b]{\textwidth}
             \centering    \includegraphics[width=\textwidth]{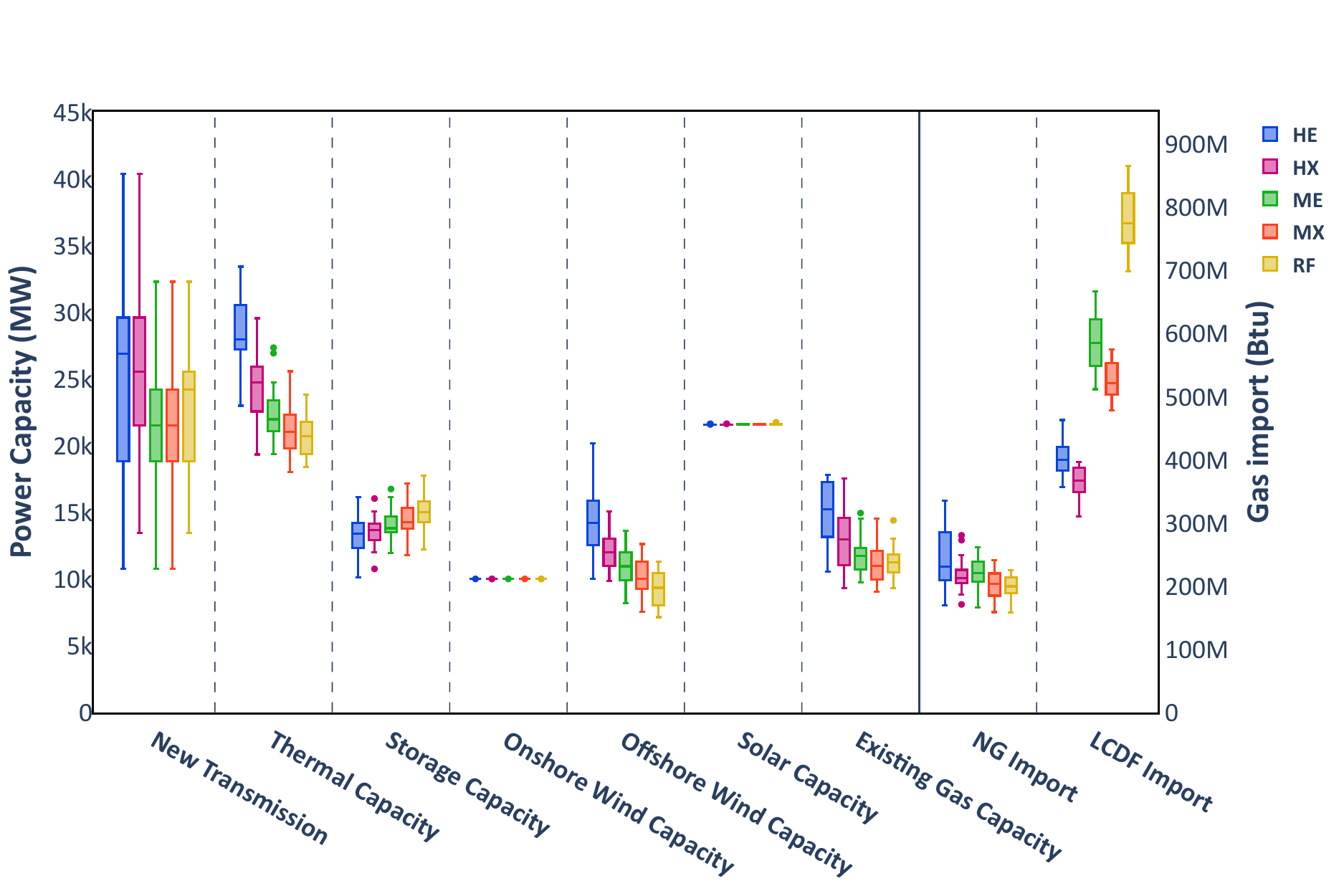}
    \caption*{{(b) 95\% decarbonization target}}
     \end{subfigure}
     \caption{Ranges of capacity and gas consumption metrics for all 20 weather years under each emissions constraint. ``Thermal capacity'' includes both new and existing gas assets. Unless otherwise specified, all capacity metrics include both existing brownfield capacity and new assets.}
     \label{SIfig:GTresults}
\end{figure}

Fig.~\ref{SIfig:GTresults} shows the ranges for capacity of key power assets and consumption of gas for all 20 weather years for each emissions constraint. Each weather year is used to calculate both the residential demand and supply-side capacity factors.

Beyond the conclusions presented for the exemplary weather year of 2003 in the main text, a handful of high-level conclusions can be drawn from the figures. Optimal deployment of transmission, thermal, storage, and offshore wind capacity is sensitive to weather across both emissions cases. The amount of thermal capacity deployed increases at higher electrification levels. The 95\% scenario requires much greater amounts of LCF than the 80\% emissions case. However, the relative proportions of each resource deployed are generally consistent across weather years. Onshore wind and solar are deployed to the maximum possible extent in both emissions scenarios across all weather years due to the land availability constraint applied to the model.

\subsection{NPV analysis of envelope improvements}\label{SIsec:NPVEnvImprov}


In order to balance the bulk system savings of envelope improvements against their demand-side installation cost, we calculate an example from our model results. Our results indicate average bulk energy system-related savings of roughly {\$2.04} billion annually in 2050 for HX over HE as a result of envelope improvements under the 95\% decarbonization goal. About 55\% of homes in the overall stock will receive envelope improvements in the HX scenario or about 3.87 million homes. If we assume the entry-level envelope improvements specified in this paper cost an average of \$20,000 per home and that the costs of these envelope improvements are incurred at once in 2050, the net present cost in 2050 is approximately \$77.4 billion. In order to assess the cost-benefit tradeoff in annual terms, we convert this capital cost into an annualized equivalent uniform series. We assume that the envelope improvements continue to deliver the same amount of bulk system savings for 40 years beyond 2050 and that the costs are distributed over the same time horizon. To account for a range of potential discount rates, we consider a 3.4\% real discount rate in line with the DOE's methodology for determining the cost-effectiveness of building energy codes \citep{DOECode2015} and a 7.1\% discount rate equivalent to that used in our supply-side modeling (see Table \ref{SItab:e-other-params}). The annualized cost is equal to the capital cost multiplied by the capital recovery factor as in Equation \ref{equation:NPV}:

\begin{equation}\label{equation:NPV}
         \$77.4 bil\cdot CRF = \$77.4 bil\cdot\frac{r(1+r)^{40}}{(1+r)^{40}-1}
\end{equation}

For \textit{r}=.034 and \textit{r}=.071, the annualized cost is \$3.57 billion and \$5.87 billion respectively. The annual bulk system savings of the envelope improvements therefore offset approximately {34.8-57.1\%} of the envelope retrofit cost.




\section{Joint Power-Gas Planning (JPoNG) Model Formulation}\label{SIsec:model-formulation}
Our model, referred to as \ModelName, determines the minimum cost planning decisions for power and gas systems considering the two systems' interdependency. The proposed model considers a range of generation and storage technologies modeled via operational and policy constraints. The formulation allows different temporal resolutions for the operation of both systems since the data availability or planning requirements can be different for each system. For example, decisions related to power generation, such as dispatch amounts and unit commitment, require hourly resolution. However, gas system operation does not involve generation decisions and only deals with transmission and storage operations for which daily resolution may be sufficient. Moreover, due to the ability of gas pipelines to provide some storage via line packing, daily resolution for scheduling gas operations could facilitate the management of intra-day variations in gas demand. In the model, the operations of both systems are coupled through two sets of constraints. The first set ensures gas flow to the power system. The second coupling constraint limits the CO$_2$ emissions incurred by consuming fossil-derived NG in both power and gas systems. 

\begin{figure*}
    \centering
    \includegraphics[width=0.9\textwidth]{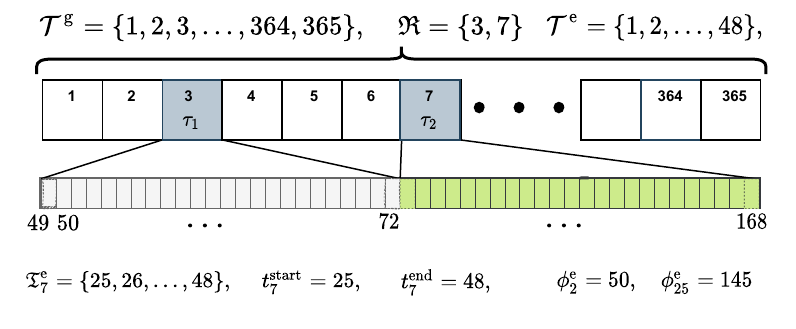}
    \caption{Illustration of the model's temporal resolution  of two representative days used for power systems operations. The top row represents planning days in a year, with days 3 and 7 as representative days forming the set $\mathfrak{R}$. The bottom row shows the hours corresponding to representative days. The set of planning periods for the gas system ($\mathcal{T}^{\text{g}}$) is the entire year. The set of planning periods for the power system ($\mathcal{T}^{\text{e}}$) is the chronologically ordered set of hours in the representative days whose mapping to their original index is given by $\phi^{\text{e}}_t, t\in \mathcal{T}^{\text{e}}$. The set of hours in their original indexing is denoted by $\mathfrak{T}^{\text{e}}_\tau$ with $t^{\text{start}}_\tau$ and $t^{\text{end}}_\tau$ signifying the starting and ending hours for the representative day $\tau$.  The representative days are chronologically ordered but are not necessarily contiguous.
    }
    \label{SIfig:planning-res}
\end{figure*}

We model power and gas system operation at an hourly and daily time resolution, respectively. For the power system, we only consider representative days of system operation to manage model tractability, while the operations of the gas system are modeled across all days of the year. Moreover, we also model the import of electric power to the system by defining an import node in the power system. The details of these systems are provided in Section ~\ref{SIsec:import-node}. The choice of daily resolution for the gas system is sufficient to capture macro-dynamics of gas flow and provides substantial computational advantages \citep{KhorramfarEtal2022}.
Fig.~\ref{SIfig:planning-res} illustrates our modeling approach for the time resolutions of both systems.

\begin{figure}[htbp]
    \centering
    \includegraphics[width=0.5\textwidth]{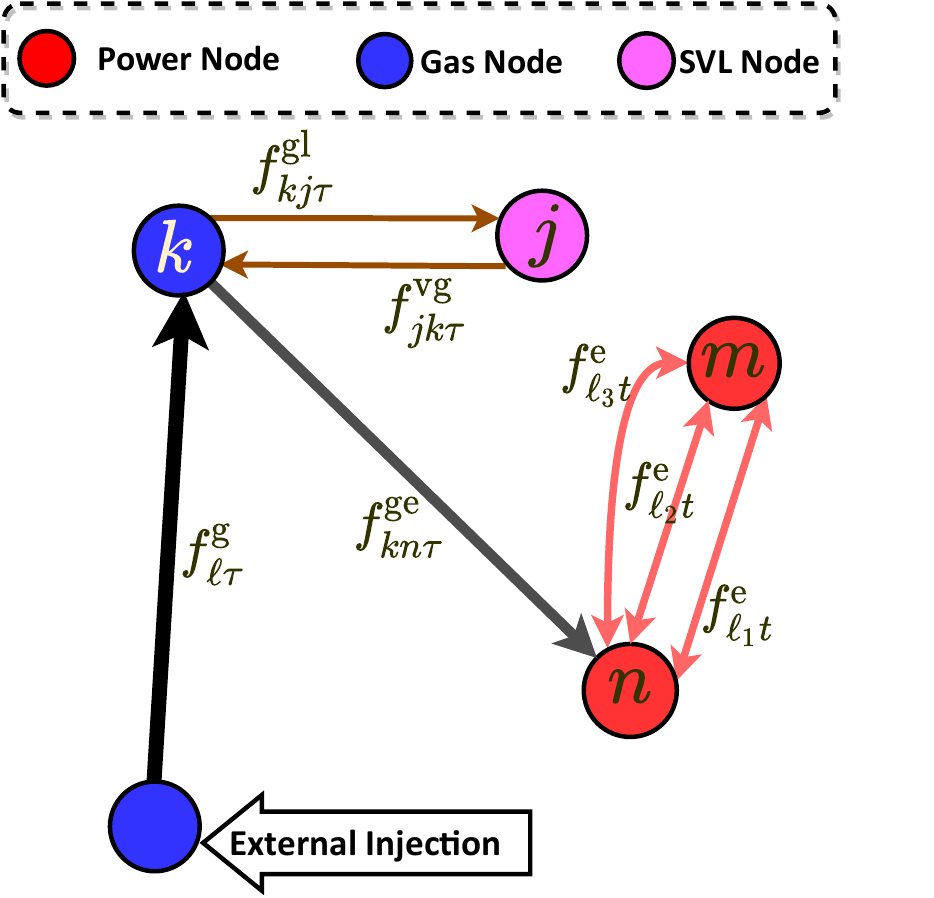}
    
    \caption{Flow variables between different nodes. In our model, power nodes can be connected by multiple bi-directional transmission lines denoted by $f^{\text{e}}_{\ell t}$.  Each power node operates a set of local gas-fired plants by drawing gas from its closest gas node. The variable $f^{\text{ge}}_{kn \tau}$ captures this flow. 
    Each gas node is connected to its closest SVL node through two unidirectional pipelines where one is from gas to SVL's liquefaction facilities denoted by $f^{\text{gl}}_{kj \tau}$; and the other one from SVL's vaporization facility to gas node denoted by $f^{\text{vg}}_{jk\tau}$. The variable $f^{\text{g}}_{\ell \tau}$ denotes the pipe flow between gas nodes. gas nodes can be connected by one or more uni-directional pipelines, but only one connection is depicted here. Candidate transmission lines and pipelines are not shown in this figure.}
    \label{SIfig:node-connections}
\end{figure}

The network representation in the model consists of three sets of nodes as depicted in Fig.~\ref{SIfig:node-connections}. The first set represents power system nodes and is characterized by different generation technologies (i.e., plant types), demand, storage, and the set of adjacent nodes by which the node can exchange electricity. The only exception for power system nodes is the \textit{import node} that, unlike other power nodes, operates on the entire year instead of representative days only. The details of the import node are provided in Section~\ref{SIsec:import-node}. The second set is gas nodes, each of which is associated with injection amount, demand, and its adjacent nodes. Storage tanks, vaporization, and liquefaction facilities, which  are commonly used in the non-reservoir storage of NG, collectively form the third set of nodes referred to as  storage-vaporization-liquefaction (SVL) nodes. We allow for the possibility of gas storage infrastructure to be located far from demand or injection points in the network, as per existing practice \citep{NGstatGuide2021} (see \ref{app-NG-topology} for detailed discussion). Accordingly, our model makes a distinction between gas and SVL nodes to account for their distinct locations. 
We also allow the gas system to use LCF, which represents a renewable source of gas fuel that is interchangeable with NG and hence can be imported and transported by the gas pipelines \citep{ColeEtal2021}. 
For ease of exposition, we separately present the power and gas systems' models as well as coupling constraints. The full description of the mathematical notation used in the formulation is described in \ref{SIsec:nomenclature}.

\subsection{Nomenclature} \label{SIsec:nomenclature}

\tablefirsthead{&\multicolumn{1}{c}{} \\ }
\tablehead{%
\\&\multicolumn{1}{c}{}\\ }
\tabletail{%
}
\tablelasttail{%
}
\small 
\begin{supertabular}{l l}
    \toprule
      \multicolumn{2}{l}{{\textbf{Sets}}}\\
      \midrule
    $\mathcal{N}^{\text{e}}$ &\hspace{-0.4cm} Power system nodes\\
    $\mathcal{P}$ &\hspace{-0.4cm}  Power plant types 
    \\
    $\mathcal{R} \subset \mathcal{P}$ &\hspace{-0.4cm} VRE power plant types 
    \\
    $\mathcal{G} \subset \mathcal{P}$ &\hspace{-0.4cm} gas-fired plant types 
    \\
    $\mathcal{CCS} \subset \mathcal{P}$ &\hspace{-0.4cm} gas-fired plant types with carbon capture technology
    \\
    $\mathcal{H} \subset \mathcal{P}$ &\hspace{-0.4cm} Thermal plant types  
    \\
    $\mathcal{Q} \subset \mathcal{P}$ &\hspace{-0.4cm} Technology with a resource availability limit\\
    $\mathcal{Q}'  $ &\hspace{-0.4cm} Set of technologies with \\
    &\hspace{-0.2cm} resource availability limits \\
    $\mathcal{T}^{\text{e}}$&\hspace{-0.4cm} Index set of representative hours for  power system\\
    $\mathfrak{R}$&\hspace{-0.4cm} Representative days\\
    $\mathfrak{T}^{\text{e}}_\tau$&\hspace{-0.4cm} Hours in $\mathcal{T}^{\text{e}}$ that are represented by day $\tau$\\
    $t^{\text{start}}_\tau,t^{\text{end}}_\tau$&\hspace{-0.4cm} First and last hour in $\mathfrak{T}^{\text{e}}_\tau$\\
     $\mathcal{L}^{\text{e}}$ &\hspace{-0.4cm} Existing and candidate transmission lines\\
     $\mathcal{L}^{\text{e}}_{nm}$ &\hspace{-0.4cm} Existing and candidate transmission lines \\
     & between node $n$ and $m$\\
    $\mathcal{S}^{\text{e}}_n$ &\hspace{-0.4cm}  All energy storage systems types\\
    $\mathcal{A}^{\text{g}}_n$ &\hspace{-0.4cm} Adjacent gas  nodes for node  $n$\\
    &\vspace{0.1cm}\\ 
    \hdashline \vspace{-0.3cm}\\
    $ \mathcal{N}^{\text{g}}, \mathcal{N}^s$ &\hspace{-0.4cm} Gas and SVL nodes\\
     $\mathcal{T}^{\text{g}}$&\hspace{-0.4cm} Days of the planning year\\
    $\mathcal{A}^s_k$ &\hspace{-0.4cm} Adjacent SVL facilities of node  $k$\\
$\mathcal{L}^{\text{g}}$ &\hspace{-0.4cm} Existing and candidate pipelines\\
    $\mathcal{L}^{\text{gExp}}_{k}$ &\hspace{-0.4cm} Existing and candidate pipelines\\
    & starting from node $k$\\
    $\mathcal{L}^{\text{gImp}}_{k}$ &\hspace{-0.4cm} Existing and candidate pipelines ending at node $k$\\
    \hspace{0.15cm}$\mathcal{L}^{\text{LCF}}$ & \hspace{-0.4cm} LCF availability levels and prices\\
            \bottomrule
\end{supertabular}

\small 

\begin{supertabular}{l l}
    \toprule
    \multicolumn{2}{l}{{\textbf{Indices}}}   \\
    \midrule 
    $n,m$ & Power system node \\
    $k$ & Gas system node\\
    $j$ & SVL facility node\\ 
    $i$ & Power generation plant type\\
    $r$ & Storage type for power network\\
    $\ell$& Electricity transmission line or gas pipeline\\
    $t$ & Time step for power system's operational periods\\
    $\tau$ & Time step for gas system's operational periods \\
    $l$ & LCF availability or price level\\
            \bottomrule
\end{supertabular}

\small 

\begin{supertabular}{l l}
    \toprule
          \multicolumn{2}{l}{{\textbf{Annualized Cost Parameters}}}\\
      \midrule
         $C^{\text{inv}}_{i}$ &\hspace{-0.5cm}  CAPEX of plants, [$\$$/plant] \\
        $C^{\text{dec}}_i$ &\hspace{-0.5cm}  Plant decommissioning cost, [$\$$/plant]\\
         ${C}^{\text{trans}}_{\ell}$ &\hspace{-0.5cm} Transmission line establishment cost, [$\$$/line] \\
        $C^{\text{EnInv}}_{r}$ &\hspace{-0.5cm} Storage establishment energy-related cost, [$\$$/MWh]\\
      $C^{\text{pInv}}_{r}$ &\hspace{-0.5cm}  Storage establishment power-related cost, [$\$$/MW]\vspace{0.1cm}\\
     \hdashline \vspace{-0.3cm}\\
     
      $C^{\text{pipe}}_{\ell}$ &\hspace{-0.5cm}  Pipelines establishment cost, [$\$$/line] \\
      $C^{\text{strInv}}_{j}$ &\hspace{-0.5cm}  CAPEX of storage tanks at SVLs, [$\$$/MMBtu]\\
      $C^{\text{vprInv}}_{j}$ &\hspace{-0.5cm}  CAPEX of vapor. plants at SVLs, [$\$$/MMBtu/hour]\\
      $C^{\text{pipeDec}}_{\ell}$ &\hspace{-0.5cm} Decommissioning cost for pipeline $\ell$ [$\$$/line]\\ 
            \bottomrule
\end{supertabular}

\small 

\begin{supertabular}{l l}
    \toprule
          \multicolumn{2}{l}{{\textbf{Annual Costs}}}\\
      \midrule
         $C^{\text{fix}}_{i}$ &\hspace{-0.5cm} Fixed operating and maintenance \\
         & cost (FOM) for plants, [$\$$] \\
        $C^{\text{EnFix}}_{r}$ &\hspace{-0.5cm} Energy-related FOM for storage, [$\$$/MWh]\\
        $C^{\text{pFix}}_{r}$ &\hspace{-0.5cm} Power-related FOM for storage, [$\$$/MW] \vspace{0.1cm}\\
        $C^{\text{trFix}}_{\ell}$ &\hspace{-0.5cm} Fixed cost of transmission line $\ell$ [$\$$/line]\\ 
        \hdashline \vspace{-0.3cm}\\  
      $C^{\text{strFix}}_{j}$ &\hspace{-0.5cm} FOM for storage tanks, [$\$$/MMBtu]\\
      $C^{\text{vprFix}}_{j}$ &\hspace{-0.5cm} FOM for vaporization plants, [$\$$/MMBtu/hour]\\
      $C^{\text{pipeFix}}_{\ell}$ &\hspace{-0.5cm} Fixed cost of pipeline $\ell$ [$\$$/line]\\ 
            \bottomrule
\end{supertabular}

\small 

\begin{supertabular}{l l}
    \toprule
          \multicolumn{2}{l}{{\textbf{Other Cost Parameters}}}\\
      \midrule
        $C^{\text{var}}_{i}$ &\hspace{-0.5cm} Variable operating and maintenance\\
        & cost (VOM) for  plants, [$\$$/MWh]\\
     $C^{\text{eShed}}$ &\hspace{-0.5cm} Unsatisfied power demand cost, [$\$$/MWh] \\
      $C^{\text{fuel}}_i$ &\hspace{-0.5cm} Fuel price for plants, [$\$$/MMBtu]\vspace{0.1cm}\\  \hdashline \vspace{-0.3cm}\\
     $C^{\text{ng}}$ &\hspace{-0.5cm} Fuel price for NG, [$\$$/MMBtu]\\
     $C^{\text{LCF}}_l$ &\hspace{-0.5cm} Price of LCF at availability interval $[U^{\text{LCF}}_l, U^{\text{LCF}}_{l-1}]$ [$\$$/MMBtu]\\
     $C^{\text{gShed}}$ &\hspace{-0.5cm} Unsatisfied gas demand cost [$\$$/MMBtu]\\
            \bottomrule
\end{supertabular}

\small 
\begin{supertabular}{l l}
    \toprule
          \multicolumn{2}{l}{{\textbf{Other Parameters for the Power System}}}\\
      \midrule
         $\rho_{nti}$ &\hspace{-0.5cm} Capacity (availability) factor for renewable plants\\
         $D^{\text{e}}_{nt}$ &\hspace{-0.5cm} Power demand, [MWh]\\
       $h_i$ &\hspace{-0.5cm} Heat rate, [MMBtu/MWh]\\
       $b_{\ell}$ &\hspace{-0.5cm} Susceptance of line $\ell\in \mathcal{L}^{\text{e}}$\\
        $\eta_i$ &\hspace{-0.5cm} Carbon capture rate, [\%]\\
       $U^{\text{prod}}_{i}$ &\hspace{-0.5cm} Nameplate capacity, [MW]\\
        $L^{\text{prod}}_{i}$ &\hspace{-0.5cm} Minimum stable output, [$\%$]\\
        $U^{\text{ramp}}_{i}$ &\hspace{-0.5cm} Ramping limit, [$\%$]\\
        $\gamma^{\text{eCh}}_r$ &\hspace{-0.5cm} Charge rate for storage\\
        $\gamma^{\text{eDis}}_r$ &\hspace{-0.5cm} Discharge rate for storage\\
        $\gamma^{\text{loss}}_r$ &\hspace{-0.5cm} hourly self-discharge rate for storage\\
        $I^{\text{trans}}_{\ell}$ &\hspace{-0.5cm} Initial capacity for transmission line $\ell$, [MW]\\
        $U^{\text{trans}}_\ell$ &\hspace{-0.5cm} Upper bound for capacity of transmission\\
        & line $\ell$, [MW]\\
        $\mathcal{I}^{\text{trans}}_{\ell}$ &\hspace{-0.5cm} 1, if  trans. line $\ell$ exists; 0, otherwise\\
        ${I}^{\text{num}}_{ni}$ &\hspace{-0.5cm} Initial number of plants\\
        $U^{\text{e}}_{\text{emis}}$ &\hspace{-0.5cm} Baseline emission of CO$_2$ in 1990 \\
        &from generation consumption, [ton]\\
       $U^{\text{CCS}}$ &\hspace{-0.5cm} Total annual carbon storage capacity, [ton]\\
        $d_n$ &\hspace{-0.5cm} Distance between node $n$ and CO$_2$ storage site \\
        $E^{\text{pipe}}$ &\hspace{-0.5cm} Electric requirement for CO$_2$ pipeline\\
        & operations [MWh/mile/ton/hour]\\
        $E^{\text{pump}}$ &\hspace{-0.5cm} Electric requirement for compression of CO$_2$\\
        $E^{\text{cprs}}$ &\hspace{-0.5cm} Number of compressors required in \\
        & the pipeline from node $n$ to the storage site\\
        &  pipelines [MWh/ton/hour]\\
        $U^{\text{prod}}_{\mathcal{Q}}$ &\hspace{-0.5cm} Production capacity for set of \\
        & plants $\mathcal{Q}\subset \mathcal{P}$, [MW]\\
        $\zeta$ &\hspace{-0.5cm} emissions reduction goal\\
        $w_t$ &\hspace{-0.5cm} Weight of the representative period $t$\\
        $\phi^{\text{e}}_t$&\hspace{-0.5cm} Mapping of representative period $t$ to its\\
        & original period in the time series\\
        $R^{\text{CRM}}$ &\hspace{-0.5cm} Capacity reserve margin rate\\
        $\gamma^{\text{CRM}}_{nit}$ &\hspace{-0.5cm} Capacity derating factor of \\
        & plant type $i$ at node $n$ at time $t$ \\
        &\\
            \bottomrule
\end{supertabular}

\small 

\begin{supertabular}{l l}
    \toprule 
          \multicolumn{2}{l}{{\textbf{Import Node Parameters}}}\\
      \midrule
      $\gamma^{\text{inflow}}_t$ & inflow rate to reservoir [MWh/h]\\
      $U^{\text{eCap}}$ & Maximum energy capacity of reservoir [MWh]\\
    $U^{\text{powCap}}$ & Maximum power capacity of reservoir [MWh]\\
      $C^{\text{fixImp}}$ & FOM cost of power capacity [\$/MWh/year]\\
            \bottomrule
\end{supertabular}

\small 

\begin{supertabular}{l l}
    \toprule
          \multicolumn{2}{l}{{\textbf{Other Parameters for the Gas System}}}\\
      \midrule
         $D^{\text{g}}_{k\tau}$ &\hspace{-0.5cm}  Gas demand, [MMBtu]\\
        $\eta^{\text{g}}$ &\hspace{-0.5cm} Emission factor for NG [ton CO$_2$/MMBtu]\\
        $U^{\text{inj}}_k$ &\hspace{-0.5cm} Upper bound for gas supply, [MMBtu]\\
        $\gamma^{\text{liqCh}}_j$ &\hspace{-0.5cm} Charge efficiency of liquefaction plant\\
        $\gamma^{\text{vprDis}}_j$ &\hspace{-0.5cm} Discharge efficiency of vaporization plant\\
        $\beta$ &\hspace{-0.5cm} Boil-off gas coefficient\\
        $I^{\text{pipe}}_{\ell}$ &\hspace{-0.5cm} Initial capacity for pipeline $\ell$, [MMBtu/day]\\
        $U^{\text{pipe}}_{\ell}$ &\hspace{-0.5cm} Upper bound capacity for pipeline $\ell$, [MMBtu/day]\\
        $\mathcal{I}^{\text{pipe}}_{\ell}$ &\hspace{-0.5cm} 1, if the pipeline $\ell$ exists; 0, otherwise\\
        $I^{\text{gStr}}_{j}$ &\hspace{-0.5cm} Initial storage capacity, [MMBtu]\\
        $I^{\text{vpr}}_{j}$ &\hspace{-0.5cm} Initial vaporization capacity, [MMBtu/d]\\
        $I^{\text{liq}}_{j}$ &\hspace{-0.5cm} Initial liquefaction capacity, [MMBtu/d]\\
        $I^{\text{store}}_{kj}$ &\hspace{-0.5cm} Initial capacity of storage facility\\
        $U^{\text{LCF}}_l$ &\hspace{-0.5cm} Available levels for LCF, [MMBtu]\\
        $U^{\text{g}}_{\text{emis}}$ &\hspace{-0.5cm} Baseline emission of CO$_2$ in 1990\\
        &from non-generation consumption, [ton]\\
        $\Omega_n$ &\hspace{-0.5cm} representative day for day $n$\\
            \bottomrule
\end{supertabular}

\small 

\begin{supertabular}{l l}
    \toprule
         \multicolumn{2}{l}{{\textbf{Investment Decision Variables}}}\\
      \midrule
         $x^{\text{op}}_{ni}\in \mathbb{R}^+$ &\hspace{-0.5cm} Number of available thermal plants $i\in \mathcal{H}$\\         
         $x^{\text{op}}_{ni}\in \mathbb{R}^+$ &\hspace{-0.5cm} Number of available VRE plants $i\in \mathcal{R}$\\  
         $x^{\text{est}}_{ni}\in \mathbb{R}^+$ &\hspace{-0.5cm} Number of new thermal plants established, $i\in \mathcal{H}$ \\
         $x^{\text{est}}_{ni}\in \mathbb{R}^+$ &\hspace{-0.5cm} Number of new VRE plants established, $i\in \mathcal{R}$ \\
         $x^{\text{dec}}_{ni}\in \mathbb{R}^+$ &\hspace{-0.5cm} Number decommissioned thermal plants, $i\in \mathcal{H}$ \\
         $x^{\text{dec}}_{ni}\in \mathbb{R}^+$ &\hspace{-0.5cm} Number decommissioned VRE plants, $i\in \mathcal{R}$ \\
         $y^{\text{eCD}}_{nr}\in \mathbb{R}^+$&\hspace{-0.5cm} Charge/discharge capacity of storage battery\\
         $y^{\text{eLev}}_{nr}\in \mathbb{R}^+$&\hspace{-0.5cm} Battery storage level\\
         $z^{\text{eInv}}_\ell\in \mathbb{B}$&\hspace{-0.5cm} 1, if transmission line $\ell$ is built; 0, otherwise\\
         $z^{\text{gInv}}_\ell\in \mathbb{B}$&\hspace{-0.5cm} 1, if pipeline $\ell$ is built; 0, otherwise\\
         $z^{\text{gDec}}_\ell\in \mathbb{B}$&\hspace{-0.5cm} 1, if pipeline $\ell$ is decommissioned; 0, otherwise\\
        $z^{\text{gOp}}_\ell\in \mathbb{B}$&\hspace{-0.5cm} 1, if pipeline $\ell$ is operational; 0, otherwise\\
            \bottomrule
\end{supertabular}

\small 

\begin{supertabular}{l l}
    \toprule
 \multicolumn{2}{l}{{\textbf{Other Decision Variables for Power System}}}\\
      \midrule 
$p_{nti}\in \mathbb{R}^+$&\hspace{-0.4cm} Generation rate, [MW]\\
         $x_{nti}\in \mathbb{R}^+$ &\hspace{-0.4cm} Number of committed plants\\
         $f^{\text{e}}_{\ell t} \in {\mathbb{R}}$&\hspace{-0.4cm} Flow rates, [MW]\\
        $\theta_{nt}\in \mathbb{R}$ &\hspace{-0.4cm} Phase angle\\
        $s^{\text{eCh}}_{ntr}\in \mathbb{R}^+$&\hspace{-0.4cm} Storage charged, [MW] \\
        $s^{\text{eDis}}_{ntr} \in \mathbb{R}^+$&\hspace{-0.4cm} Storage discharged, [MW] \\
        $s^{\text{eLev}}_{ntr} \in \mathbb{R}^+$&\hspace{-0.4cm} Storage level, [MWh] \\
        $s^{\text{rem}}_{n\tau r} \in \mathbb{R}$&\hspace{-0.4cm} Storage carry over during\\
        & day $\tau$ for storage type $r\in \mathcal{S}^{\text{sL}} $\\
        $s^{\text{day}}_{n\tau r} \in \mathbb{R}^+$&\hspace{-0.4cm} Storage level at the beginning of\\
        &day $\tau$ for storage type $r\in \mathcal{S}^{\text{sL}}$\\
        $\kappa^{\text{capt}}_{nt} \in \mathbb{R}^+$&\hspace{-0.4cm} Captured CO$_2$ [ton/h] \\
        $\kappa^{\text{pipe}}_{n} \in \mathbb{R}^+$&\hspace{-0.4cm} CO$_2$ pipeline capacity [ton/h] \\
         $a^{\text{e}}_{nt}\in \mathbb{R}^+$ &\hspace{-0.4cm} Amount of load shedding, [MWh]\\
         $\mathcal{E}^{\text{e}}$ &\hspace{-0.4cm} Total emission from power system\\
         &\\
            \bottomrule
\end{supertabular}

\small 

\begin{supertabular}{l l}
    \toprule
 \multicolumn{2}{l}{\textbf{Import Node Variables}}\\
      \midrule  
$s^{\text{cap-imp}}_t$ & Energy capacity of reservoir\\
    $x^{\text{imp}}$ & Maximum power capacity \\
            \bottomrule
\end{supertabular}

\small 
\begin{supertabular}{l l}
    \toprule
 \multicolumn{2}{l}{\textbf{Other Decision Variables for NG System (all in MMBtu)}}\\
      \midrule  
         $x^{\text{gStr}}_{j}\in \mathbb{R}^+$ &\hspace{-0.4cm} Installed additional  storage capacities\\
          $x^{\text{vpr}}_{j}\in \mathbb{R}^+$ &\hspace{-0.4cm} Installed additional vaporization capacities\\
        $f^{\text{g}}_{\ell\tau} \in {\mathbb{R}}^+$&\hspace{-0.4cm} Flow rates\\
        $f^{\text{ge}}_{kn\tau} \in {\mathbb{R}^+}$&\hspace{-0.4cm} Flow rates from gas nodes to power nodes\\
        $f^{\text{gl}}_{kj\tau} \in {\mathbb{R}^+}$&\hspace{-0.4cm} Flow rates from gas nodes to \\
        & \quad liquefaction plants\\
        $f^{\text{vg}}_{jk\tau} \in {\mathbb{R}^+}$&\hspace{-0.4cm} Flow rates from vaporization plants\\
        & \quad to gas nodes\\
         $g_{k\tau }\in \mathbb{R}^+$&\hspace{-0.4cm} gas supply (injection)\\
        $s^{\text{gStr}}_{j\tau}\in \mathbb{R}^+$&\hspace{-0.4cm} Storage capacities \\
        $s^{\text{vpr}}_{j\tau},s^{\text{liq}}_{j\tau} \in \mathbb{R}^+$&\hspace{-0.4cm} Vaporization and liquefaction amounts\\
         $a^{\text{g}}_{k\tau}\in \mathbb{R}^+$ &\hspace{-0.4cm} Amount of load shedding\\
        $a^{\text{LCF}}_{k\tau l}\in \mathbb{R}^+$ &\hspace{-0.4cm} Amount of LCF consumed\\
        $y^{\text{LCF}}_l \in\mathbb{B}$ &\hspace{-0.5cm} 1, if LCF consumption exceeds level $l$\\
        $\lambda^{\text{LCF}}_l \in[0,1]$ &\hspace{-0.5cm} Share of LCF consumption at interval $[U^{\text{LCF}}_l, U^{\text{LCF}}_{l-1}]$\\
        $\mathcal{E}^{\text{g}}$ &\hspace{-0.5cm} Total emission from gas system\\
        \bottomrule
\end{supertabular}

\subsection{Power System Model}
\noindent\textbf{Objective Function:}
\begin{subequations}
\label{elec-obj}
\begin{align}
      \min & \sum_{n \in \mathcal{N}^{\text{e}}} \sum_{i \in \mathcal{P}}  (C^{\text{inv}}_i x^{\text{est}}_{ni}+C^{\text{fix}}_{i} x^{\text{op}}_{ni} + \sum_{r \in \mathcal{S}^{\text{e}}_n}(C^{\text{pInv}}_r+C^{\text{pFix}}) y^{\text{eCD}}_{nr})\notag\\
       &+\sum_{n \in \mathcal{N}^{\text{e}}} \sum_{r \in \mathcal{S}^{\text{e}}_n}(C^{\text{EnInv}}_r+C^{\text{EnFix}}) y^{\text{eLev}}_{nr} \label{elec-obj-1}\\
        & +\sum_{n \mathcal{N}^{\text{e}}} \sum_{i \in \mathcal{P}}  C^{\text{dec}}_i x^{\text{dec}}_{ni} \label{elec-obj-2}\\
    &  +\sum_{n \in \mathcal{N}^{\text{e}}}\sum_{i \in \mathcal{P}} \sum_{t \in \mathcal{T}^{\text{e}}}w_t p_{nti}C^{\text{var}}_{i}  \label{elec-obj-3}\\
      &+ \sum_{l\in \mathcal{L}^{\text{e}}}  C^{\text{trans}}_{\ell} z^{\text{eInv}}_\ell \notag\\
      &+\sum_{\ell \in \mathcal{L}^{\text{e}}:I^{\text{trans}}_\ell=1} C^{\text{trFix}}_\ell +\sum_{\ell \in \mathcal{L}^{\text{e}}:I^{\text{trans}}_\ell=0} C^{\text{trFix}}_\ell z^{\text{eInv}}_\ell    \label{elec-obj-5} \\ 
      &+\sum_{n\in \mathcal{N}^{\text{e}}} d_n C^{\text{inv}}_{\text{CO}_2}\kappa^{\text{pipe}}_{n}+ C^{\text{str}}_{\text{CO}_2} \sum_{n\in \mathcal{N}^{\text{e}}} \sum_{t\in \mathcal{T}^{\text{e}}} w_t \kappa^{\text{capt}}_{nt}\label{elec-obj-6}\\ 
         &+  \sum_{n \in \mathcal{N}^{\text{e}}}\sum_{i \in \mathcal{P}} \sum_{t \in \mathcal{T}^{\text{e}}}w_t p_{nti} (C^{\text{fuel}}_{i} h_i)  \label{elec-obj-7}\\
    &+\sum_{n \in \mathcal{N}^{\text{e}}} \sum_{t \in \mathcal{T}^{\text{e}}}w_t C^{\text{eShed}}_n a^{\text{e}}_{nt}
    \label{elec-obj-8}\\
    &+C^{\text{fixImp}}x^{\text{imp}}
    \label{elec-obj-9}
\end{align}
\end{subequations}

The objective function~\eqref{elec-obj} minimizes the  total investment and operating costs incurred in the power system. The first term~\eqref{elec-obj-1} is the investment and fixed operation and maintenance (FOM) costs for generation and storage. The term~\eqref{elec-obj-2} captures the cost of plant retirement or decommissioning. The variable operating and maintenance  (VOM) cost is represented by the term~\eqref{elec-obj-3}. 
The network expansion and FOM costs are included in term~\eqref{elec-obj-5}. The cost of CO$_2$ transport and storage infrastructure required to accompany CCGT-CCS power generation is incorporated by term~\eqref{elec-obj-6} which also captures the cost associated with establishing CO$_2$ pipelines and storage. Here, we conservatively assume that each CO$_2$ pipeline connects a power node to the storage site, which ignores the possibility of meshed network design for CO$_2$ transport. The cost of gas consumption for non NG-fired power plants (i.e., nuclear) is ensured by term~\eqref{elec-obj-7}.  The term~\eqref{elec-obj-8} penalizes the load shedding in the power system. The last term~\eqref{elec-obj-9} incurs the fixed cost for the power capacity of the import node. As mentioned, we only report the costs associated with non-import power node.  

\noindent\textbf{Investment:} For every $n\in \mathcal{N}^{\text{e}},i\in \mathcal{P}$
\begin{subequations}
\begin{align}
      & x^{\text{op}}_{ni} = I^{\text{num}}_{ni}-x^{\text{dec}}_{ni}+x^{\text{est}}_{ni} &  \label{elec-c1}\\
\end{align}
\end{subequations}
Constraints~\eqref{elec-c1} calculate the number of operating plants.  

\noindent\textbf{Generation, Ramping, and Load Shedding:}
For every $n \in \mathcal{N}^{\text{e}},t\in \mathcal{T}^{\text{e}}$
\begin{subequations}
\begin{align}
    &L^{\text{prod}}_{i} x_{nti} \leq  p_{nti} \leq   U^{\text{prod}}_{i}x_{nti}&\hspace{-2cm} i\in \mathcal{H} 
    \label{elec-c4}\\
    &\vert p_{nti} -p_{n,(t-1),i}\vert \leq   U^{\text{ramp}}_{i} U^{\text{prod}}_{i} (x_{nti}-x^{\text{up}}_{nti})+\notag\\
    &\max(L^{\text{prod}}_{i}, U^{\text{ramp}}_{i})U^{\text{prod}}_{i}x^{\text{up}}_{nti}   & \hspace{-2cm} i\in \mathcal{H} 
    \label{elec-c5}\\
     &p_{nti} \leq \rho_{nti}U^{\text{prod}}_{i} x^{\text{op}}_{ni} &\hspace{-2cm} i \in \mathcal{R}  \label{elec-c6}\\
     & a^{\text{e}}_{nt}\leq D^{\text{e}}_{n\phi^{\text{e}}_t}& \label{elec-c7} 
\end{align}
\end{subequations}

The generation limits are imposed in constraints~\eqref{elec-c4}. 
Constraints~\eqref{elec-c5} are the ramping constraints that limit the generation difference of thermal units in any consecutive time periods to a ramping limit in the right-hand side of the equation. The generation pattern of VREs is determined by their hourly profile in the form of capacity factor; constraints~\eqref{elec-c6} limit the generation of VRE to hourly capacity factor (i.e. $\rho_{nti}$) of maximum available capacity (i.e. $U^{\text{prod}}_{i}x^{\text{op}}_{ni}$). Constraints~\eqref{elec-c7} state that the load-shedding amount can not exceed demand. Note that we use the mapping $\phi^{\text{e}}_t$ to access the demand in the corresponding hour of a representative day.

\noindent\textbf{Power Balance Constraints:}
For every $n\in \mathcal{N}^{\text{e}}, t\in \mathcal{T}^{\text{e}}$

\begin{align}
        &\sum_{i \in \mathcal{P}}p_{nti} +\sum_{m\in \mathcal{N}^{\text{e}}}\sum_{l\in \mathcal{L}^{\text{e}}_{nm}}\sign(n-m) f^{\text{e}}_{\ell t}+
        \sum_{r\in \mathcal{S}^{\text{e}}_n} (s^{\text{eDis}}_{ntr}-s^{\text{eCh}}_{ntr}) \notag \\ 
        &+ a^{\text{e}}_{nt}=D^{\text{e}}_{n \phi^{\text{e}}_t}+d_n E^{\text{pipe}}\kappa^{\text{pipe}}_{n}+ E^{\text{cprs}} E^{\text{pump}}\kappa^{\text{capt}}_{nt}
    & \hspace{-2cm} 
    \label{elec-c8}
\end{align}

Constraints~\eqref{elec-c8} ensure that for each node and for each planning period the generation, the net flow, the net storage, and the load shedding amount should be equal to the net demand. The net demand is defined in the right-hand side where the first term is the baseline demand, the second term is the electricity consumption by CO$_2$ pipelines and the last term is the electricity used by compressors.  
The notation $\sign(n-m)$ is the \textit{sign} function that takes value -1 if $n<m$, value 1 if $n>m$, and 0 otherwise.
We use this function to ensure that $f^e_{\ell t}$ appears with opposite signs (i.e., negative of positive signs) in the balance equations of the nodes connected by transmission line $\ell$. 

\noindent\textbf{Network Constraints:} 
For every $l\in \mathcal{L}^{\text{e}},t \in \mathcal{T}^{\text{e}}, \text{ and } n,m\in \mathcal{N}^{\text{e}}_\ell$

\begin{subequations}
\begin{align}
    & \vert f^{\text{e}}_{lt }\vert \leq I^{\text{trans}}_{\ell}&\hspace{-2cm}  \text{if } \mathcal{I}^{\text{trans}}_\ell=1  
    \label{elec-c9}\\
      & \vert f^{\text{e}}_{lt }\vert \leq U^{\text{trans}}_\ell z^{\text{eInv}}_{\ell}& \hspace{-2cm}  \text{if } \mathcal{I}^{\text{trans}}_\ell=0  \label{elec-c10}\\
\end{align}
\end{subequations}
Flow for the existing transmission lines is limited by constraints~\eqref{elec-c9}. Constraints~\eqref{elec-c10} limit the flow in candidate transmission lines only if it is already established (i.e., $z^{\text{t}}_{\ell}$=1). Throughout the paper, we use $M$ to denote a big number. 

\noindent\textbf{Storage Constraints:} For every $n\in \mathcal{N}^{\text{e}}, r\in \mathcal{S}^{\text{e}}_n$
\begin{subequations}
\begin{align}
 s^{\text{eLev}}_{n t^{\text{start}}_\tau r} &=(1-\gamma^{\text{loss}}_{r}) (s^{\text{eLev}}_{n,t^{\text{end}}_\tau r}-s^{\text{rem}}_{n\tau r})+\notag \\
 & \gamma^{\text{eCh}}_r s^{\text{eCh}}_{nt^{\text{start}}_\tau r}-\frac{s^{\text{eDis}}_{n t^{\text{start}}_\tau r}}{\gamma^{\text{eDis}}_r},\qquad  \tau \in \mathfrak{R} & \label{elec-c15}\\
  s^{\text{eLev}}_{n, t-1, r} &=(1-\gamma^{\text{loss}}_{r}) (s^{\text{eLev}}_{ntr})+ \gamma^{\text{eCh}}_r s^{\text{eCh}}_{ntr}-\frac{s^{\text{eDis}}_{n t r}}{\gamma^{\text{eDis}}_r}&\notag \\
  & \hspace{3cm} t\in \mathcal{T}^{\text{e}}\backslash \{t^{\text{start}}_\tau \lvert \  \tau \in \mathfrak{R} \}   \label{elec-c16}\\
  s^{\text{day}}_{n,\tau+1,r} & = (1-24\gamma^{\text{loss}}_{r})s^{\text{day}}_{n,\tau,r}+s^{\text{rem}}_{n\Omega_{\tau},r}, \quad \tau \in \mathcal{T}^{\text{g}}\backslash 365 \label{elec-c17}\\
s^{\text{day}}_{n,1,r} & = (1-24\gamma^{\text{loss}}_{r})s^{\text{day}}_{n,\tau,r}+s^{\text{rem}}_{n\Omega_{\tau},r},\quad \tau = 365& \label{elec-c18}\\
s^{\text{day}}_{n\tau r} & = s^{\text{eLev}}_{nt^{\text{end}}_\tau r}-s^{\text{rem}}_{n\tau r}, \quad \tau \in \mathfrak{R} \label{elec-c19}\\
s^{\text{rem}}_{n\tau r}&=0, \quad \tau \in \mathfrak{R}, r\in \mathcal{S}^{\text{sS}} \label{elec-c20}\\
    & s^{\text{eCh}}_{ntr}\leq y^{\text{eCD}}_{nr}  &  \label{elec-c21}\\
    &s^{\text{eCh}}_{ntr}\leq y^{\text{eCD}}_{nr}  &  \label{elec-c21-2}\\
    & s^{\text{eLev}}_{ntr}\leq y^{\text{eLev}}_{nr} &  \label{elec-c22}
\end{align}
\end{subequations}

Recall (see Fig.~\ref{SIfig:planning-res}) that representative days are not necessarily consecutive; hence our formulation accounts for the carryover storage level between representative days, which is particularly important when modeling LDES. Li et al. \citep{LiEtal2022} enforce the beginning and ending storage levels of each representative day to 50\% of the maximum storage level. Here, we use a similar approach for short-duration batteries in which we time-wrap the beginning and ending hours of a day. That is, we assume the same charging state for the beginning and ending hours of a day, implicitly precluding energy carryover between representative days. For LDES, however, we use the method proposed in \citep{GenX2017,KotzurEtal2018} in which the unrestricted variable $s^{\text{rem}}_{n\tau r}$ models the carryover from a representative day $\tau$ to the next.

Constraints~\eqref{elec-c15} model battery storage dynamics for the initial hours of each representative day. Constraints~\eqref{elec-c16} model the storage balance for the remaining hours.  The energy transfer across two consecutive representative days is modeled via constraints~\eqref{elec-c17}. The storage in the first and last representative days is related by constraints~\eqref{elec-c18}. Constraint~\eqref{elec-c19} only applies to representative days and ensures that the storage at the beginning of a day is equal to the storage level at the last hour of the day minus the storage carryover. The storage carryover for short-duration batteries is prevented by constraints~\eqref{elec-c20}. Finally, the charge/discharge limits on storage level are imposed in constraints~\eqref{elec-c21}  to \eqref{elec-c22}. Analogous to similar studies on power system expansion \citep{SepulvedaEtal2021,LiEtal2022}, we do not account for use-dependent storage capacity degradation.




\noindent\textbf{Resource Availability Constraints:} 
\begin{align}
   & \sum_{n\in \mathcal{N}^{\text{e}}}\sum_{i\in \mathcal{Q}}  U^{\text{prod}}_i x^{\text{op}}_{ni} \leq U^{\text{prod}}_{\mathcal{Q}} & \mathcal{Q}\in \mathcal{Q}'
   \label{elec-c24}
\end{align}

We consider resource availability limits for the development of VRE sources. In comparison to thermal plants, the siting of renewable resources is a major challenge due to the relatively large land area footprint per MW, the spatial heterogeneity in their resource availability and land availability limits due to such non-energy considerations as preserving the natural landscape \citep{E3Team2020}. Therefore, constraint~\eqref{elec-c24} limits the installed capacity of a certain set of power plants to their maximum availability limit. The parameter $\mathcal{Q}$ denotes a generation technology class for which there is a resource availability limit. These classes include solar, onshore wind, offshore wind, and nuclear and are represented by set $\mathcal{Q}'$. Note that each technology class can include multiple plant types. For example, nuclear technology can include existing and new nuclear plant types.  

\noindent\textbf{Carbon Capture and Storage (CCS) Constraints:} 
For every $n\in  \mathcal{N}^{\text{e}}, t\in \mathcal{T}^{\text{e}}$ 
\begin{subequations}
\begin{align}
   &\kappa^{\text{capt}}_{nt} = \eta^{\text{g}} \eta_i h_ip_{nti} & i\in \mathcal{CCS} \label{elec-c25}\\
   &\kappa^{\text{capt}}_{nt}\leq \kappa^{\text{pipe}}_{n} \label{elec-c26}\\
   &\sum_{n\in \mathcal{N}^{\text{e}}}\sum_{t\in \mathcal{T}^{\text{e}}} \kappa^{\text{capt}}_{nt} \leq U^{\text{CCS}}& \label{elec-c27}
\end{align}
\end{subequations}

The constraint~\eqref{elec-c25} computes the amount of captured carbon in gas-fired power plants equipped with CCS technology. Constraint~\eqref{elec-c26} determines the CO$_2$ pipeline capacity. Finally, constraint~\eqref{elec-c27} limits the total amount of captured CO$_2$ to the annual CO$_2$ storage capacity. 

\noindent\textbf{Capacity Reserve Margin (CRM):}

\begin{subequations}
\begin{align}
   &\sum_{n \in \mathcal{N}^{\text{e}}} \left(\sum_{i \in \mathcal{P}} \gamma^{\text{CRM}}_{nit} U^{\text{prod}}_i x^{\text{op}}_{ni} + \sum_{r\in \mathcal{S}^{\text{e}}_n} (s^{\text{eDis}}_{ntr}-s^{\text{eCh}}_{ntr}) \right) \geq (1+R^{\text{CRM}}) \sum_{n \in \mathcal{N}^{\text{e}}} D^{\text{e}}_{n\phi^{\text{e}}_t}, & t\in \mathcal{T}^{\text{e}} \label{elec-c28}
\end{align}
\end{subequations}
The CRM constraint ensures that the installed capacity in the system plus the net discharging power from storage technologies exceeds a certain level of aggregated load across the region for all time periods. CRM is usually defined for the peak hour load, but we impose it for all time periods as our model is intended to be used for future infrastructure planning where the load projections are uncertain even if they are modeled as deterministic parameters. 

\noindent\textbf{Import node constraints:}
Assuming the $n'$ is the import node, the set of associated constraints are:
\begin{subequations}
\begin{align}
   & s^{\text{cap-imp}}_0 = 0.7 U^{\text{eCap}}\label{elec-29}\\
   & s^{\text{cap-imp}}_t = s^{\text{cap-imp}}_{t-1} + \gamma^{\text{inflow}} - p_{n'ti}& \mathfrak{T}^e_\tau, \tau\in \mathcal{T}^g, i \in \mathcal{P}\\
   & s^{\text{cap-imp}}_{8760}=0.7U^{\text{eCap}}\\
   & s^{\text{cap-imp}}_{2881} \leq 0.55U^{\text{eCap}} \label{elec-30}\\
   & f^{\text{e}}_{\ell, \tau_1+h}=f^{\text{e}}_{\ell, \tau_2+h} & 
 t_1,t_2 \in \mathcal{T}^{\text{g}} \text{ if } \Omega_{\tau_1} = \Omega_{\tau_2}, h\in \{1,2,\ldots, 24 \} \label{elec-31}\\
 &p_{n'ti} \leq U^{\text{powCap}} \label{elec-32} \\
\end{align}
\end{subequations}

Constraints~\eqref{elec-29}-\eqref{elec-30} impose the energy capacity limits and equations. The details of energy capacity limits are provided in Table~\ref{SItab:res-avail}. The operations of the import node are carried out for the entire year, rather than over representative days like other power nodes. Therefore, in constraints~\eqref{elec-31} we enforce the flow between the import node and other power nodes to be the same for any days that are represented by the same day. Finally, the constraints~\eqref{elec-32} restrict the production capacity to its maximum limit.

\subsection{Gas System Model}
We now model the objective function and constraints pertaining to the gas system in the \ModelName.

\noindent\textbf{Objective Function:}
\begin{subequations}\label{ng-obj}
\begin{align}
    \min & \sum_{l\in \mathcal{L}^{\text{g}}} \left (C^{\text{pipe}}_{\ell} z_{\ell}^{\text{g}} + c^{\text{pipeDec}}_\ell z^{\text{gDec}}\ell + c^{\text{pipeFix}}_\ell z^{\text{gOp}}_\ell \right)   \label{ng-obj-1}\\
    &+ \sum_{k \in \mathcal{N}^{\text{g}}} \sum_{\tau \in \mathcal{T}^{\text{g}}}  C^{\text{ng}} g_{k\tau}\label{ng-obj-2}\\
    &+ \sum_{j \in \mathcal{N}^s}(C^{\text{strInv}}_j x^{\text{gStr}}_{j} +C^{\text{vprInv}}_j x^{\text{vpr}}_{j}) \label{ng-obj-3}\\
    &+ \sum_{j \in \mathcal{N}^s}\left(C^{\text{strFix}}(I^{\text{gStr}}_{j}+x^{\text{gStr}}_j)+C^{\text{vprFix}}(I^{\text{vpr}}_{j}+x^{\text{vpr}}_j) \right)\label{ng-obj-4}\\
    &+ \sum_{k \in \mathcal{N}^{\text{g}}} \sum_{\tau \in \mathcal{T}^{\text{g}}} (C^{\text{gShed}} a^{\text{ng}}_{k\tau })\label{ng-obj-5}\\
    &+ \sum_{l \in \mathcal{L}^{\text{LCF}}\backslash 1} C^{\text{LCF}}_l (U^{\text{LCF}}_{l}-U^{\text{LCF}}_{l-1}) \lambda^{\text{LCF}}_{l} \label{ng-obj-6}
\end{align}
\end{subequations}

The objective function~\eqref{ng-obj} minimizes the total investment and operating costs incurred in the gas system. The first term~\eqref{ng-obj-1} is the investment, strategic decommissioning and FOM costs for new, existing, and operational pipelines, respectively. The second term~\eqref{ng-obj-2} is the cost of procuring gas from various sources to the system. For example, New England procures its NG from Canada and its adjacent states such as New York. 
Term~\eqref{ng-obj-3} and ~\eqref{ng-obj-4} handle the investment and FOM costs associated with gas storage, respectively. The term~\eqref{ng-obj-5} is the penalty for gas load shedding. {The last term~\eqref{ng-obj-6} captures the cost of LCF consumption.}

\noindent\textbf{Gas Balance Constraint:}  For every $k\in \mathcal{N}^{\text{g}}, \tau \in \mathcal{T}^{\text{g}} $
\begin{align}
   &g_{k\tau} -\sum_{l \in \mathcal{L}^{\text{gExp}}_{k}} f^{\text{g}}_{\ell\tau}+\sum_{l \in \mathcal{L}^{\text{gImp}}_{k}} f^{\text{g}}_{\ell\tau}-\sum_{n\in \mathcal{A}^{\text{e}}_k} f^{\text{ge}}_{kn\tau } \notag\\
   &+\sum_{j\in \mathcal{A}^s_k} (f^{\text{vg}}_{jk \tau}-f^{\text{gl}}_{kj\tau})+ a^{\text{LCF}}_{k\tau}+a^{\text{g}}_{k\tau}=D^{\text{g}}_{k\tau} &  \label{ng-c1}
\end{align}

Constraints~\eqref{ng-c1} state that for each node and period, the imported gas (i.e., injection), flow to other gas nodes, flow to power nodes, flow to and from storage nodes, load satisfied by LCF, and unsatisfied gas load should add up to demand. Unlike power flow, the flow in pipelines is assumed to be unidirectional, as is typical for most long-distance transmission pipelines involving booster compressor stations \citep{VonWaldEtal2022} , for which we ignore the relatively small electricity demand.  

\noindent\textbf{Flow on Representative Days:}

\begin{align}
   &f^{\text{ge}}_{kn\tau_1 } = f^{\text{ge}}_{kn\tau_2 }& \tau_1, \tau_2 \in \mathcal{T}^{\text{g}} \text{ if } \Omega_{\tau_1} = \Omega_{\tau_2} \label{ng-c2-1}
\end{align}

Given the set of representative days used to model power system operations (see Fig~\ref{SIfig:planning-res}), constraint~\eqref{ng-c2-1} ensures that gas consumption by the power system for all the days represented by the same day is identical.

\noindent\textbf{Gas and LCF Supply Constraints:}
For every $k\in \mathcal{N}^{\text{g}}, \tau \in \mathcal{T}^{\text{g}}$
\begin{subequations}
\begin{align}
   & L^{\text{inj}}_k \leq g_{k\tau}+a^{\text{LCF}}_{k\tau}\leq U^{\text{inj}}_k & \label{ng-c2}\\
  & \lambda^{\text{LCF}}_l \leq y^{\text{LCF}}_l & l\in \mathcal{L}^{\text{LCF}} \label{ng-c3}\\
   &y^{\text{LCF}}_l \leq y^{\text{LCF}}_{l-1} &l\in \mathcal{L}^{\text{LCF}} \backslash 1 \label{ng-c3-2}\\
 &\sum_{k\in \mathcal{N}^{\text{g}}}\sum_{\tau \in \mathcal{T}^{\text{g}}}  a^{\text{LCF}}_{k\tau} = \sum_{l\in \mathcal{L}^{\text{LCF}}\backslash 1}\lambda^{\text{LCF}}_l (U^{\text{LCF}}_{l}-U^{\text{LCF}}_{l-1}) &  \label{ng-c3-3}
\end{align}
\end{subequations}
The gas fuel import limits are imposed in constraints~\eqref{ng-c2}. {Constraints~\eqref{ng-c3} impose the share of consumption in the LCF availability at interval $[U^{\text{LCF}}_{l-1},U^{\text{LCF}}_{l}]$ is only positive if the consumption level exceeds $U^{\text{LCF}}_{l-1}$. Constraints~\eqref{ng-c3-2} ensures that the LCF consumption can not exceed level $l$ before exceeding level $l-1$. The total consumption of LCF across all availability levels is calculated in constraints~\eqref{ng-c3-3}. }

\noindent\textbf{Flow Constraints:}
For every $\ell \in \mathcal{L}^{\text{g}}, \tau \in \mathcal{T}^{\text{g}}, j\in \mathcal{N}^s$
\begin{subequations}
\begin{align}
    & f^{\text{g}}_{\ell \tau } \leq U^{\text{pipe}}_{\ell} z^{\text{gOp}}_{\ell}&  \label{ng-c6}\\
    &\sum_{k \in \mathcal{N}^{\text{g}}:j\in \mathcal{A}^s_k} f^{\text{gl}}_{kj \tau} =s^{\text{liq}}_{j\tau} & \label{ng-c7}\\
    &\sum_{k \in \mathcal{N}^{\text{g}}:j\in \mathcal{A}^s_k } f^{\text{vg}}_{jk \tau}=s^{\text{vpr}}_{j\tau}  & \label{ng-c8}
\end{align}
\end{subequations}
The constraints~\eqref{ng-c6}  limit the flow between gas nodes for operational pipelines, respectively. The flow to liquefaction facilities is calculated in constraints~\eqref{ng-c7}. Similarly, the flow out of vaporization facilities is modeled via constraints~\eqref{ng-c8}. 

\noindent\textbf{Storage Constraints:} 
For every $j\in \mathcal{N}^s, \tau \in \mathcal{T}^{\text{g}}$
\begin{subequations}
\begin{align}
   &s^{\text{gStr}}_{j\tau} = (1-\beta) s^{\text{gStr}}_{j,\tau-1}+\gamma^{\text{liqCh}}_j s^{\text{liq}}_{j\tau}-\frac{s^{\text{vpr}}_{j\tau}}{\gamma^{\text{vprDis}}_j}  &  \label{ng-c9}\\
    &s^{\text{vpr}}_{j\tau}\leq  I^{\text{vpr}}_{j}+x^{\text{vpr}}_j  &  \label{ng-c10}\\
    &s^{\text{gStr}}_{j\tau}\leq  I^{\text{gStr}}_{j}+x^{\text{gStr}}_j  &  \label{ng-c11}
\end{align}
\end{subequations}
Constraints~\eqref{ng-c9} ensure the storage balance. Constraints~\eqref{ng-c10} and \eqref{ng-c11} limit the capacity of vaporization and storage tanks to their initial capacity plus the increased capacity, respectively.

\noindent\textbf{Operational Pipelines:} 
\begin{subequations}
\begin{align}
   &z^{\text{gOp}}_\ell =  \mathcal{I}^{\text{pipe}} + z^{\text{gInv}}_\ell - z^{\text{gDec}}_\ell  &  \label{ng-c12}
    \end{align}
\end{subequations}
Pipeline $\ell$ is operational if either it is existing and not decommissioned, or newly established. 

\subsection{Coupling Constraints}
The following constraints are coupling constraints that relate operational decisions of the power and gas systems together.
\begin{subequations}
\begin{align}
    & \sum_{k \in \mathcal{A}^{\text{e}}_n} f^{\text{ge}}_{k n\tau } =   \sum_{t \in \mathfrak{T}^{\text{e}}_\tau} \sum_{i \in \mathcal{G}} h_i p_{nti} &\hspace{-2cm} n\in \mathcal{N}^{\text{e}}, \tau \in \mathfrak{R}\label{coup-1}\\
 &\mathcal{E}^{\text{e}} = \sum_{n\in \mathcal{N}^{\text{e}}}\sum_{t\in \mathcal{T}^e}\sum_{i \in \mathcal{G}} w_t(1-\eta_i)\eta^{\text{g}} h_i p_{nti}&\notag \\
 &\mathcal{E}^{\text{g}} =\sum_{k \in \mathcal{N}^g}\sum_{\tau \in \mathcal{T}^g} \eta^{\text{g}}( D^{\text{g}}_{k\tau}-  a^{\text{LCF}}_{k\tau}-a^{\text{g}}_{k\tau})\notag \\
 &  \mathcal{E}^{\text{e}}+\mathcal{E}^{\text{g}} \leq (1-\zeta) (U^{\text{e}}_{\text{emis}} +U^{\text{g}}_{\text{emis}}) &\label{coup-2}
\end{align}
\end{subequations}
The first coupling constraints~\eqref{coup-1} capture the flow of gas to the power network for each node and at each time period. 
The variable $\mathcal{E}^{\text{e}}$ accounts for CO$_2$ emission due to the consumption of gas in the power system. The variable  $\mathcal{E}^{\text{g}}$ computes the emission from the gas system by subtracting the demand from LCF consumption and gas load shedding. The second coupling constraint~\eqref{coup-2} ensures that the net CO$_2$ emissions associated with the power-gas system are below a pre-specified threshold value, which is defined based on a baseline (e.g., historical) emissions level. The first term is the emissions due to non-power gas consumption (i.e., gas consumption in the gas system such as space heating, industry use, and transportation). Since the model does not track whether LCF is used to meet non-power gas demand or for power generation, the first term computes gross emissions from all gas use presuming it is all fossil and then subtracts emissions benefits from using LCF.

Here we treat LCF as a carbon-neutral fuel source \citep{ColeEtal2021}, and thus the combustion emissions associated with its end-use are equal to the emissions captured during its production. 
The second term captures the emission from NG-fired power plants.
Alternatively, the emission constraints can be applied only to the power system as in \citep{SepulvedaEtal2021} or applied separately to each system as in \citep{VonWaldEtal2022}.

\section{Supply-side Model Data}\label{SIsec:input-data}
This section describes how we obtained parameters for both networks for the New England Case study. We start with the power system and explain the data preparation process. We then expound on the steps we took to construct the gas network and associated data inputs. In this study, our region is New England. However, the modeling framework and the solution approach are applicable to other regions to any spatial resolution.

\begin{figure}[htbp]
    \centering
    \includegraphics[width=0.6\textwidth]{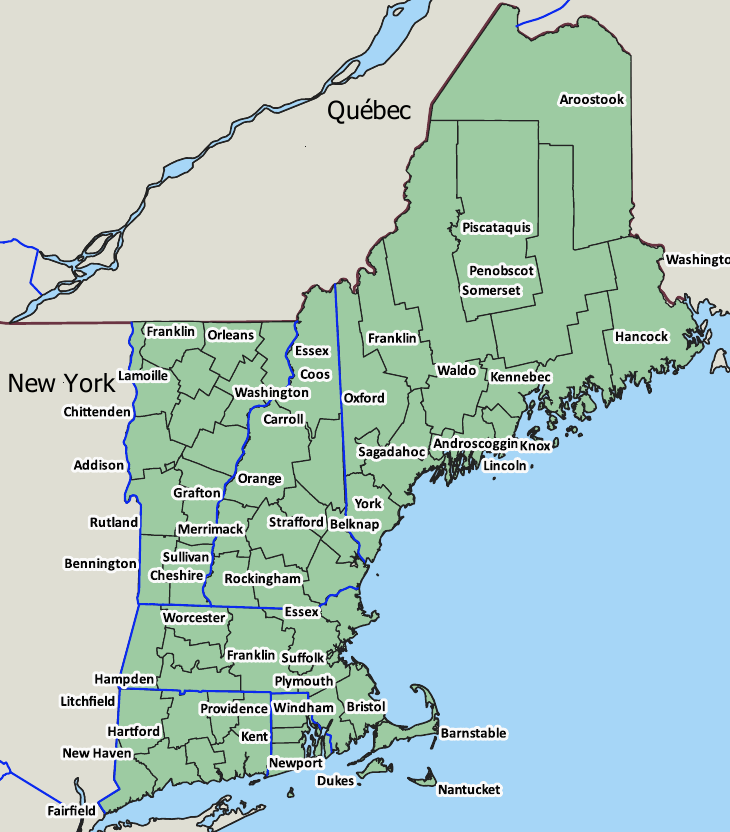}
    \caption{New England counties}
    \label{SIfig:NE-counties}
\end{figure}

Fig.~\ref{SIfig:NE-counties} shows the name and boundaries of New England counties. The region currently has 67 counties with populations ranging from seven thousand to more than 1.6 million. 

\subsection{Power System Data}
Our parameters for the power network are based on the \textit{US Test System} developed by \textit{Breakthrough Energy} \citep{PowerSimPaper2020}. The dataset contains high-temporal and spatial resolution load and VRE data for the entire US ``base$\_$grid'' in the year 2016.
It also contains load data for the year 2020 and projections for the year 2030. The dataset provides a test system with detailed information for existing buses, substations, plants, branches, and generation profiles for existing renewable energy capacity, including solar and land-based wind. We only use a subset of these datasets as explained in the following sections.

\noindent\textbf{General Note about Tables:} The first column of tables~\ref{SItab:exis_plant_params}, \ref{SItab:new_plant_params}, \ref{SItab:storage-params}, \ref{SItab:e-other-params},  and \ref{SItab:ng-other-params} shows the associated symbol in the formulation. The notations with \textit{tilde} are \textit{crude} cost values whose manipulated forms (e.g., regional update, annualization) are used in the numerical model. For example,  $\tilde{C}^{\text{inv}}_i$ in Table~\ref{SItab:new_plant_params} denote the value of ${C}^{\text{inv}}_i$ before annualization.

\subsubsection{Power Nodes}\label{SIsec:Powernodes}
\begin{figure}[htbp]
    \centering
    \includegraphics[width=0.6\textwidth]{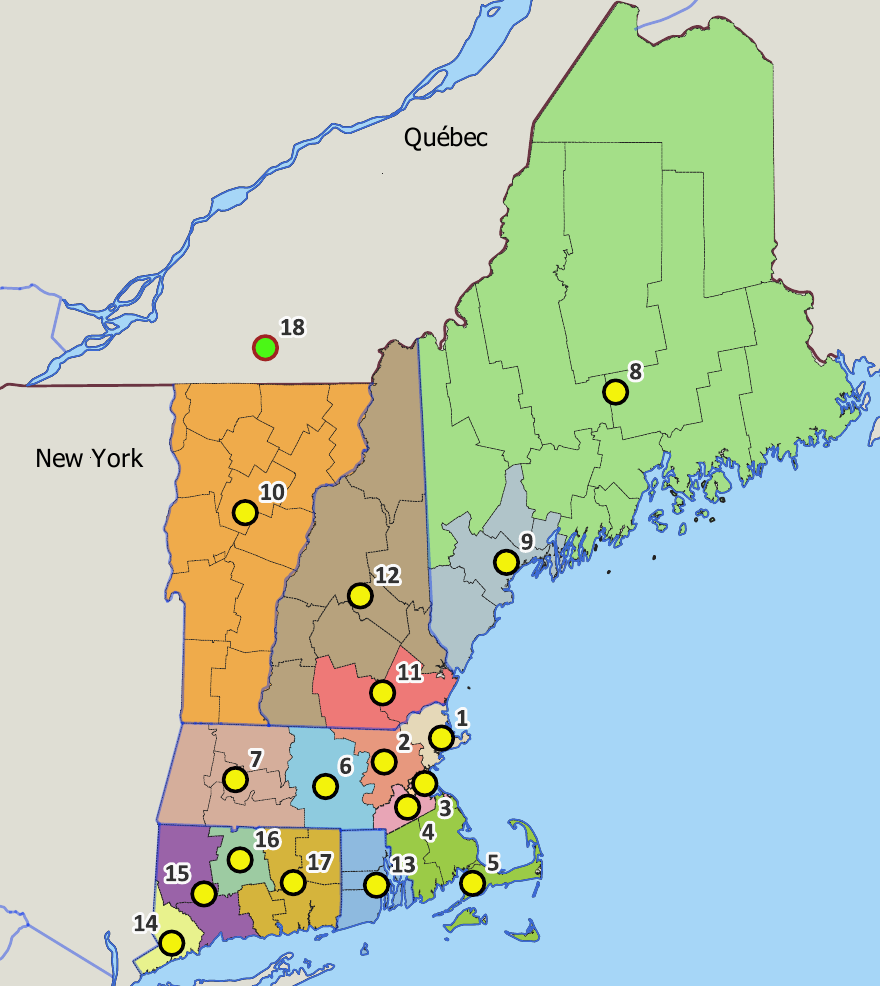}
    \caption{Grouping of New England counties into load zones, and the location of power nodes including the import node in Quebec}
    \label{SIfig:power-nodes}
\end{figure}

The power network has 18 nodes, which, except for the import node 18, are located in New England regions and correspond to a group of counties with a more than 600,000 aggregate population. The locations of nodes inside New England are set to the centroid of the counties they represent. The number of nodes is selected to sufficiently capture the network effect without significantly compromising the problem scalability. All counties of a group are in the same state and form a contiguous landmass. Fig~\ref{SIfig:power-nodes} shows the grouping of counties and the power node associated with each group. The import node 18 is located in the Quebec province of Canada. 

\subsubsection{Import Node}\label{SIsec:import-node}
Roughly 14\% of New England power consumption is imported \citep{ISONE-Mix2022}, mostly from Quebec, and the amount is expected to significantly increase by 2050 \citep{DimanchevEtal2020}. Therefore, we define an import node in Quebec as a power system node, albeit with some differences from other power nodes that are located inside New England. Node 18 in Fig.~\ref{SIfig:power-nodes} shows the import node's location. Our modeling assumptions for the import node are based on \citep{DimanchevEtal2020}. The study represents the entire Quebec system as a single bathtub that combines all reservoir and run-of-the-river systems. This leads to a reservoir with a maximum energy capacity ($U^{\text{eCap}}$) of 175.5 TWh and a power capacity ($U^{\text{powCap}}$) of 41.1 GW. According to the study, the minimum stable output is 27\% of Quebec's maximum power capacity. Also, it is assumed that the energy capacity of the reservoir should be at 70\%, 55\%, and 70\% of its maximum capacity on January 1, May 1, and December 31, respectively. The ramp rate of power production is given at 14\% of its maximum power capacity. The reservoir only incurs yearly FOM cost at \$29312 per MW. Our model does not allow the investment of new power plants or storage facilities in the node. 

There is currently a high-voltage transmission line between Quebec and a location in central Massachusetts which we assumed to be node 6 \citep{DimanchevEtal2020}. The maximum flow capacity of this line is 2 GW. There is another transmission line project with a capacity 1.2 GW under construction between Quebec and southwest Maine \citep{NECEC-project}. We assumed that this line is connected to node 9. 
The load and inflow rates for 2050 are provided in \citep{DimanchevEtal2020}. Around 90\% of Quebec's power load is already satisfied by hydropower and the rest mainly by solar and onshore wind generation. Since we only consider hydropower generation, we assume the power load of node 18 to be 90\% of its projected load for 2050.


\subsubsection{Transmission Lines}\label{SIsectrans-lines}
In the \textit{Breakthrough Energy} \citep{PowerSimPaper2020} data, we consider the base$\_$grid and start by filtering the data for New England region, which corresponds to zones 1 to 6 in the dataset. We then filter for high voltage buses \citep{ReEDS2019} i.e., those with voltage greater than or equal to 345kV. This process results in 188 nodes, each of which represents a high-voltage bus. The filtering process results in 192 transmission lines with known susceptance and maximum flow limit.
We use this data to identify the existing lines between power nodes. We first assign each of the 188 buses to its nearest county, and then keep the existing lines between any two counties belonging to  different power nodes.
This process resulted in 30 existing transmission lines, with total capacity of 103 GW, in the power network. 

\begin{figure}[htbp]
    \centering
    \includegraphics[width=0.6\textwidth]{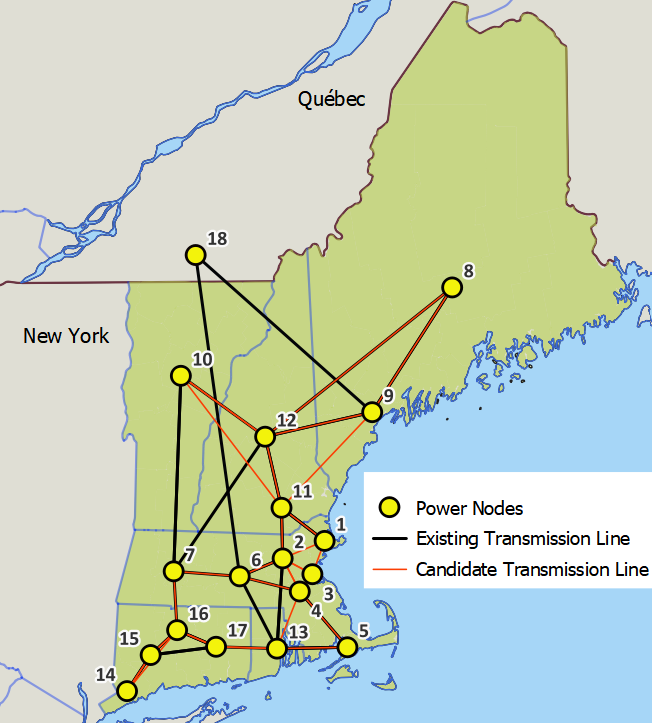}
    \caption{Existing and candidate transmission lines in the power network}
    \label{SIfig:all-trans-lines}
\end{figure}

Once the existing transmission lines are identified, we create candidate lines. We assume that each node can be connected to the two nearest  nodes  via candidate transmission lines, thus creating 34 candidate lines. The susceptance and maximum flow limit of these lines is set to their average in the set of existing lines. Fig.~\ref{SIfig:all-trans-lines} shows all the existing and candidate transmission lines in the current data set. It is worth noting that multiple transmission lines connect some node pairs, but only one is depicted in the figure.

\subsubsection{Power Plants}\label{SIsec:power-data}

The ``base$\_$grid'' data contains power plant information at each bus. We first remove power plant types of ``dfo'' (distillate fuel oil) and ``coal'' as their share is not substantial in the generation fleet of the historical year (i.e., 2016), and there are plans to completely phase out those plant types by 2050. We then calculate the existing generation capacity at each bus from each plant type by considering ``in-service'' plants with generation capacity greater than 10MW for ``ng'' (i.e., gas-fired plants) and ``nuclear'' type and greater than 2MW for VREs. Note that ``ng'' plants are treated as a lump since the Breakthrough Energy data set does not provide a breakdown between the capacity of combined-cycle and open-cycle plants. We then assign each plant to the nearest node and eventually aggregate the generation capacity of each plant type for each of the 17 nodes. Tables~\ref{SItab:exis_plant_params} and \ref{SItab:new_plant_params} present the technical assumptions for the existing and new power plants used in this study. Table~\ref{SItab:plant-node} presents the nameplate capacities for various generation technologies at each power node. Most of the parameters for the existing plants are derived from the Breakthrough Energy data set \citep{PowerSimData}. The footnote text in Table~\ref{SItab:exis_plant_params} presents details on the value of each parameter. The technical assumptions for new plants are largely derived from National Renewable Energy Laboratories (NREL) Annual
Technology Baseline 2021 edition (ATB 2021) for the year 2045 \citep{ATB2021}. Values of parameters not available in ATB 2021 are adopted from the corresponding existing power plants or obtained from Sepulveda et al. \citep{SepulvedaEtal2021}. Details of each parameter are provided in the footnote text in Table~\ref{SItab:new_plant_params}.

\begin{table}[ht]
\centering
\caption{Aggregate nameplate capacity of existing plants at each node (MW)}
    \footnotesize
    \setlength\tabcolsep{2.5pt}
\begin{tabular}{l|rrrrrrrrrrrrrrrrr}
\toprule
        & 1    & 2  & 3    & 4    & 5   & 6    & 7    & 8   & 9   & 10  & 11   & 12  & 13   & 14   & 15  & 16  & 17 \\
        \midrule 
ng      & 1327 & 80 & 3549 & 1807 & 23 & 848 & 630  & 805   & 0 & 1024 & 1526   & 1526 & 1811 & 1359 & 1235 & 959 \\
solar   & 25   & 58 & 9    & 46   & 102 & 154  & 75  & 0   & 0   & 57  & 0    & 0   & 10   & 0    & 0   & 15   & 10 \\
wind    & 2    & 0  & 2    & 0    & 8   & 3    & 56   & 686 & 213 & 115 & 0   & 185 & 51   & 0    & 0   & 5   & 0   \\
hydro   & 0    & 38 & 0    & 0    & 0   & 4    & 1822 & 406 & 199  & 280  & 33   & 219 & 0    & 68   & 48  & 8  & 0   \\
nuclear & 1226 & 0  & 0    & 0    & 617 & 0    & 0    & 0   & 0 & 0   & 0    & 0   & 0    & 0    & 0   & 0   & 1889\\
\bottomrule
\end{tabular}

\label{SItab:plant-node}
\end{table}


\begin{table}[ht]
\caption{Parameter Values for Existing Plants}
    \centering
    \scriptsize
    \setlength\tabcolsep{3pt}
    \begin{tabular}{ll|rrrrrrrr}
    \toprule
      Symbol &  Parameter$\backslash$Type                     & gas     & solar    & wind   & hydro   & nuclear  \\
\midrule       
& FOM [\$/kW/year]$^1$             & 21     & 23       & 43         & 78       & 145      \\
$C^{\text{var}}_i$& VOM [\$/MWh]$^2$                 & 5      & 0        & 0          & 0        & 2        \\
$\eta_i$& CO$_2$ Capture [$\%$]$^3$        & 0      & -        & -          & -      & -   \\
$h_i$ & Heat Rate    [MMBtu/MWh]$^4$     & 8.7    & 0        & 0          & 0   & 10.6        \\
$\tilde{C}^{\text{dec}}_i$ & Decom. cost per plant [$\$$]$^5$ & 5.0e6  & 4.5e4    & 1e6        & -            & 3.0e8 \\
$U^{\text{prod}}_i$ & Nameplate capacity [MW]$^6$      & {137}    & {4.7}     & {47}         & {15}      & 933  \\
$L^{\text{prod}}_i$ & Min stable output [$\%$]$^7$       & 31     & 0        & 0          & 0       & 42      \\
$U^{\text{ramp}}_i$ & Hourly Ramp rate [$\%$]$^8$      & 96     & -        & -          & -        & 25      \\
$C^{\text{fix}}_i$ & FOM per plant   [\$/MW/year]$^{9}$ &3.6e6&1.45e5     & 1.8e6      & 1.8e6  & 1.4e8 \\
\bottomrule
    \end{tabular}
    \label{SItab:exis_plant_params}
    
    \footnotesize{$^{1}$ and $^{2}$ from \citep{ATB2021} in year 2019,\qquad $^3$ from \citep{eiaEmisRate2021},\qquad $^4$  approximated from linear and quadratic coefficient of heat rate curve provided in \citep{PowerSimData},\qquad $^5$ estimated from \citep{Raimi2017} except for nuclear which is obtained from \citep{NukeDecom}. Decommissioning of hydro plants is not considered,\qquad $^6$ and $^7$ approximated from \citep{PowerSimData},\qquad $^8$ estimated from 30-min ramp rate in \citep{PowerSimData},\qquad $^{9}$ $r1*1000*r6$ where $r1$ is the FOM value in $r1$ and $r6$ is nameplate capacity provide in row$6$
    }
\end{table}

\begin{table}[ht]
\caption{Parameter Values for New Plants}
    \centering
    \footnotesize
    \setlength\tabcolsep{2pt}
    \begin{tabular}{ll|rrrrrrr}
    \toprule
  Symbol &  Parameter$\backslash$Type                        &OCGT       & CCGT       & CCGT   & solar & wind     & wind-off. & nuclear  \\
  &&&& -CCS &-new & -new&-new&-new\\
\midrule
&CAPEX [\$/kW]$^1$       & 780      & 935      & 2167     & 672       & 808      & 2043          & 6152     \\
&FOM [\$/kW/year]$^2$            & 21       & 27       & 65       & 15        & 35       & 74            & 145      \\
$C^{\text{var}}_i$& VOM [\$/MWh]$^3$               & 5        & 2        & 6        & 0         & 0        & 0             & 2        \\
$\eta_i$ &CO$_2$ Capture [$\%$]$^4$   & 0      & 0    & 90     & -         & -        & -             & -        \\
$h_i$&Heat Rate    [MMBtu/MWh]$^5$   & 9.72      & 6.36      & 7.16    & 0         & 0        & 0             & 10.46     \\
&Lifetime [year]          & 30       & 30       & 30       & 30        & 30       & 30            & 30       \\
$U^{\text{prod}}_i$& Nameplate capacity [MW]$^6$    & 237      & 573      & 400      & 10         & 10       & 10           & 360      \\
$\tilde{C}^{\text{inv}}_i$& CAPEX per Plant [$\$$]$^7$ & 1.85e8 &5.36e8 &8.67e8 & 6.72e6 & 8.01e6 &2.04e7 & 2.21e9 \\
$L^{\text{prod}}_i$& Minimum stable output [$\%$]$^8$ & 25       & 33       & 50       & 0         & 0        & 0             & 50      \\
$U^{\text{ramp}}_i$& Hourly ramping rate [$\%$]$^{9}$               & 100      & 100      & 100      & -       & -      & -           & 25       \\
$C^{\text{fix}}_i$& FOM per plant [\$/yr]$^{10}$ & 5.0e6 & 1.55e7 & 2.6e7 & 1.5e5  & 3.5e5 & 5.22e7      & 5.22e7 \\
\bottomrule
    \end{tabular}
    \label{SItab:new_plant_params}
    \footnotesize{$^{1-5}$ from \citep{ATB2021} in year 2045. For CCGT-CCS, ``Conservative'' technology class is considered. For ``wind-new'' and wind-offshore ``Moderate-Class4'' technology class is considered. For all others ``Moderate'' cost assumption is considered,\qquad $^6$ from \citep{SepulvedaEtal2021} for ``OCGT'', ``CCGT'', CCGT-CCS, and ``nuclear-new''. For VRE, a modular capacity of 10MW is considered, \qquad $^7$ $r1*1000*r6$ where $r1$ is the CAPEX value in row$^1$ and $r6$ is nameplate capacity provide in row$^6$, \qquad $^8$ from \citep{SepulvedaEtal2021}, \qquad $^{9}$ from \citep{SepulvedaEtal2021}, \qquad $^{10}$ $r2*1000*r6$ where $r2$ is the FOM value in $r1$ and $r6$ is nameplate value provide in $r7$
    }
\end{table}

\subsubsection{Offshore Wind Siting}
The development of large-scale offshore wind power generation is currently ongoing only for Massachusetts and Rhode Island \citep{BOEM-lease-regions}, hence we allow the establishment of offshore wind plants only at nodes 1, 2, 3, 4, 5, and 13. \\

\subsubsection{Power Load}\label{SIsec:powerloaddata}
We obtain the residential load for each of the electrification scenarios from the bottom-up approach introduced in the paper and elaborated further in Section~\ref{SIsec:ResStock}. To account for non-residential load, we consider the 2050 hourly load profiles provided as part of the NREL's Electrification Future Study (EFS) Load Profile dataset \citep{EFSload2022}. The repository contains hourly load profiles for various electrification (Reference, Medium, High) and technology advancement (Slow, Moderate, Rapid) scenarios. The load profiles are provided for several years, including 2050, and are further disaggregated by state, sector (residential, commercial, etc.), and subsectors (space heating and cooling, water heating, etc.). For all scenarios, we consider the aggregated state-level hourly demand profile for the \textit{High} electrification level with \textit{Moderate} technology advancement. The state-level non-residential load is further disaggregated to the power node level based on the population share of each node throughout the entire state.

\subsubsection{Regional Cost Multipliers}
We incorporate regional capital cost multipliers provided in \textit{ReEDS Model Documentation} \citep{ReEDS2019}, summarized in Table~\ref{SItab:regional-mults}, to distinguish the capital costs of new power plants in the different model regions. These multipliers are applied to the baseline capital costs reported in Table~\ref{SItab:new_plant_params}. Subsequently, the capital costs are annualized to be included in the single-stage investment planning model using the following formula for the annual cost fraction: $ \frac{\omega}{1-(\frac{1}{1+\omega})^{{lt}}}$. Here, $lt$ is the lifetime of the specific technology and $\omega$ corresponds to the discount rate of 7.1\%. Thus, the annualized CAPEX for new power plants is obtained by multiplying the  CAPEX by the annual cost fraction and regional multiplier. For every other investment cost (i.e., transmission lines, pipelines, storage, etc.), no regional cost multiplier is considered and we only multiply the CAPEX by the annual fraction factor to get the annualized CAPEX.

\begin{table*}[ht]
\caption{Regional CAPEX multipliers for new plant types}
\label{SItab:regional-mults}
\centering
    \scriptsize
    \setlength\tabcolsep{2pt}
\begin{tabular}{c| rrrrrrr}
\toprule
State/Technology &OCGT   & CCGT   & CCGT & solar & wind & wind-off. & nuclear \\
 &&& -CCS &-new & -new&-new&-new\\
\midrule 
Connecticut (CT)      & 1.25 & 1.3  & 1.3    & 1.15      & 1.4      & 1.1             & 1.1         \\
Massachusetts (MA) & 1.1  & 1.1  & 1.1    & 1.05      & 1.35     & 1.1             & 1.05        \\
Maine (ME) & 1.25 & 1.3  & 1.3    & 1.1       & 1.35     & 1.1             & 1.1         \\
New Hampshire (NH)   & 1.1  & 1.1  & 1.1    & 1.05      & 1.35     & 1.1             & 1.05        \\
Rhode Island (RI) & 1.2  & 1.25 & 1.25   & 1.1       & 1.35     & 1.1             & 1.05        \\
Vermont (VT) & 1.1  & 1.1  & 1.1    & 1.05      & 1.35     & 1.1             & 1.05 \\
\bottomrule
\end{tabular}
\end{table*}

\subsubsection{Plant Decommissioning Cost}
{The decommissioning costs} for power plants and pipelines are not annualized. Instead, 
we assume a gradual retirement process starting from the mid-year date of 2040 until the planning year 2050. Since the model only considers a time horizon of a single year, we divide the decommissioning costs for these assets by 10. The distribution of the decommissioning cost over multiple years allows the model to decommission the asset if it is not utilized and its fixed cost is higher than the distributed decommissioning cost. Otherwise, the model keeps the idle asset without decommissioning it.

\subsubsection{Power Storage}
Energy storage is likely to be an essential part of the future power systems dominated by VRE supply. While many storage technologies are proposed or under development, we model Li-ion batteries. The technology provides short-term storage, usually around four hours. We sourced Li-ion battery cost and performance assumptions from \citep{ATB2021} as summarized in Table~\ref{SItab:storage-params}.

\begin{table*}[ht]
    \centering
        \caption{power storage parameters}
    \label{SItab:storage-params}
    \begin{tabular}{ll|c c c}
         Symbol &  Parameter & Li-ion   \\
         \midrule
     $\tilde{C}^{\text{EnInv}}$&    Energy capital cost [$\$$/kW]\\
       $\tilde{C}^{\text{pInv}}$&  Energy power cost [$\$$ /kWh]& 156$^4$ \\
        $\gamma^{\text{eCh}}_r$&  Charge efficiency & 0.92$^5$ \\
        $\gamma^{\text{eDis}}_r$&  Discharge efficiency & 0.92$^6$  \\
       $C^{\text{EnInv}}_r$&   Energy related FOM ($\$$/kWh/year) & 3.22$^7$ \\
        $C^{\text{pInv}}_r$&   Power related FOM ($\$$/kW/year) & 3.9$^8$ \\
    $\gamma^{\text{selfD}}_{r}$ & hourly self-discharge rate & 2.08e-5$^9$ \\
         & Lifetime & 15$^{10}$ \\
         \bottomrule
    \end{tabular}
    
\footnotesize{columns $^1$ and $^2$ from \citep{MITEI2022FES},\\ $^{3}$ and $^4$ from \citep{ATB2021} in year 2045 averaged over ``Advance'', ``Moderate'' and ``Conservative'' estimates, \qquad $^5$ and $^6$ from \citep{ATB2021} where the round-trip efficiency is provided at 85\%, \qquad  $^7$ and $^8$ 2.5$\%$ of energy capital and power cost (row 1 and 2), respectively  \citep{ColeFrazier2021}, \qquad $^9$ from \citep{MITEI2022FES} the monthly self-discharge provided at 1.5\%, \qquad $^8$ from \citep{ATB2021}}
\end{table*}

    

\subsubsection{Reserve Margin}
{The capacity derating factor}  $\gamma^{\text{CRM}}_{nit}$ for VRE plants is set to their capacity factors. For other plants, the value is set to 1. The capacity reserve margin rate $R^{\text{CRM}}$ is set at 15\%, which is in the range of 13-17\% recommended by North American Electric Reliability Corporation (NERC) \citep{ReimersEtal2019}. 

\subsubsection{Transmission Line FOM}
{The fixed cost for transmission lines} is estimated from \citep{EPRI2009}, in which the operating and maintenance cost for a transmission line of 134 kV is given as about 20\% of the investment cost. Therefore, we estimate the FOM cost as 20\% of the CAPEX over its lifetime. We currently consider CAPEX of a line at 7398 $\$$/MW/mile over 30 years so that the annual FOM will be $C^{\text{trFix}}_\ell \frac{7398\times 0.2}{30} \approx 49.3 \$$/MW/mile for all $\ell \in \mathcal{L}^{\text{e}}$. Investment and FOM costs of a line can be calculated by multiplying these values by the length and maximum capacity of a line.

\subsubsection{CCS parameters}
We based our estimation of CCS parameters on \citep{TeletzkeEtal2018, BlondesEtal2013,BoardEtal2019}. We assume that the collected carbon is stored in Appalachian Basin at a Marcellus region located in the middle of Pennsylvania \citep{BlondesEtal2013}. We then calculate the distance between each node and the storage site. The total capacity of the Appalachian Basin is 1278 Mt (megaton). Assuming the Basin is operable for 100 years, the total annual carbon storage capacity $U^{\text{CCS}}$ becomes 12.78 Mt. Other parameters are calculated as follows:
\begin{itemize}
    \item $C^{\text{inv}}_{\text{CO2}}$: Reference \citep{BoardEtal2019} provides CAPEX for 10 and 100 miles pipelines. We consider 100-mile pipelines as all distance values are greater than 200 miles. The CAPEX and FOM for 100 miles are 225 $\$$M and 1.3 $\$$M (million dollars), respectively. With 30 years of a lifetime for CO$_2$ pipelines and WACC=7.1\%, the CAPEX becomes 18.31 $\$$M. The FOM is given at 1.3 $\$$M, so the per mile investment and FOM is (18.31e6+1.3e6)/100 = 196e3 $\$$/mile. The reference assumes that the capacity of the pipeline is 10 Mt/y (megaton per year). Therefore, the levelized investment and FOM becomes 196e3/10e6 = 0.0196 $\$$/mile/ton.
    \item $C^{\text{str}}_{\text{CO2}}$: The Appalachian basin is an aquifer type storage. The CAPEX is given at 4.3 $\$$M in \citep{BoardEtal2019} for a storage site of type aquifer with 7.3 Mt capacity per year. The CAPEX consists of injection site screening and evaluation, injection equipment, and drilling of 6 wells. Assuming 30 years of lifetime, the annualized CAPEX becomes 350e3 $\$$. The FOM is given at 600e3 $\$$, so the levelized investment and FOM is (350e3+600e3)/7.3e6 = 0.13 $\$$/ton. 
    \item $E^{\text{pipe}}$: For a pipeline of 100 miles long with a capacity 10 Mt/year, reference \citep{BoardEtal2019} gives the electric requirement at 32,000 MWh/year or equivalently 32/8760e6 = 0.00365e-6 [MWh/mile/ton/hour].
    \item $E^{\text{pipe}}$: For a pipeline with a capacity of 10 Mt/year, each pump consumes 4190 MWh electricity annually which accounts for 4190/(8760e6)=0.478e-6 MWh/ton/hour. 
    \item $E^{\text{cprs}}$: Compression pump are located every 3.3 miles along the pipeline \citep{BlondesEtal2013}. Therefore, the value is $d_n$/3.3, where $d_n$ is the distance of the node from the carbon storage location in miles.
\end{itemize}

We assume that CAPEX for compression pumps is negligible. Also note that these cost estimations for CCS storage can be an underestimation as we levelized the investment and FOM costs and do not consider other cost parameters such as labor, FOM of compression pumps, and fugitive emission amount which are listed in \citep{BlondesEtal2013}.

\subsubsection{Other Parameters}
Other economic and technical assumptions for the power network are presented in Table~\ref{SItab:e-other-params} and Table~\ref{SItab:res-avail}. 

\begin{table*}[ht]
\caption{Other parameters for power network}
    \centering
    \begin{tabular}{ll|r}
    \toprule
     Symbol &  Parameter & Value\\
    \midrule 
    & Weighted average cost of capital (WACC)$^{1}$ & $7.1\%$\\
 $\tilde{C}^{\text{trans}}_\ell$&    Transmission line investment cost [$\$/$MW/mile]$^2$ &7398 \\
    &Li-ion Battery lifetime [year]$^3$ & 30\\
    &Transmission line lifetime [year]$^4$ & 30\\
    & Uranium price [$\$$/MMBtu]$^5$ & 0.72\\
    \end{tabular}
    \label{SItab:e-other-params}
    
    \footnotesize{$^1$ from \citep{SepulvedaEtal2021}, \qquad  $^2$ average of costs reported in Table 9 of \citep{DimanchevEtal2020},\qquad $^3$ from \citep{ATB2021}, \qquad  $^4$ from \citep{ATB2021} in year 2045, $^5$ from \citep{SepulvedaEtal2021} in year 2045
    }
\end{table*}

\begin{table}[ht]
\caption{Resource Availability Data}
    \centering
    \begin{tabular}{l|r}
    \toprule
     $\mathcal{Q}$ & Maximum Available Value\\
    \midrule 
    {``solar'', solar-UPV} & 22 GW\\
    {``wind'', ``wind-new''} & 10 GW\\
    {wind-offshore} & 280 GW\\
    {``nuclear'', ``nuclear-new''} & 3.5 GW\\
    ``hydro'' & 2.6 GW
    \end{tabular}
    \label{SItab:res-avail}
    
    \footnotesize{Resource availability amounts are obtained from \citep{E3Team2020} except for ``hydro'' which is obtained from \citep{DimanchevEtal2020} such that 853 MW is run-of-river and 1768 MW is pumped storage hydro power
    }
\end{table}


\subsection{Gas System Data}
The gas network consists of two types of nodes: 1) Gas nodes that have injection capacity and load and 2) SVL nodes. We further distinguish between gas nodes whose load is zero but can be used to inject gas in the system and those that have gas load. The nodes without load are created in the boundary of the region to represent the connection of gas nodes to the outside region. We call the former set of nodes ``boundary nodes'' and the latter ``load nodes.'' Note that this distinction is for the exposition of the data input and has no modeling implications. Each SVL node consists of Storage, Vaporization, and Liquefaction (SVL) facilities, each with its own capacity. This section provides details of each node type and other parameters associated with the gas system. 

\subsubsection{NG Nodes}\label{app-NG-topology}
We construct the gas network based on the data available on the Energy Information Administration (EIA) website  \citep{eiaWebsite2022_NG_pipe}. The pipeline data provides information on the interstate and major pipelines, and their start and ending counties. Each pipeline is a uni-directional means of transferring gas with a  daily capacity limit. We first filter all the pipelines whose ending counties are one of the New England counties. We consider each of these counties as a load node. Some load nodes are connected to regions outside New England via pipelines. To capture the import of gas to the region via pipelines, we create boundary nodes in locations where a pipeline connects a load node to an outside region. This process resulted in six boundary nodes and 17 load nodes. The import of gas fuel to the region is realized either through pipelines from neighboring regions or coastal import facilities from overseas \citep{eiaWebsite2022_NG_pipe,NGstatGuide2021}. We consider the boundary nodes as ground injections. The injection to the  system can also be realized by LNG facilities located in Everett and Cape Ann, Massachusetts \citep{NGstatGuide2021} that are equipped with vaporization plants. Our model does not differentiate between LNG and NG in terms of cost, consumption, or any other characteristics. 

NG is usually stored in its liquefied form called LNG. LNG storage usually involves three types of facilities:
\begin{itemize}
    \item Storage tank: smaller tanks located inland that are filled by truck supply from the import facilities and are used to regulate the pressure of the gas system – these facilities have tanks and vaporization facilities \citep{MallapragadaEtal2018}. The capacity of a storage facility is measured in energy/volume level (i.e., MMBtu or MMScf)
    \item Liquefaction: a facility that receives NG from a pipeline system and liquefies it at a temperature of around -162 Celsius \citep{MallapragadaEtal2018, Mesko1996}. The capacity of the liquefaction facility is measured in terms of flow rate (i.e., MMBtu/period or MMScf/period)
    \item Vaporization: The facility evaporates the LNG leaving the storage facility by warming it with seawater or air to produce gas that is injected into the pipeline network. In certain areas, trucks are used to transport LNG from a major storage site to a smaller storage site where they are later evaporated \citep{MallapragadaEtal2018}. For ease of transportation, storage and vaporization facilities are built at the same location. The capacity of the vaporization facility is measured in terms of flow rate (i.e., MMBtu/day or MMScf/day).
\end{itemize}

\begin{figure}[htbp]
    \centering
    \includegraphics[width=0.6\textwidth]{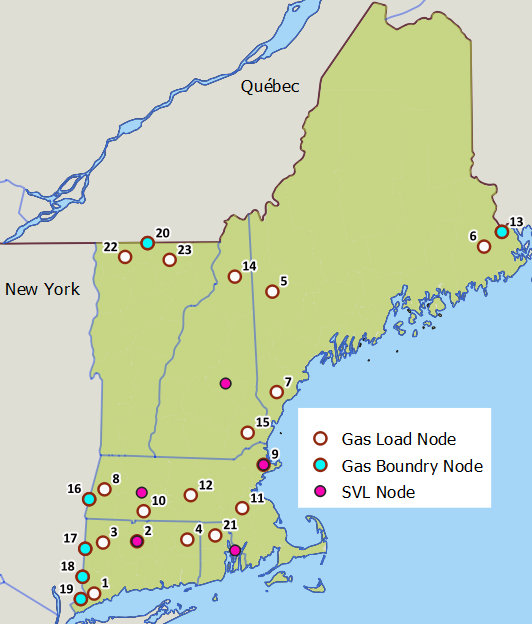}
    \caption{Location of each type of node in the gas system. Note that two SVL and load nodes share the same location. SVL nodes are not numbered here for clarity of the figure. }
    \label{SIfig:NG-nodes-all}
\end{figure}

We construct the SVL network based on the data provided in \citep{NGstatGuide2021}. We assume that all three facilities are located at the same location. There are currently five  liquefaction and 43 storage facilities in New England. The exact location of these facilities is not provided, but the source provides a map of the region showing the approximate locations of storage tanks. Given that there are only five liquefaction facilities, we first cluster storage tanks into five locations and assume a liquefaction and a vaporization facility at each location. This assumption effectively approximates the practice of moving LNG via trucks from the centralized liquefaction facilities to the distributed storage (and vaporization) facilities. Each of these locations is a node in the SVL network. The total liquefaction, vaporization and storage capacities for the New England region are given in \citep{NGstatGuide2021}. To account for variations in the capacities, we unevenly distribute these capacities over five SVL.
This process resulted in six boundary nodes, 17 load nodes, and five SVL nodes as depicted in Fig.~\ref{SIfig:NG-nodes-all}. Note that node 9 also

\subsubsection{Pipelines}
We use the pipeline data from \citep{eiaWebsite2022_NG_pipe} to identify the existing pipelines between load nodes and between boundary nodes and load nodes. Each existing pipeline has a known daily capacity limit. We construct candidate lines for boundary nodes by creating a pipeline between them to three of their nearest load nodes. 

\begin{figure}
    \centering
    \includegraphics[width=0.6\textwidth]{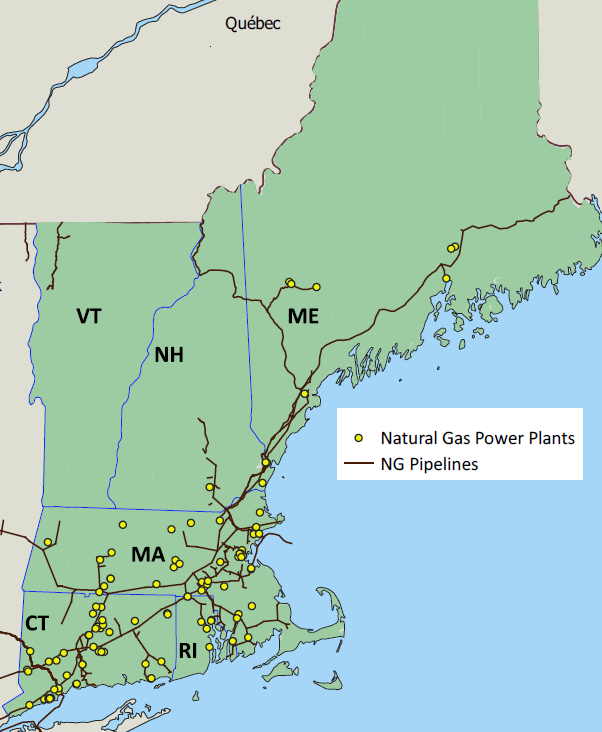}
    \caption{Existing NG pipelines and gas-fired power plants in New England}
    \label{SIfig:pipe-power-plants}
\end{figure}

The transfer of gas between different nodes is assumed to be realized as follows:

\begin{itemize}
    \item Boundary and load nodes to load nodes: The existing interstate pipelines between boundary and load nodes are provided in \citep{eiaWebsite2022_NG_pipe}. We estimate the existing intrastate pipelines based on the map provided in \ref{SIfig:pipe-power-plants} with capacities associated with the interstate pipeline ending in the node from which the intrastate pipeline originates. For candidate pipelines, we assume that each node is connected to 2 nearest non-adjacent, non-boundary load nodes, resulting in 46 candidate pipelines. The capacity of candidate pipelines is set to the average capacity of the existing ones. 
    \item Gas load node to power node: The location of gas-fired power plants as well as gas pipelines is provided in \citep{EIA-layer-maps}. Fig.~\ref{SIfig:pipe-power-plants} shows that each power plant is connected to one major pipeline. We use this fact to assume that each power node is already connected to its nearest NG node through distribution pipelines. The pipeline capacity available to these neighboring gas nodes thus limits the amount of flow between gas and power nodes, hence we do not consider a limit to pipelines connecting gas and power nodes.
    \item Gas load nodes and SVL nodes: We assume that each gas load node is already connected to its nearest SVL node through two sets of pipelines. The first set carries gas from the gas node to the liquefaction facility in an SVL, and the second set of pipelines transports gas from the vaporization facility in an SVL to a gas node. 
\end{itemize}

\begin{figure}
    \centering
    \includegraphics[width=0.6\textwidth]{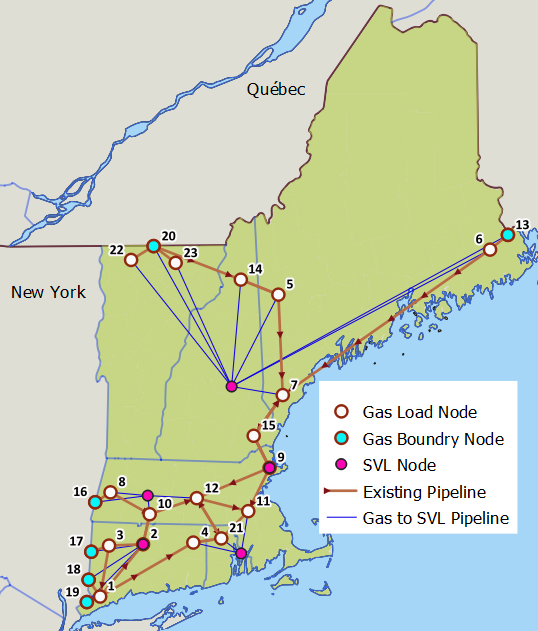}
    \caption{Existing pipelines between three types of gas nodes. The connections between load and SVL nodes are shown by a single line for clarity of the figure.}
    \label{SIfig:exis-pipelines}
\end{figure}

\begin{figure}
    \centering
    \includegraphics[width=0.6\textwidth]{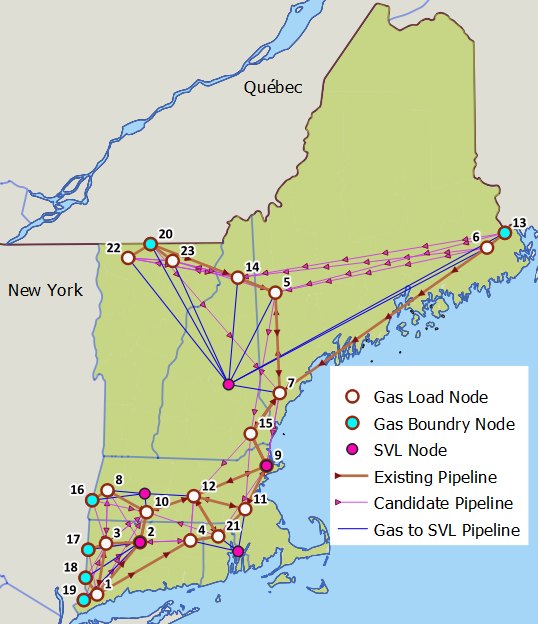}
    \caption{Existing and candidate pipelines between three types of gas nodes}
    \label{SIfig:all-NG-pipelines}
\end{figure}

\begin{figure}
    \centering
    \includegraphics[width=0.7\textwidth]{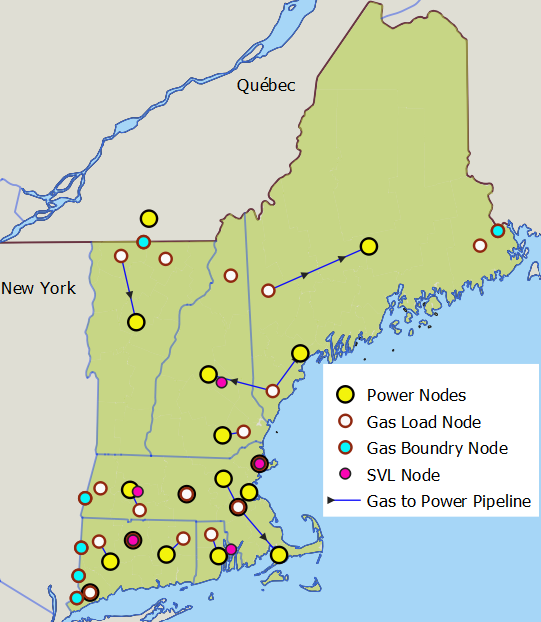}
    \caption{Pipelines between power nodes and gas load nodes}
    \label{SIfig:NG2power-lines}
\end{figure}

\begin{figure}
    \centering
    \includegraphics[width=0.8\textwidth]{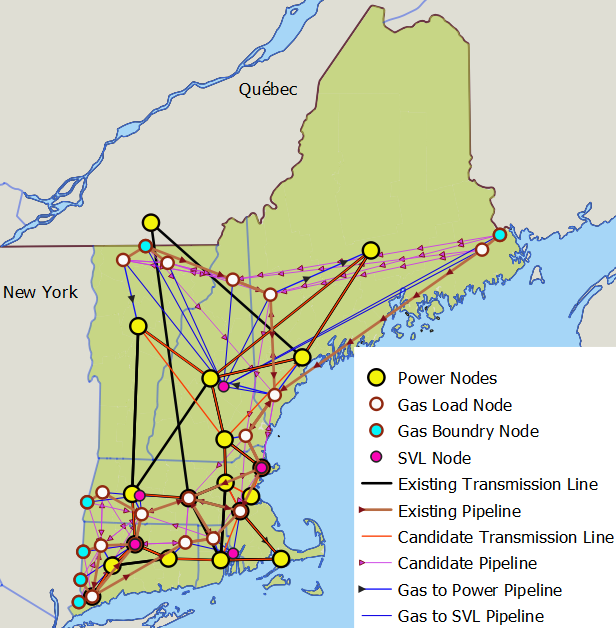}
    \caption{All existing and candidate connections in New England case study}
    \label{SIfig:all-connections}
\end{figure}

Fig.~\ref{SIfig:exis-pipelines} illustrates the existing pipelines of all node types of the gas system. Currently, in the year 2022, there are 
25 pipelines that connect boundary and load nodes. The procedure that we considered to create candidate lines resulted in 46 potential connections that are depicted in Fig.~\ref{SIfig:all-NG-pipelines}. The power nodes draw gas from their nearest load nodes as illustrated in Fig.~\ref{SIfig:NG2power-lines}. Finally, Fig.~\ref{SIfig:all-connections} shows all the connections we consider in \ModelName.

\subsection{LCF Availability} \label{SI:LCF_avail}
The supply curve for LCF modeled in this study is based on limited available analysis in the literature on the potential cost and availability of LCF for the Northeast region. In developing the LCF supply curve, we primarily focused on supply and cost of bio-methane or renewable natural gas.\\
We start by estimating the availability of biogenic LCF in New England. The annual production potential of New England for 2040 is given in \citep{NE_RNG_report2019} for three scenarios, namely low, high, and technical resource potential. We use the technical resource potential in 2040 reported by this study, at 250.8 tBty \citep{NE_RNG_report2019}. The other source of LCF to be used in New England is the production in other states as well as the neighboring regions. New York is a likely neighboring state that has plans for biogenic LCF and may be able to partially provide for New England's LCF needs. We used the maximum production potential of biogenic LCF in New York State in 2040, estimated by another study at 272.3 tBTu under a maximum growth scenario \citep{NY_RNG_report2022}. We further assume that consumption of biogenic LCF among subregions in the US Northeast (New England and New York) is proportional to their current natural gas consumption. In 2022, New England’s share was 41\% of the total natural gas consumed in the US Northeast \citep{eiaWebsite2022_NG_pipe}. Accordingly, the combined biogenic LCF potential for the US Northeast is 250.8+272.3=523.1 tBtu, of which 0.41*523.1=214.5 tBtu is assumed to be available for consumption in New England across all sectors.\\

The biogenic supply of LCF is priced based on a recent study on LCF supply for Massachusetts that estimated that the first 33\% of biogenic LCF could be produced at 10\$/MMBtu while the rest would be at around 25\$/MMBtu (estimated from Fig. 3 of \citep{MA_RNG_report2022}). Collectively, the above biogenic supply of 214.5 TBtu and the two price levels define the first two bins of the supply curve as shown in  Table~\ref{SItab:LCF_avail}.

Any amount of LCF in excess of the projected biogenic supply of 214.5 tBtu is assumed to be sourced from synthetic methane production, which involves coupling biogenic CO2 sources with renewable H$_2$. This process will require electricity input that was not explicitly considered in our analysis. As a conservative estimate, we priced the LCF supply above 214.5 TBtu at 50\$/MMBtu. Interestingly, across the modeled high electrification scenarios (HE, HX), we find that LCF consumption stands at 70-277 TBtu indicating limited to no reliance on synthetic methane supply under high electrification of buildings sector.

\begin{table}[htbp]
    \centering
    {\begin{tabular}{|l|l l l|}
    \toprule
        Availability Intervals (tBtu)  &  [0, 70.8] &(70.8, 214.5] &(214.5, 1000]\\
         Price (\$/MMBtu) & 10 &25& 50\\
         \bottomrule
    \end{tabular}}
    \caption{{LCF availability and price levels. About 33\% of total capacity is produced at 10 \$/MMBtu, and the rest at 25\$/MMBtu. Any amount beyond that, is assumed to be produced by synthetic fuel at a price 50\$/MMBtu. The last interval is capped by 1000 tBu as a large upper bound for modeling purposes. }}
    \label{SItab:LCF_avail}
\end{table}


\subsubsection{Pipelines FOM Costs}
The FOM cost of gas pipelines is estimated from \citep{HafnerLuciani2022handbook} in which it is stated that 5-10\% (on average 7.5\%) of a pipeline's cost across its lifetime is operating cost, of which 72\% is the fixed cost. {Since we assume a 50-year lifetime for new pipelines \citep{Chohan2023}, the FOM cost for pipelines is calculated as $C^{\text{pipeFix}}_\ell = \frac{20e6\times 0.075*0.72}{50}=21600 \$$/mile for all $\ell \in \mathcal{L}^{\text{g}}$.}

\subsubsection{Pipelines Decommissioning Cost}
A cost estimate for decommissioning of pipelines in the US Gulf of Mexico between 1995 to 2015 is provided in \citep{Kaiser2017PipeDecom} is provided at around 3e5 $\$$/mile for offshore pipelines. {However, we could not find any substantial study estimating the decommissioning cost for the onshore pipelines. Furthermore, unlike offshore pipelines, onshore pipelines can be repurposed to be used by other energy carriers such as hydrogen. Therefore, it is not clear whether the decommissioning of pipelines will carry any cost at the end of their lifetime value, hence we set the decommissioning cost to zero for all pipelines.}  

\subsubsection{Other Parameters}
Other economical and technical assumptions for the gas system are presented in Table~\ref{SItab:ng-other-params}.

\begin{table*}
\caption{Other parameters for NG/SVL network}
    \centering
    \begin{tabular}{ll|r}
    \toprule
    Parameter & Value\\
    \midrule 
  $\tilde{C}^{\text{strInv}}_j$&  Storage tank CAPEX$^1$ [$\$$/MMBtu]& 729.1\\
    $\tilde{C}^{\text{vprInv}}_j$&  Vaporization CAPEX$^2$ [$\$$/MMBtu] & 1818.31\\
    $C^{\text{strFix}}_j$&  Storage tank FOM$^3$ [$\$$/MMBtu]& 3.6\\
   $C^{\text{vprFix}}_j$& Vaporization FOM$^4$ [$\$$/MMBtu/d] & 327.3\\
   $\gamma^{\text{ligCh}}_j$ & Liquefaction charge efficiency ($\%$) & 100\\
   $\gamma^{\text{vprDis}}_j$ & Vaporization discharge efficiency$^5$ ($\%$) & 98.9\\
 $C^{\text{gShed}}_j$  & Gas load shedding cost [$\$/$MMBtu] & 1000\\
  $\eta^{\text{g}}$&  Emission factor for NG $^{6}$ [ton/MMBtu]& 0.053 \\
 $C^{\text{alt}}$ & LCF price [\$/MMBtu]$^{7}$ &20\\
  & SVL lifetime [year] & 30\\
    $\tilde{C}^{\text{pipe}}_\ell$& Pipeline investment cost [$\$$/mile]$^{8}$ & 20e6\\
   $C^{\text{ng}}$& NG price [$\$/$MMBtu]$^{9}$  & 5.45 \\
     &  Pipeline lifetime$^{10}$ [year] & 50\\
   \bottomrule
    \end{tabular}
    \label{SItab:ng-other-params}
    
    \footnotesize{$^1$ from Table 1 of \citep{LNGcapex2019},\qquad $^2$ from Table 1 of \citep{LNGcapex2019},\qquad $^3$ 0.5$\%$ of CAPEX according to \citep{Mesko1996},\qquad $^4$ 1.8$\%$ of CAPEX according to \citep{Mesko1996},\qquad $^5$ estimated from \citep{MallapragadaEtal2018},
    \qquad $^{6}$ from \citep{eiaWebsite2022_NG_emissions},
    \qquad $^{7}$ from \citep{ColeEtal2021},
    \qquad $^{8}$ approximated from ``Pipeline projects'' provided in \citep{eiaWebsite2022_NG_pipe} for gas pipeline projects. We consider the completed 
new pipeline projects in New England between 2011 to 2020 and divide the total cost to the length of the pipeline, \qquad $^9$ from \citep{SepulvedaEtal2021}
\qquad $^{10}$ from \citep{Chohan2023}
    }
\end{table*}


\subsubsection{NG Load}\label{SIsec:NGloaddata}

Similar to the power system, the gas load for the residential sector is obtained from the bottom-up approach proposed in this paper.
The load for gas is generally disaggregated into five sectors, including residential, commercial, industrial, vehicle fuel consumption, and electric power customers \citep{eiaWebsite2022_NG_pipe}. The gas consumption for electric power generators is a decision variable in our model, so our input for gas demand involves the gas demand in the remaining three sectors. 
The monthly state-level gas load for all five sectors is available on the EIA website,  ``Natural Gas Data'' page \citep{eiaWebsite2022_NG_pipe}. The EIA website provides no information on the distribution of the monthly demand over its days (i.e., the load shapes). Therefore, we consider industrial, commercial, and vehicle fuel consumption in 2017 and uniformly distribute the monthly load across their corresponding days. We then scale the values based on the annual industrial, commercial and transportation gas consumption in 2050 under high electrification, moderate technology advancement scenario of NREL's \textit{Electrification Futures Study} \citep{EFSload2022}. Finally, we aggregate loads for all sectors and subsectors to obtain the gas consumption of each state in 2050 under the two electrification scenarios. 
Once the daily gas demand is obtained for each state, we disaggregate the demand over each node in the proportion of the population each node represents in 2019 \citep{CensusData2019}, which implicitly assumes that there is no change in the proportion of the population by 2050.

\subsection{Emission Amounts}\label{SIsec:emission-amount}
All New England states have set a goal to reduce the emission of GHG in 2050 by at least 80$\%$  below a baseline year, which is 1990 for all states except Connecticut \citep{Brattle2019}. The 80\% goal has been subject to discussion in recent years to increase, with Vermont already setting the target at 95\%. The total CO$_2$ emission for the New England states was 171.2 metric tons (mt) in 1990, of which 43.9mt was electricity energy-related emission and 23.6mt was NG energy-related emission \citep{eiaCO2emis}. The remaining emission was caused by consuming coal and petroleum that we do not consider in this study. Based on these figures, $U^{\text{g}}_{\text{emis}} =$ 23.6e6 and $U^{\text{e}}_{\text{emis}}=$43.9e6.



\section{Capacity factor calculations}\label{SIsec:CFs}
In our capacity factor modeling, we use weather data most fit-for-purpose for modeling the individual resource types, primarily using inputs specifically created for and calibrated to the usage in production modeling. We do not include climate change impacts in the capacity factor modeling.

\subsection{Solar capacity factors}\label{SIsec:methodsCFsolar}
We calculate solar capacity factors for the 2001-2020 weather years using the pvlib library in Python \citep{PVLib2018}. We leverage solar data from the National Solar Radiation Database, which is more fit for purpose than the ERA5 data used in our demand-side analysis. Our input parameters follow the assumptions of Brown \& O’Sullivan \citep{Brown2020}, except we do not model inverter efficiency losses (which are already considered in our cost parameters). We use the default incidence angle modifier assumptions in the pvlib physical model and assume an anti-reflective coating index of 1.3. Capacity factors are calculated for each weather location used in the demand-side analysis.
Due to the cold climate of the study region, we use NREL’s snow coverage model to approximate the hourly impact of snow coverage on solar capacity factors. We use snow accumulation data from the ERA5 data used in the demand analysis. The data is provided in millimeters of water equivalent, which we convert to inches of snow accumulation by using the one-equation model presented in \citep{Hill2019}. While the one-equation model is shown to have a negative bias, it is useful as a rough approximation in our analysis, particularly because snow depth is only one component of the model for snow-related capacity derating. The snow coverage model is described in \citep{Ryberg2017}, which calculates the fraction of the panel covered in snow given the hourly snowfall, outdoor temperatures, panel geometry, and an empirically defined coefficient describing the rate that snow slides off the panel. We assume the capacity derating is proportional to the fraction of panel surface covered by snow. Because the model is specific to fixed-tilt systems and we model systems with single-axis tracking, we approximate the slide rate for a panel with a tilt equal to the latitude of the panel site. We do not consider the likelihood that climate change reduces the frequency and intensity of snowfall events.

Capacity factors for each zone are calculated by averaging the capacity factors for the sites located within them. We use the same sites as in the demand analysis, shown in Fig.~\ref{SIfig:weathermap}. For zones that do not include a site, capacity factors are taken from the site closest to the zone’s centroid. 
Our solar capacity factors averaged 20.3\% across the zones for 20 years. This is in agreement with NREL estimations of capacity factors in the region for 2019 \citep{ATB2022}.

\subsection{Onshore and offshore wind capacity factors}\label{Sisec:methodsCFWind}
\begin{figure}
    \centering
    \includegraphics[width=\textwidth]{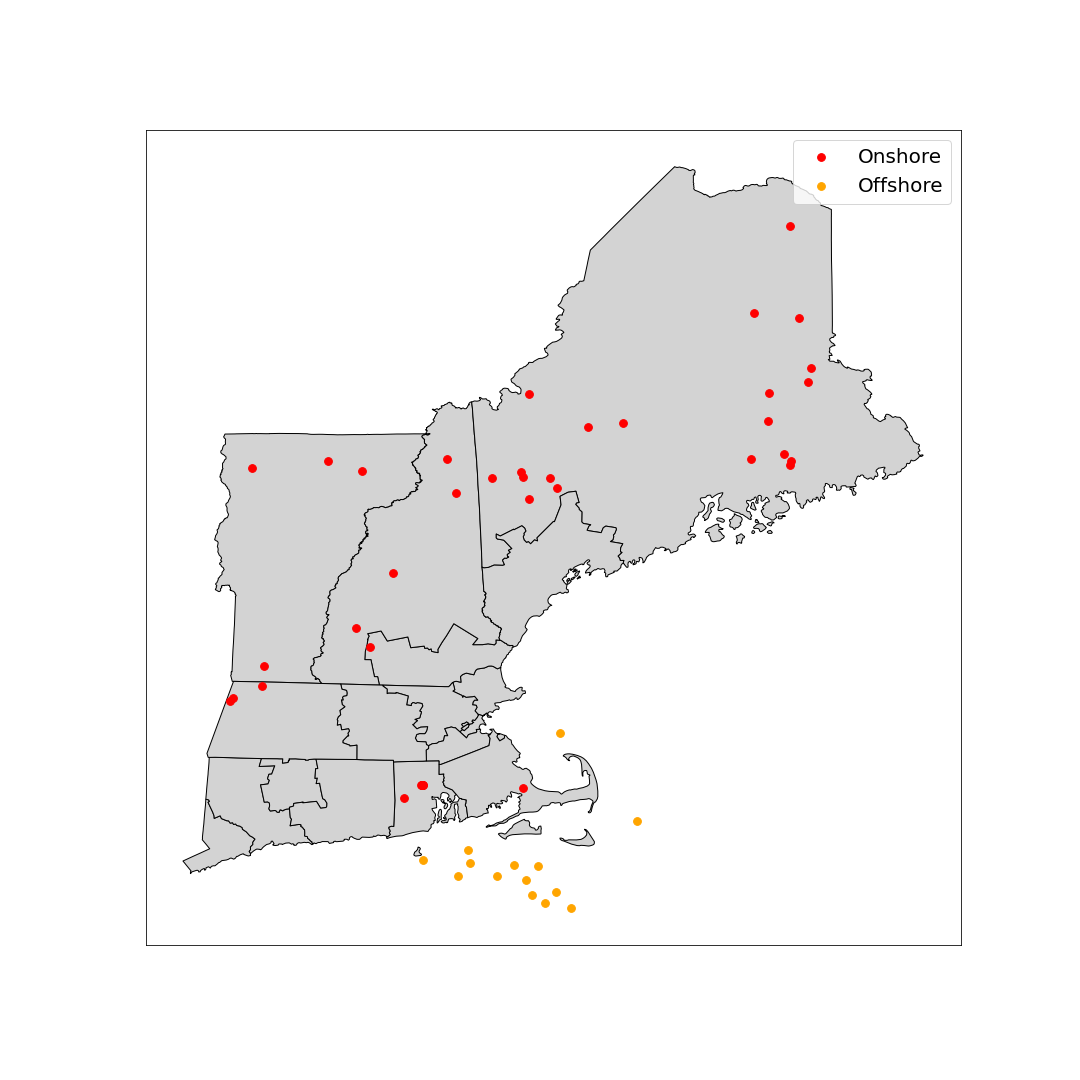}
    \caption{Map of onshore and offshore wind locations used to calculate zone-level capacity factors for weather years 2001-2020.}
    \label{SIfig:windmap}
\end{figure}

We calculate wind capacity factors for the 2001-2020 weather years using publicly-available wind speed data provided by ISO-NE \citep{ISONEW-2-yr-data}. The data is provided for a set of existing and hypothetical onshore and offshore plant locations, shown in Fig.~\ref{SIfig:windmap}.

For each site, we calculate wind capacity factors. We use power curves to convert wind speeds into power output. For our modeled wind plants, we assume an onshore hub height of 100 m and an offshore hub height of 150 m. The ISO-NE wind dataset includes wind speeds at differing heights for each location which often do not match these standard heights. We adjust the speeds according to power-law approximations of the wind profile. The typical exponent value used in the onshore wind modeling literature is 1/7. For offshore, we assume an exponent of 0.11 \citep{Hsu1994}. We note that these adjustments generally only change the speeds by 1-2\%. We calculate our capacity factors as in Brown and Botterud \citep{BrownBotterud2021}. However, because we lack pressure data at hub height, we assume the air density at hub height is approximately equal to the standard air density used in the power curves. The onshore power curve is borrowed from the ``Gamesa: G126/2500'' turbine used in Brown and Botterud and the offshore power curve is taken from \citep{MusialEtal2016}. Both curves are shown in Fig. \ref{SIfig:powercurves} below. In addition to the high-speed cutoffs shown in the power curves, we assume that the capacity factor for a given site is zero if the temperature at hub height is less than -30 C. To approximate hourly temperature at hub height, we use surface temperature data from the ERA5 dataset used in the demand-side analysis and then assume a decrease of $6.5^oC$ per 1000m of altitude gain, a lapse rate common in atmospheric science. Because the temperature data does not reflect the effects of climate change, which will likely result in warmer temperatures at hub height and less frequent cut-offs, this method is conservative. We uniformly reduce the capacity factors by 19\% to reflect other system losses and downtime in alignment with Brown and Botterud \citep{BrownBotterud2021}.

\begin{figure}
    \centering
    \includegraphics{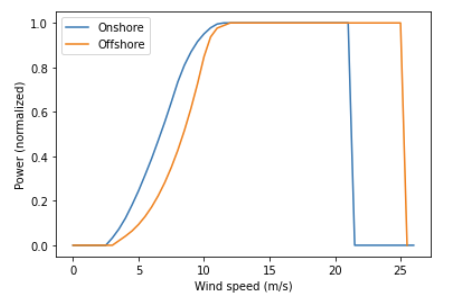}
    \caption{Power curve for onshore wind comes from data for the Gamesa:G126/2500 turbine as presented in Brown and Botterud \citep{BrownBotterud2021}. Power curve for offshore wind is taken from the NREL Reference 6MW offshore turbine model proposed in \citep{MusialEtal2016}}.
    \label{SIfig:powercurves}
\end{figure}

We assign capacity factors for onshore wind to each zone in a manner analogous to the solar capacity factors, either averaging the capacity factors of the sites located within the zone or taking data from the site closest to the zone. Offshore wind sites are assigned to each state based on the state associated with the site in the ISO-NE dataset, which generally corresponds to the states granted the relevant BOEM leases or, for hypothetical sites, the state closest to the site. We only calculate offshore wind capacity factors for Massachusetts and Rhode Island, as they are the only states with existing BOEM leases at the time of our analysis and therefore the most likely to have significant operating offshore generation in 2050. Each coastal zone within a given state is assumed to have the same offshore wind capacity factors. The non-coastal zones do not allow for offshore wind generation as a candidate resource.
Our average onshore wind capacity factors across the 20 years vary from 24.0\% to 49.5\% depending on the zone, with an average of 39.3\%. Our offshore wind capacity factors average 49.8\% across the states. The values are generally in alignment with expectations for these technologies \citep{WiserEtal2022,ATB2021}.

\section{Bottom-Up Building Energy Model}\label{SIsec:bottomup}

\subsection{ResStock Modeling}\label{SIsec:ResStock}
To generate the most granular level of load data in this study, we employ NREL’s ResStock tool \citep{ResstockComstock2022}. We use ResStock to produce two key outputs: heating-and-cooling-related thermal loads, which are the thermal energy required to maintain comfort in the living space, and non-heating-and-cooling-related fuel demands such as gas demand for water heating. At the time of conducting the analysis for this paper, the most recent release version of ResStock (2.5) did not support hourly heating-and-cooling-related thermal loads as a model output. Therefore, we used a clone of the develop branch on ResStock’s GitHub, which with more recent updates included these features. 

Our modeling relies on archetypes that collectively represent the residential building stock. Because each archetype represents a group of many identical homes that may receive varying upgrades, we define so-called \textit{sub-archetypes} to represent each of these variants. The \textit{baseline sub-archetype} is the original version of the archetype, with no upgrades. The \textit{electrified sub-archetypes} capture different possible mixes of heat pump sizing and envelope upgrades applied to the baseline sub-archetype (which we refer to as \textit{upgrade packages}). The electrified sub-archetypes are as follows:

\begin{enumerate}
    \item Summer-sized heat pump, no envelope upgrades
    \item Summer-sized heat pump, with envelope upgrades
    \item Winter-sized (``whole-home'') heat pump, no envelope upgrades
    \item Winter-sized (``whole-home'') heat pump, with envelope upgrades
\end{enumerate}

It should be noted that summer-sized heat pumps use the existing system as a backup heating source during the coldest periods. This is typically a hybrid heating system because most of the stock has existing fuel-based heating.

\subsubsection{Building Stock Turnover Modeling}\label{SIsec:stockprojections}
The input parameters of ResStock include housing stock data that is presented in the form of probability distributions for a range of various interdependent building characteristics. In order to reflect the 2050 housing stock, we alter the default ResStock distributions to approximate the projected housing stock for 2050. We generalize the projected housing stock changes in Massachusetts, described  in ``Building Sector Report of the Massachusetts 2050 Decarbonization Roadmap Study'' \citep{BuildingRoadmap}, to the entirety of  New England. When normalized for floor area, the Roadmap projects that the cohort of residential buildings constructed after the present day will be 23\% of the stock in 2050. We modify the ResStock probability distributions to reflect this growth by proportionally increasing the likelihood that the ResStock model samples an archetype of 2010s vintage, the most recent vintage available in ResStock. This implicitly increases the proportions of homes in our 2050 baseline stock that have newer construction characteristics, such as high-quality insulation. The proportions of older vintages in the remainder of the stock are assumed unchanged.  In addition to overall stock turnover, we also incorporate projected changes in building type – for example, the Roadmap projects increasing rates of construction for large multifamily buildings in the coming decades. {However, we do not include representation of routine building upgrades (e.g. appliance replacement, window replacement) that may occur in existing homes between the present day and 2050.}

\subsubsection{Generating thermal load data}\label{SIsec:thermalloads}
Thermal loads for heating and cooling in the baseline sub-archetypes are necessary inputs for our modeling of the electrified sub-archetypes. ResStock provides data for heating and cooling thermal loads if and only if heating and cooling systems are present, which is not the case for every baseline sub-archetype in New England. To retrieve these loads, we use ResStock to apply an upgrade to every baseline sub-archetype such that it has a heating and cooling system. In particular, we apply an upgrade of a typical single-speed air-source heat pump (in ResStock: ``HVAC Heating Efficiency$\vert$ASHP, SEER 15, 9.0 HSPF''). ResStock properly sizes the systems such that they meet thermal comfort requirements. Rather than outputting thermal loads, ResStock outputs the heating and cooling energy supplied by the system (\textit{delivered loads}). We accordingly assume that the delivered loads are roughly equivalent to the thermal loads. ResStock also provides the fuel demands corresponding to the new air-source heat pump upgrade, however, we opt to use a different method as the ResStock method largely relies on inefficient electric resistance backup heating. Instead, we develop a heat pump model that more accurately approximates industry standard sizing methods for cold-climate regions such as New England (details in Section \ref{SIsec:HPmodel}).

\subsubsection{Envelope improvements}\label{SIsec:envimprov}
Our scenarios consider envelope improvements that refer to all post-construction upgrades made to the building exterior to reduce heat loss and improve thermal efficiency. Examples of these activities include adding insulation and sealing air leaks. We use ResStock to generate thermal loads in the presence of basic envelope upgrades for both the baseline and electrified sub-archetypes. We approximately align these improvements with the improvements specified in the ``ECM2 – Medium Efficiency'' package outlined in the Building Sector Report of the Massachusetts 2050 Decarbonization Roadmap Study \citep{BuildingRoadmap}, which includes the lowest degree of envelope improvements considered in the report, and is thus what we assume to be a basic level of envelope retrofit. ResStock offers a discrete list of improvement options that do not necessarily match a level shown in the Roadmap. We show the comparisons in Table \ref{SItab:envimprov} .

\begin{table}[]
    \centering
\caption{Comparison of upgrades specified in the Massachusetts Decarbonization Roadmap versus the equivalents that we modeled in ResStock.}
    \begin{tabular}{l|c c}
    \toprule 
    \textbf{Upgrade} & \textbf{MA Decarbonization Roadmap} & \textbf{ResStock equivalent} \\
    \midrule 
    Roof/Ceiling insulation & R-60 & R-60 \\
Wall (fill) insulation & R-15 & R-13 \\
Sheathing insulation & N/A & R-5 \\
Rim joist insulation & N/A & R-13 \\
Foundation wall insulation & N/A & R-10 \\
Infiltration reduction & 0.4 CFM/sf at 0.3 in. wc. & 2.25 ACH50 \\
\bottomrule
    \end{tabular}
    \label{SItab:envimprov}
\end{table}

 The Roadmap only lists a generic ``wall insulation'' characteristic of R-15, while ResStock takes separate insulation values for wall fill, sheathing, rim joist, and foundation wall insulation. Hence, we select ResStock insulation values such that the combination of these components across the wall section will equal roughly R-15. A given improvement is only applied to a home if it is related to a characteristic the home actually has (e.g., a basement) and the existing option is less efficient. The Roadmap's ECM2 package also includes Energy Recovery Ventilation (ERV), which we do not model as it was not a capability of ResStock at the time we conducted our analysis.

\subsubsection{Water heating improvements}\label{SIsec:waterheater}
    In addition to an air-source heat pump, we assume that electrified sub-archetypes also receive a heat-pump water heater (HPWH). We model this in ResStock by selecting a 66-gallon HPWH with a Uniform Energy Factor (UEF) of 3.35. Beyond space and water heating, we do not consider the electrification of other residential end uses. However, space and water heating make up the vast majority of residential electrifiable energy demand \citep{RECS2015}.

\subsection{Heat pump model}\label{SIsec:HPmodel}
We use the thermal loads provided in ResStock as inputs to our modeling of heating and cooling demand for the electrified sub-archetypes. ResStock provides thermal loads for a single living space representing the entirety of the home. Therefore, our models assume the home will be heated by a single heat pump, although in reality some households may opt for multiple smaller heat pumps. 

\subsubsection{Data underlying heat pump model}\label{SIsec:NEEPdata}
Our ASHP model is based on applying statistical linear regression on the large ``Cold-Climate Air Source Heat Pump'' dataset from NEEP \citep{NEEP}. The NEEP dataset includes data for thousands of cold-climate heat pumps that are submitted by manufacturers and certified by AHRI (Air-Conditioning, Heating, and Refrigeration Institute). Key characteristics of heat pump performance and the associated fuel demands rely on the operating conditions. In our models, we primarily consider the effects of outdoor air temperature (i.e., ambient temperature). This is common in the literature \citep{WaiteModi2020,VaishnavFatimah2020}. The NEEP dataset provides capacity and efficiency values at multiple temperatures for all ASHPs listed in the dataset.
We preprocess the dataset to form the basis of our regression model. First, we filter for models with Heating Seasonal Performance Factor (HSPF) equal to 10, the level currently required to qualify for electrification incentives in Massachusetts \citep{MassSave}. In order to avoid extrapolating beyond the range of our regression data, we then filter for models for which the manufacturer provided data for performance below -15 $^o$C (5 $^o$F). We then remove duplicate data which leads to some heat pumps being over-represented in the dataset. {This results in our final model being based on a dataset of 161 different heat pump models}. We note that because some heat pumps in the dataset are likely to be more popular than others, this does not result in an aggregate of the most likely heat pump to be adopted, but rather an approximation of the “average” HSPF 10 heat pump on the market.

\subsubsection{Coefficient of Performance (COP) Model}\label{SIsec:COPmodel}
The hourly COP of the heat pump determines the ratio of the useful thermal energy supplied to (or removed from) the space to the electricity consumption in a given hour.

\begin{equation}
\text{COP}_{\text{heat}}=\frac{\vert \dot{Q}_{\text{out}} \vert}{\dot{E}_{\text{in}}}
\end{equation}

\begin{figure}
    \centering
    \includegraphics[width=.9\textwidth]{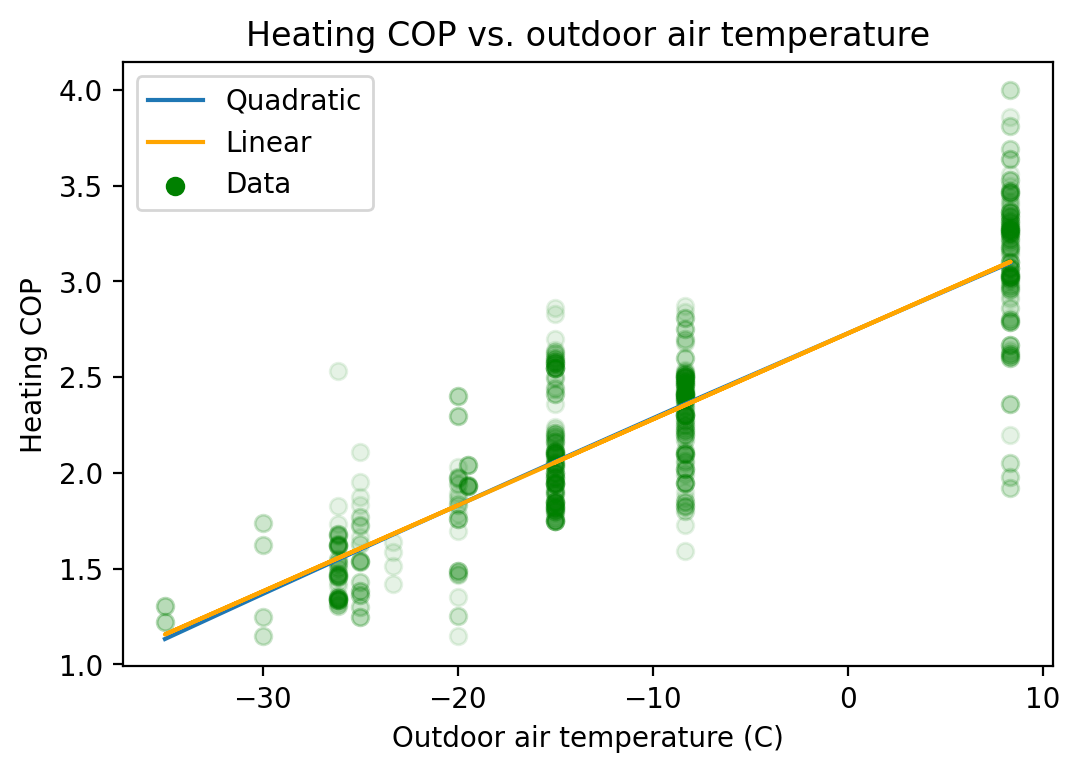}
    \caption{Regression models of the NEEP data for heating COP versus outdoor air temperature. We chose the linear model for our analysis. The model extrapolates beyond the range of the data displayed.}
    \label{SIfig:COPheat-vs-temp}
\end{figure}

\begin{figure}
    \centering
    \includegraphics[width=\textwidth]{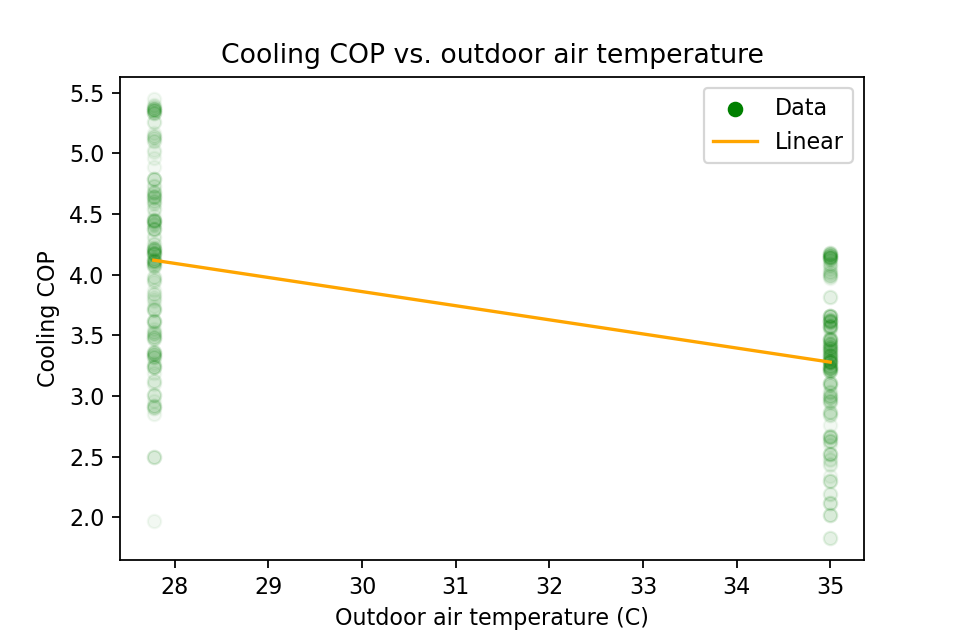}
    \caption{Regression models of the NEEP data for cooling COP versus outdoor air temperature. The model extrapolates beyond the range of the data displayed.}
    \label{SIfig:COPcool-vs-temp}
\end{figure}

The coefficient of performance (COP) for heat pumps in heating mode decreases in colder temperatures. Similarly, warmer temperatures adversely impact  the COP of heat pumps in cooling mode.  We compute the hourly COPs as a function of the hour’s outdoor air temperature. In addition to temperature, COP varies with the ``part-load ratio'' which is the amount of energy the heat pump supplies relative to its maximum capacity at the operating outdoor air temperature. To simplify the model, we do not consider part-load performance, an assumption made in similar studies \citep{VaishnavFatimah2020}.
The temperature-vs.-COP function is based on a least-squares linear regression of the COP and temperature values listed in the NEEP dataset. The NEEP dataset lists multiple COPs for a given temperature depending on the part-load ratio of the heat pump. We use the COP values for heat pump operation at maximum capacity. There are separate COP curves for heating and cooling:

\begin{equation}
\text{COP}_{\text{heat,h}}=f(T_h)=0.045T_h+2.73
\end{equation}

\begin{equation}
\text{COP}_{\text{cooling,h}}=f(T_h)=-0.116T_h+7.35
\end{equation}

where $T_h$ is the parameter for the hourly temperature in degrees Celsius. Our linear model for heating mode, as shown in Fig.~\ref{SIfig:COPheat-vs-temp}, is inherently a simplification of reality. According to performance testing, the dependence of COP on temperature is non-linear \citep{Shoukas2022}. For our heating COP dataset, linear and quadratic fits result in essentially equivalent curves, suggesting that additional model complexity would not necessarily result in better fit at the expense of tractability. Similarly, Fig.~\ref{SIfig:COPcool-vs-temp} shows the regression model for COP in cooling mode.

\subsubsection{Heating Capacity Derating Model}\label{SIsec:capmodel}
\begin{figure}
    \centering
    \includegraphics[width=0.9\textwidth]{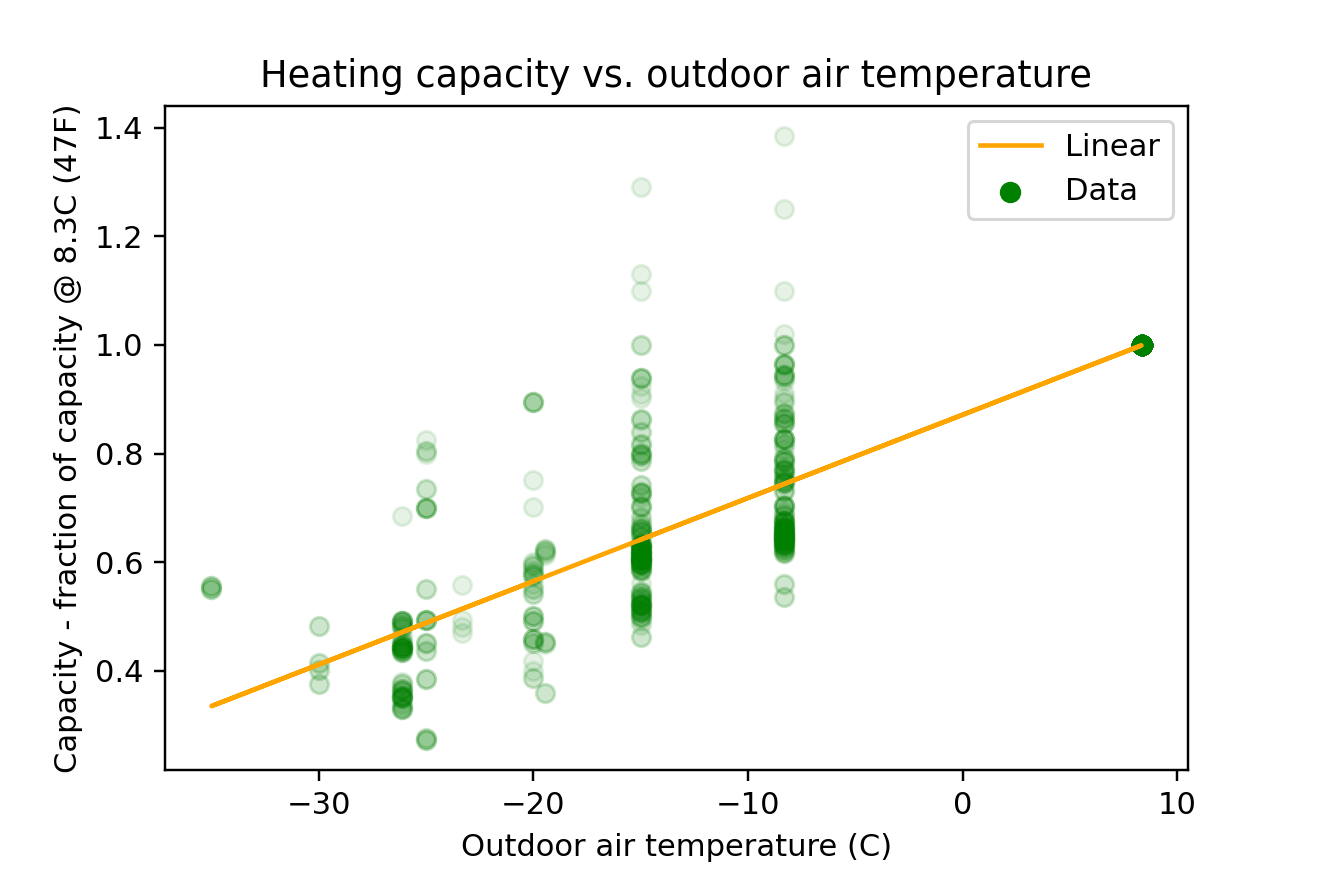}
    \caption{Regression model of the NEEP data for heating capacity versus outdoor air temperature. Because the heat pumps in our model can have different capacities, we normalize the curve to the capacity at 8.3$^o$C (47$^o$F). The model extrapolates beyond the range of the data displayed.}
    \label{SIfig:Capheat-vs-temp}
\end{figure}

The \textit{capacity} of a heat pump refers to the maximum heating or cooling output it can produce. Similar to COP, the capacity depends on temperature such that colder temperatures generally reduce the heating capacity of a heat pump relative to warmer temperatures, and vice versa. Colder temperatures typically imply higher heating loads within a building; because this coincides with reduced heating capacity in the heat pump, there is a temperature range for which heating supply cannot match heating load. The modeling of capacity derating enables us to identify hours in which the heat pump’s capacity is less than the heating load, hence backup heating may be necessary (discussed in more detail in Section \ref{SIsec:Auxheating}). Therefore, our model incorporates temperature-related capacity derating for heating.
In addition to COP values at each temperature, the NEEP dataset contains heating capacity values. For the ASHP models represented in the dataset, there are \textit{maximum} and \textit{minimum} capacity values provided for each temperature. Such a range of capacities can be present in a variable speed system, which has a compressor speed that can be modulated via controls, meaning there can be a range of capacities for a given temperature. We assume that for a given heat pump model represented in the dataset, the capacity value at a given temperature is the maximum value listed in the dataset. To obtain a model for the capacity derating, we take a regression of the capacity values across the dataset for all models, normalized compared to the capacity at 8.3 $^o$C (47 $^o$F) of each respective model. This enables us to obtain a slope that represents the average percentage loss of heating capacity for every degree Celsius drop in temperature, for all heat pumps represented in the dataset. 
The capacity value for a given hour, $C_h$, is calculated as

\begin{equation}
C_h(T_h)=(1-0.0153\cdot(T_{\text{sizing}}-T_h))\cdot C_{\text{sizing}}
\end{equation}
We first define a 
\textit{sized capacity}, $C_{sizing}$, the capacity at which the heat pump is sized at the \textit{sizing temperature}, $T_{sizing}$. The method for sizing is discussed in Section \ref{SIsec:sizing}, and differs depending on the sizing method. We assume the decrease in the capacity below $T_{sizing}$ is proportional to the slope obtained from the regression in our capacity derating model. 
Note that the logic also applies in reverse, e.g., if we assume that the heat pump is running at an outdoor air temperature 10 degrees above the sizing temperature, the heat pump has a capacity 15.3\% higher than the capacity at the sizing temperature. Similar to the COP model, our linear model for heating capacity abstracts away from potential non-linearities in the temperature dependence of heating capacity, which is empirically the case \citep{Shoukas2022}. Our model is depicted visually in Fig.~\ref{SIfig:Capheat-vs-temp}.

As a way to simplify our model, we do not consider capacity derating for cooling. The NEEP capacity data suggests that cooling capacity declines only modestly with increases in temperature. Additionally, given the dominance of heating loads over cooling loads in our study area, cooling capacity derating is less relevant to determining important hourly demand phenomena such as peak electricity loads.

\subsubsection{Sizing Model}\label{SIsec:sizing}
A key consideration in installing a heat pump is its \textit{size}, that is, the capacity. In our analysis, we define two possible sizing methods during an installation: sizing a smaller heat pump, primarily for cooling (summer sizing) and sizing a larger heat pump for heating the whole home (winter sizing). Both systems provide some amount of heating in the winter, with the winter-sized system intended to heat throughout the year. For each archetype, we size heat pumps for summer and winter approximately according to current and proposed \textit{ACCA S} methods, a set of industry-standard guidelines for HVAC sizing \citep{ACCAProposed2022}. As inputs to our sizing methods, we leverage data for the typical meteorological year (TMY) \citep{Wilcox2008} to produce archetype loads and temperatures from ResStock corresponding to typical weather conditions. We do this as part of an attempt to approximate the ACCA S sizing methods’ usage of long-run weather averages to determine the design conditions. Our sizing methods differ from ACCA S in that we use the delivered heating or cooling loads (see Section \ref{SIsec:thermalloads}) rather than so-called ``design loads,'' both of which are modeled outputs, but generated through different methods. We do this because ResStock did not provide design loads for winter-sized heat pumps as an output at the time of our analysis.\newline

\hspace*{-.65cm}\textbf{Winter sizing}

In the case of winter sizing, the size of the installed heat pump is primarily determined by the heating load. For each archetype, we first take the 99th percentile hourly heating load from the year of TMY data and “size” the heat pump such that it can provide this capacity $C_{sizing}$ at the sizing temperature, $T_{sizing}$. In the case of winter sizing, we use the 1st percentile temperature. Because the 99th percentile load often occurs at temperatures above the 1st percentile temperature, many of our archetypes effectively meet 100\% of the load across the typical meteorological year.
The cooling capacity is tied to the heating capacity. Generally, the nameplate heating and nameplate cooling capacity of a heat pump are similar, where the nameplate heating capacity is the capacity at 8.3 $^o$C (47 $^o$F) and the nameplate cooling capacity is the capacity at 35 $^o$C (95 $^o$F). Analysis of our dataset shows that these values are on average within about 4\% of one another. Therefore, we assume the heating and cooling nameplate capacities are equal. Because we do not model cooling capacity derating, we effectively set the temperature-invariant cooling capacity as equal to the derated capacity of the heating system at 8.3 $^o$C (47 $^o$F), that is, $C$(8.3$^o$C).
Because we only use a single capacity curve, our heat pump models and the associated sizing methods do not consider the potential issue of “short-cycling,” which is rapid on-off switching that may occur when the heating or cooling demand is below the heat pump's minimum capacity at the operating temperature. Although variable speed systems offer a range of capacities at any given temperature, short cycling may still occur, particularly in winter-sized systems, leading to humidity control issues that would generally be desired to be avoided in the sizing calculations. We assume the combination of our sizing methods and heat pump operation does not result in excessive short-cycling that would necessitate different installation configurations.\newline

\hspace*{-.65cm}\textbf{Summer sizing}

In the case of summer sizing, the heating capacity of the installed heat pump is tied to the summertime cooling load (kW). {For each archetype, we calculate the 99th percentile cooling load using the TMY weather data. We then size the cooling capacity of the heat pump to provide 1.3x this value (kW) at $T_{sizing}$ where $T_{sizing}$ is the 99th percentile temperature. We then set the nameplate heating capacity equal to the cooling capacity.} The 130\% factor approximates the ACCA S method for sizing a heat pump primarily used for cooling in a heating-dominated climate. Hence, we use it as the basis of our method for ''summer-sized'' systems.

\subsubsection{Backup heating for ''summer-sized'' systems}\label{SIsec:Auxheating}
In our primary modeling scenarios, we consider the usage of backup heating systems, particularly in the case of summer-sized systems that are not sized to meet heating loads in the winter. The possible configurations of a backup heating system are typically constrained by the nature of the existing heating system and the type and size of heat pumps installed. We refer to summer-sized systems for which the backup system is fueled by non-electric fuels as a ``hybrid'' system. We define our configurations based on those used in NREL’s End-Use Savings Shapes (EUSS) project \citep{EUSS2022}.\newline

\hspace*{-.65cm}\textbf{Configuration 1: Ducted home w/ existing backup}

For electrified sub-archetypes where the corresponding baseline sub-archetype has an existing ducted system, we assume the home will have installed a ducted heat pump downstream of the existing furnace or boiler that will serve as backup. The existing system may be fueled by gas, fuel oil, electricity, or such other fuels as wood or propane. In this case, the heat pump and the existing system cannot run at the same time. Installers often define a switchover temperature, $T_{switchover}$ below which the heat pump becomes relatively inefficient and has reduced capacity, where the heat pump is deactivated and the backup system meets the entire load. We define $T_{switchover}$ as $41^oF$ ($5^oC$) in summer-sized systems, in line with NREL’s assumption for an existing backup system in the EUSS study. The efficiency of the backup system is treated as temperature-invariant and is taken from the features of the corresponding baseline sub-archetype, which include the nominal efficiency of the existing system. Although true for only a small portion of homes in New England, the existing system may be electric resistance.\newline

\hspace*{-.65cm}\textbf{Configuration 2: Ductless home w/ existing backup}

For archetypes where the existing heating system is ductless, such as those with water-based heating distribution systems, the heat pump and existing system generally will not interfere with one another and can run simultaneously. In this case, we assume sensors and controls have been installed such that, when the heat pump is unable to meet the heating load, the backup system runs to make up the difference between the heating load and the heat pump’s capacity. This is likely to be necessary for summer-sized systems during the winter.\newline

\hspace*{-.65cm}\textbf{Configuration 2a: Ductless or Ducted home w/ electric backup}

While our scenarios in the main text assume all summer-sized heat pumps are configured with the existing sustem as backup, our model also includes logic to reflect newly-installed electric resistance backup heating, as evaluated in section \ref{SIsec:ElecBackup}. In this case, we assume the electric resistance coils have been installed downstream of the heat pump in a ducted system or separate from the existing system in a ductless system (e.g., by installing electric baseboard heating). The heat pump and backup system can run simultaneously as in Configuration 2. We treat the electric resistance backup as having a temperature-invariant efficiency of 100\%.

\subsection{Aggregation modeling}\label{SIsec:aggmethod}
\subsubsection{Aggregation method}
For each archetype, a weight $W$ is applied, equal to the number of homes the archetype represents in the residential stock. Additionally, each archetype in a given zone $z$ has the same weight. We determine the overall number of homes represented by a given archetype by dividing the projected number of homes $H_z$ in the zone by the number of archetypes in the zone, $A_z$:

\begin{equation}
    W_z = H_z/A_z
\end{equation}

$A_z$ is equal to roughly 400 for every zone in our analysis. 

For each archetype in this analysis, there is a single baseline sub-archetype and four electrified sub-archetypes considered, each with its own weight. Let $W_u$ correspond to the weight of a sub-archetype with a given upgrade package in the set of possible upgrade packages $U$. One such $u$ is the baseline home $b$, represented by weight $W_b$. The electrified sub-archetypes correspond to the remaining weights $W_u$:

\begin{enumerate}
    \item $W_s$: Summer-sized heat pump, no envelope upgrades
    \item $W_{se}$: Summer-sized heat pump, with envelope upgrades
    \item $W_w$:  Winter-sized heat pump, no envelope upgrades
    \item $W_{we}$:  Winter-sized heat pump, with envelope upgrades
\end{enumerate}

Let a given archetype in zone $z$ be archetype $i$, which is in the set of all archetypes in zone $z$, $I_z$. For a given archetype $i$, the sum of the weights of all its sub-archetypes is equal to the weight of the archetype, which is equal to the shared weight of all the archetypes in the zone:

\begin{equation}
W_i=\sum_{u} W_{i,u}= W_{i,b}+W_{i,s}+W_{i,se}+W_{i,w}+W_{i,we}= W_z \quad \forall \: i \in I_z
\end{equation}

The weights for the sub-archetypes can be altered for any given archetype to reflect the level of heat pump deployment and mixing of sizing methods in the portion of the stock represented by the archetype. 
The profiles for each sub-archetype consist of the heating-and-cooling-related and all other demands. The hourly profiles for each fuel type $f$ for zone $z$ are equal to the summation of the profiles for all sub-archetypes in the zone multiplied by their respective weights. If $L_{f,i,u,h}$ is the demand for fuel $f$ for archetype $i$ with upgrade package $u$ in hour $h$, then the hourly demand for fuel $f$ in zone $z$ is:

\begin{equation}
\sum_{i=1}^{n} W_{i,b} L_{f,i,b,h} + W_{i,s} L_{f,i,s,h} + W_{i,se} L_{f,i,se,h} + W_{i,w} L_{f,i,w,h} + W_{i,we} L_{f,i,we,h} = L_{f,h}
\end{equation}

\subsubsection{Heat pump deployment scenarios}\label{SIsec:deploymentscenarios}
We base our heat pump deployment scenarios on the Massachusetts Clean Energy and Climate Plan (CECP) for 2050, with our Medium Electrification scenarios corresponding to the CECP “Hybrid” scenario and the High Electrification scenarios corresponding to the CECP “High Electrification” scenario \citep{MA-climate-plan}. The CECP adoption projections include separate adoption rates for systems of different sizes and are listed in units of heating systems. On advice from the Massachusetts Executive Office of Energy \& Environmental Affairs, we assume the number of homes that have adopted heat pumps is equal to the number of heat pumps in the overall heating stock. The CECP scenarios do not include projections for adoptions of envelope improvement for 2050; hence, for our envelope upgrade scenarios, we assume that 70\% of all electrified homes receive envelope upgrades. 
The deployment scenarios manifest themselves as the weights. In our modeling, we assume all regions have levels of adoption in line with the CECP scenarios. We also assume the heat pump deployment scenarios are such that every archetype receives the same proportions of upgrade packages (i.e., proportions of electrified and baseline sub-archetypes), with the exception of archetypes for which the baseline sub-archetype already has a heat pump, which do not receive upgrades at all.  For example, if we assume 75\% of homes in Zone z get new winter-sized heat pumps with envelope improvements:

\begin{equation}
W_{i,we} = 0.75 W_i = 0.75 W_z
\end{equation}

If Wz is 400 homes, then 300 of the homes represented by archetype $i$ are represented by sub-archetype $i,we$. If there are 50 archetypes total for Zone $z$ that do not already have a heat pump in their baseline sub-archetype, then 300 homes represented by each archetype receive upgrade package $we$ and in total 15,000 homes in the zone receive upgrade package $we$.
Our reference scenario is taken from NREL’s Electrification Futures study’s “High Electrification – Moderate Technology Advancement” scenario \citep{EFSETA2022}. Similar to the CECP, it presents projections in terms of heating system stock numbers; however, it does not include information on sizing. As our most conservative case, we assume heat pumps deployed in this scenario are summer-sized (including existing backup) and that no envelope upgrades are applied. Table \ref{SItab:techscenarios} shows the scenarios in a table.

\begin{table}[htp]
\centering
\caption{Electrification scenarios in tabular form. Units in percent of total homes in the region.}
\begin{tabular}{l|lllll}
\toprule
      & RF & ME & MX & HE & HX \\
      \midrule
Summer-sized & 6.2     & 41.2     & 12.4  & 16.8   & 5.0     \\
Summer-sized + Eff Improv   & 0     & 0     & 28.8    & 0  & 11.8 \\
Winter-sized & 0 & 16.1 & 4.8 & 62.1 & 18.6 \\
Winter-sized + Eff Improv & 0 & 0 & 11.3 & 0 & 43.4 \\
\bottomrule
\end{tabular}
\label{SItab:techscenarios}
\end{table}

For a small number of archetypes in our projected 2050 stock, the baseline sub-archetype already has a heat pump modeled in ResStock, which by default fits under Configuration 2a (see Section \ref{SIsec:Auxheating}). We assume these homes contribute to the projections of adoption.  We lump these into the “summer-sized” category in Fig. 1 in the main text, although they would exhibit electricity demands in excess of the “winter-sized” systems due to high reliance on electric resistance heating.

\subsubsection{Population and household data}\label{SIsec:popdata}
Our analysis calls for projections of the number of homes in 2050. Generally, state-level projections for home growth through 2050 are unavailable. We assume the number of homes grows proportionally to households or population, depending on data availability. Massachusetts provides household count projections as part of its Decarbonization Roadmap \citep{BuildingRoadmap}. For other states, we generally source state-level population growth projections from state agencies \citep{PopNH,PopRI,PopCT,PopCooper}. We also assume that each county’s number of homes will grow at a rate equal to the state-wide projected growth. In some states, projections only extend to 2040. In these cases, we linearly extrapolate the growth through 2050. Depending on the zone, each archetype represents between 600 and 2,000 homes.

\subsubsection{Modeling of present-day demand}\label{SIsec:presentdaymodel}

In addition to the reference case, it is useful to have a current baseline against which to compare the hourly demands of the future scenarios. We refer to this case as the “present-day” case. Although historical aggregate demands such as annual demands are available for the residential sector \citep{EIA-residential-consumption}, residential hourly demands generally are not. Hence, we attempt to emulate these demands using the same workflow as the electrification scenarios. Similar to the future scenarios, we simulate 400 archetypes per zone, however, we use ResStock’s default housing dataset representative of the 2018 stock. We use weather data for 2001-2020 without any climate change adjustments applied. We determine our archetype weights using home count data for 2020. In this data, there is no modeling of electrification impacts. Similar to the demand modeling for 2050, where the 2001-2020 weather data is used to represent individual possible weather pattern realizations that are then adjusted for climate change, the present-day demand data can be seen as representative of demands for the 2020 baseline under a portfolio of 2001-2020 weather patterns. Implicitly, these results assume there are no climate change effects that would significantly impact weather patterns between 2001-2020. {We also note that these figures do not consider the energy demand impacts of the COVID-19 pandemic, as they are simulated values.}

\subsection{Demand-side weather and climate modeling}
\subsubsection{Weather data}\label{SIsec:weatherdata}
\begin{figure}
    \centering
    \includegraphics[width=0.5\textwidth]{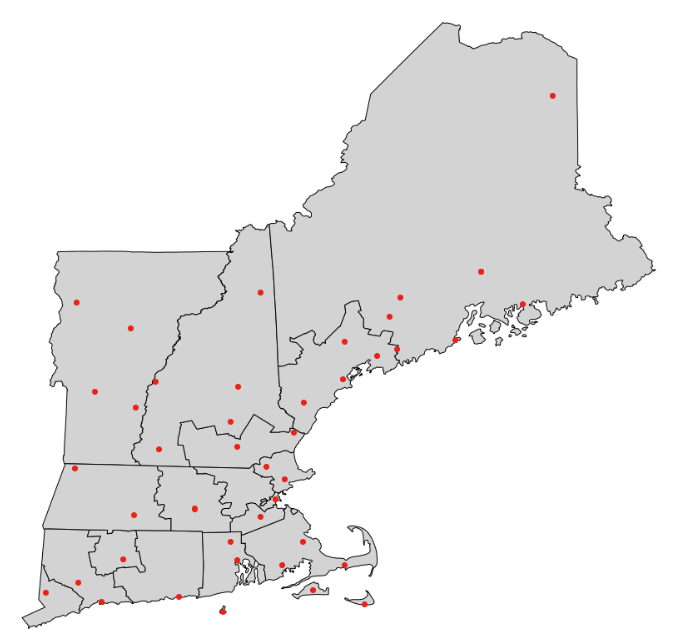}
    \caption{Map of locations of weather data used in the analysis.}
    \label{SIfig:weathermap}
\end{figure}

Hourly weather data is a key input to the bottom-up method. We collect hourly actual meteorological year (AMY) data for 2001-2020 in 44 locations across New England corresponding to the weather stations in the DOE’s TMY3 dataset, shown in Fig.~\ref{SIfig:weathermap}. Although the model can support higher levels of spatial granularity, {in this analysis we are limited to using weather data from the locations within the TMY3 dataset because our heat pump sizing calculations require TMY data as an input before we can run any building energy simulations using AMY data.} Each archetype is assigned the weather data closest to the county in which the archetype lies.

We source the TMY data used in this analysis from the DOE TMY3 dataset \citep{BianchiFontanini2021}. Our AMY data was provided by OikoLab, which furnish an API for more efficiently accessing large amounts of weather reanalysis data produced by the ERA5 project \citep{OikoLab}. The ERA5 data is also openly available on the ERA project website \citep{era5}. {We elaborate more on our rationale for selecting the ERA5 data below.}

\hspace{0cm}
\begin{flushleft}
{\underline{Selecting the ERA5 dataset}}
\end{flushleft}

{Our weather data needs for the bottom-up analysis are particularly constrained. We are looking for multi-decadal dataset at an hourly resolution, that also contained the several hourly weather parameters needed for building energy demands via the ResStock tool. In addition, we sought a dataset with high spatial granularity, which would be useful for capturing weather heterogeneity in a relatively small region like New England.}

\begin{figure}
    \centering
    \includegraphics[width=\textwidth]{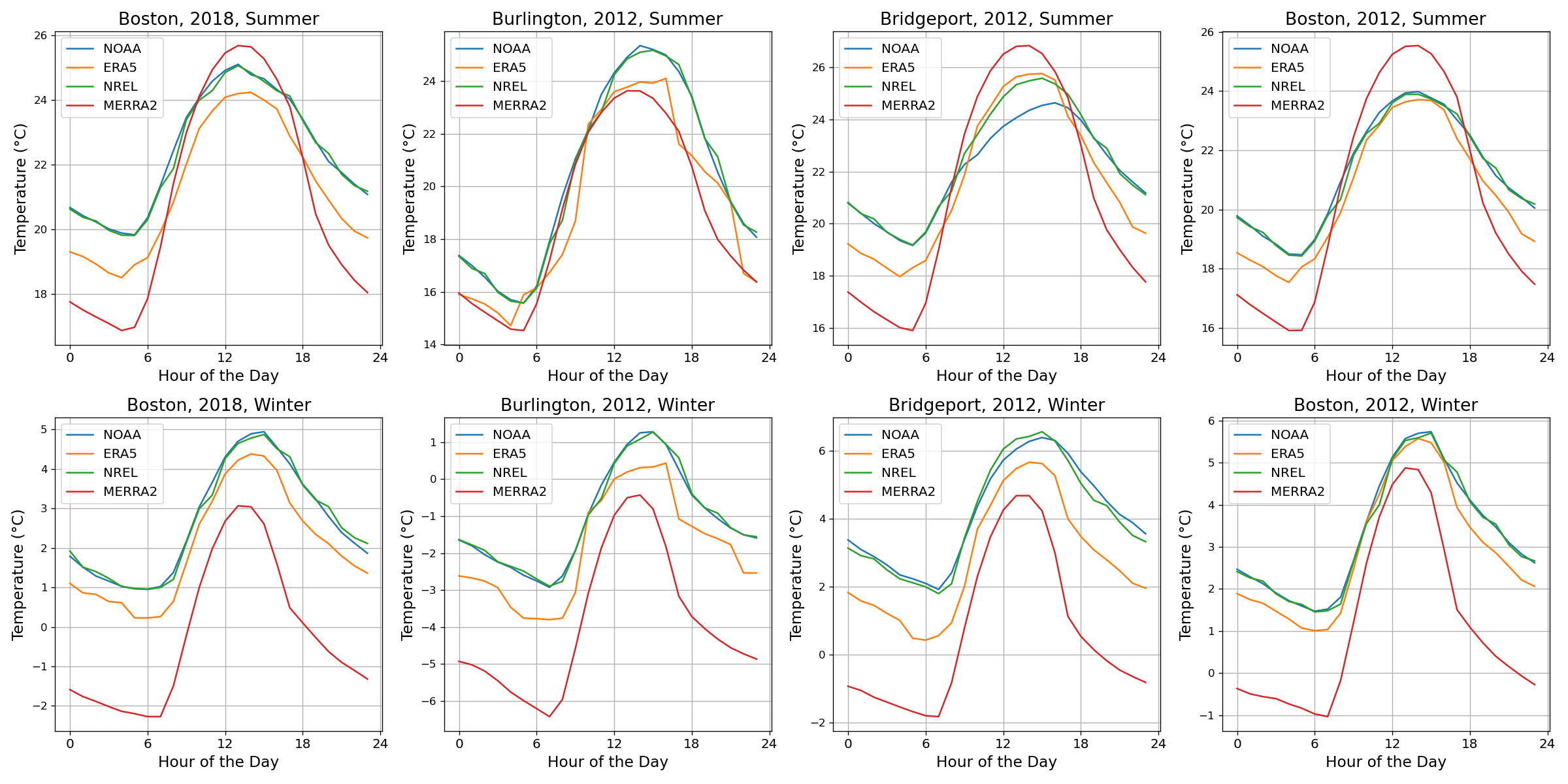}
    \caption{Comparison of different weather dataset temperature series against NOAA observational data for varying locations and periods. Profiles show average temperatures at each hour of the day for the season indicated. NOAA data from \citep{NOAAdata}, NREL data from \citep{ResstockComstock2022}, MERRA2 data from \citep{MERRA2017}}    
    \label{SIfig:weathercomparison}
\end{figure}

{We generally lack the expertise to assess the comparative merit or accuracy of different historical weather datasets based on the parameters alone, an area for which there is a rapidly growing literature among climate scientists 
\citep{Tarek2020,Gleixner2020,Graham2019}. However, because outdoor air temperature is the most impactful weather-related determinant of demand, we compared the temperature profiles among two datasets that we identified as meeting our constraints, the ERA5 \citep{era5} and MERRA2 \citep{MERRA2017} datasets. In Fig.~\ref{SIfig:weathercomparison}, we show comparisons of the two datasets as benchmarked against observation data provided by NOAA \citep{NOAAdata} for a handful of locations in different subregions of New England (namely: Boston, MA; Burlington, VT; and Bridgeport, CeiT). We also compare the datasets to the weather data used by NREL in the End Use Load Profiles project \citep{ResstockComstock2022}. To capture the differences across different time periods, we compare the accuracy of the average daily temperature profiles for the winter months (January, February, November, and December) and summer months (May, June, July, and August) in both 2012 and 2018.}

{Immediately noticeable is that the NREL weather data fits the NOAA data the best. This is because this dataset is constructed primarily from the NOAA data, with some interpolation applied for missing observations \citep{ResstockComstock2022}. However, the NREL data could not be a candidate dataset for our analysis because NREL has only published the data with limited features and for just two weather years. Among the ERA5 and MERRA2 datasets and across the locations, it can be seen that the ERA5 temperature profile fits significantly better, regardless of location or year, suggesting that ERA5 is the better data source for this particular study area. For the locations and seasons shown, the ERA5 data exhibits an average absolute temperature error of 0.4 to 1.3 C, whereas the MERRA2 data exhibits an error of 1.7 to 3.5 C.}\\

{The full set of temperature distributions for Boston across the 20 weather years used in our electrification scenarios (based on the morphed ERA5 data) are shown in Fig. \ref{SIfig:tempdists} below.}

\begin{figure}
    \centering
    \includegraphics[width=\textwidth]{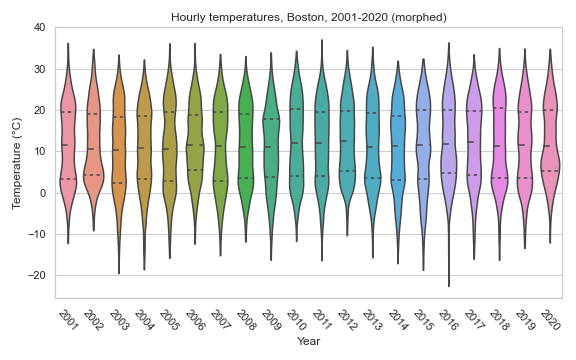}
    \caption{{Temperatures for Boston across the 20 weather years, including the morphing to approximate climate change in 2050. The width of the violin is proportional to the number of hours at the given temperature. The half-line represents the median, and the dashed lines represent the quartiles. The extremes indicate the maximum and minimum annual temperatures.}}
    \label{SIfig:tempdists}
\end{figure}

\subsection{Morphing method}\label{SIsec:morphingmethod}

The 2050 time horizon of our analysis necessitates consideration of climate change effects on demand patterns. Increases in temperatures and changes in other meteorological conditions will likely affect demand \citep{VanEtal2019,DirksEtal2015}. Building energy models like the EnergyPlus model underlying ResStock require several specific weather variables at hourly resolution. Climate models generally do not produce predictions at hourly resolution for these variables, an issue that is broadly recognized in the literature \citep{CraigEtal2022}, instead providing predictions at the daily or monthly resolution. The “morphing method,” first introduced in \citep{BelcherEtal2005}, is a particularly common method for overcoming this disconnect in the literature, combining hourly historical weather data with monthly climate models to produce realistic weather patterns that reflect the long-run effects of climate change \citep{JentschEtal2013,MachardEtal2020}. The morphing method offsets and scales the hourly values within each month to reflect the monthly average changes projected by the climate model. It also manipulates the hourly values to reflect changes in the monthly maxima and minima when they are available. We use the morphing method to apply the effects of climate change to our baseline 2001-2020 weather data. As described in \citep{BelcherEtal2005}, the selection of the morphing operation depends on the nature and units of the underlying variable. For example, it is appropriate to “shift” mean temperatures because they are in absolute units of C, but relative humidity, which is provided in percentages and cannot go below zero, is more amenable to a “scale” operation. We select the CESM2 model \citep{DanabasogluEtal2020} as the basis for the morphing because of its popularity, because its data is accessible through the CMIP6 project \citep{ONeillEtal2016}, and because it provides the necessary variables as outputs. We select the SSP3-70 emissions scenario as an approximation of medium-to-high warming. In determining the long-run changes from the climate model, we compare the average outputs of the model between the periods 2015-2023 and 2046-2054. As opposed to simply comparing the outputs for 2020 to outputs for 2050, taking averages of multi-year periods smooths out any changes that may be the result of the interannual weather variation simulated in CESM2 rather than long-run climatic changes. We note that because our baseline weather data forced through the morphing method is taken from any year 2001-2020, we inherently assume no differences among the 2001-2020 weather years due to the effects of climate change.

\begin{table}[htbp]
\centering
\caption{EnergyPlus/ResStock variable to CMIP6 standard variable mappings}
\label{table:mappings}
\begin{tabular}{|p{0.35\linewidth}|p{0.35\linewidth}|p{0.25\linewidth}|}
\hline
\textbf{EnergyPlus/ResStock variable} & \textbf{CMIP6 Standard variable name(s) used in morphing} & \textbf{Monthly morphing operation applied} \\ \hline
Dry bulb temperature & tas, tasmin, tasmax & ``Shift'' mean temperatures, ``stretch'' distance between max \& min temperatures \citep{BelcherEtal2005} \\ \hline
Wind speed & sfcWind & ``Stretch'' as in \citep{DickinsonBrannon2016} \\ \hline
Relative humidity & hurs & ``Stretch'' as in \citep{DickinsonBrannon2016} while keeping bounded between 0-100\% \\ \hline
Atmospheric pressure & ps & ``Shift'' as in \citep{BelcherEtal2005} \\ \hline
Total sky cover & clt & ``Shift'' as in \citep{BelcherEtal2005}, while keeping bounded between 0-100\% \\ \hline
Global horizontal radiation, diffuse horizontal radiation, direct normal radiation horizontal IR radiation & rsds & ``Stretch'' based on global horizontal radiation as in \citep{JentschEtal2008} due to projections only existing for this variable \\ \hline
\end{tabular}
\end{table}

A variable required by EnergyPlus but not provided in the CESM2 model is the dew point temperature. We approximate it from the morphed dry bulb temperature and relative humidity values using the MetPy package in Python \citep{MetPy2022}.

\bibliography{Bibliography2.bib}


\end{document}